\documentclass{aastex631}
\usepackage{graphicx}
\usepackage{txfonts}
\usepackage{xcolor}
\usepackage{float}
\usepackage{comment}
\usepackage{pgf}   
\usepackage{enumitem}
\usepackage{eurosym}
\def\degr{\hbox{$^\circ$\,}}
\def\arcmin{\hbox{$^\prime$}}
\def\arcsec{\hbox{$^{\prime\prime}$}}

\def\gsim{\mathrel{\hbox{\rlap{\lower.55ex \hbox {$\sim$}}
                   \kern-.3em \raise.4ex \hbox{$>$}}}}
\def\lsim{\mathrel{\hbox{\rlap{\lower.55ex \hbox {$\sim$}}
                   \kern-.3em \raise.4ex \hbox{$<$}}}}

\def\he2{\hbox{He\,{\sc ii} $\lambda$4686}}

\def\RL1{\hbox{{$R_{L_{1}}$}}}

\begin{document} 
   \title{The BlackGEM telescope array I: Overview\thanks{www.blackgem.org}}
       \author[0000-0002-4488-726X]{P.J.~Groot}
    \affiliation{Department of Astrophysics/IMAPP, Radboud University,
    P.O. Box 9010, 6500 GL, Nijmegen, The Netherlands}
    \affiliation{Department of Astronomy, University of Cape Town,
    Private Bag X3, Rondebosch, 7701, South Africa}
    \affiliation{ South African Astronomical Observatory, P.O. Box 9,
    Observatory, 7935, South Africa}
    \affiliation{The Inter-University Institute for Data Intensive
    Astronomy, University of Cape Town, Private Bag X3, Rondebosch,
    7701, South Africa}

    \author[0000-0002-6636-921X]{S.~Bloemen}
    \affiliation{Department of Astrophysics/IMAPP, Radboud University,
    P.O. Box 9010, 6500 GL, Nijmegen, The Netherlands}

    \author{P.M.~Vreeswijk}
    \affiliation{Department of Astrophysics/IMAPP, Radboud University,
    P.O. Box 9010, 6500 GL, Nijmegen, The Netherlands}

    \author[0000-0002-2626-2872]{J.C.J.~van~Roestel}
    \affiliation{Anton Pannekoek Institute for Astronomy, University of Amsterdam, P.O. Box 94249, 1090 GE Amsterdam, The Netherlands}  

    \author[0000-0001-5679-0695]{P.G.~Jonker}
    \affiliation{Department of Astrophysics/IMAPP, Radboud University,
    P.O. Box 9010, 6500 GL, Nijmegen, The Netherlands}

    \author[0000-0002-0752-2974]{G.~Nelemans}
    \affiliation{Department of Astrophysics/IMAPP, Radboud University,    P.O. Box 9010, 6500 GL, Nijmegen, The Netherlands}
    \affiliation{Instituut voor Sterrenkunde, KU Leuven, Celestijnenlaan 200D, 3001 Leuven, Belgium}
    \affiliation{SRON, Netherlands Institute for Space Research, Niels Bohrweg 4, 2333 CA, Leiden, The Netherlands}

    \author[0000-0001-7901-9545]{M.~Klein-Wolt}
    \affiliation{Department of Astrophysics/IMAPP, Radboud University,
    P.O. Box 9010, 6500 GL, Nijmegen, The Netherlands}

    \author{R.~Lepoole}
    \affiliation{Leiden Observatory, Leiden University, P.O. Box 9513,
    NL-2300 RA Leiden, The Netherlands}

    \author[0000-0003-3114-2733]{D.L.A.~Pieterse}
    \affiliation{Department of Astrophysics/IMAPP, Radboud University,
    P.O. Box 9010, 6500 GL, Nijmegen, The Netherlands}

    \author{M.~Rodenhuis}
    \affiliation{NOVA, Netherlands Research School for Astronomy,
    P.O. Box 9513, NL-2300 RA Leiden, The Netherlands}

    \author{W.~Boland}
    \affiliation{NOVA, Netherlands Research School for Astronomy,
    P.O. Box 9513, NL-2300 RA Leiden, The Netherlands}

    \author[0000-0002-5288-312X]{M.~Haverkorn}
    \affiliation{Department of Astrophysics/IMAPP, Radboud University, P.O. Box 9010, 6500 GL, Nijmegen, The Netherlands}

    \author[0000-0003-1822-7126]{C.~Aerts}
    \affiliation{Department of Astrophysics/IMAPP, Radboud University,
    P.O. Box 9010, 6500 GL, Nijmegen, The Netherlands}
    \affiliation{Instituut voor Sterrenkunde, KU Leuven,
    Celestijnenlaan 200D, 3001 Leuven, Belgium}
    \affiliation{Max Planck Institute for Astronomy, K\"onigstuhl 17,
    69117, Heidelberg, Germany}

    \author{R.~Bakker}
    \affiliation{TechnoCenter, Faculty of Science, Radboud University,
    P.O. Box 9010, 6500 GL, Nijmegen, The Netherlands}

    \author{H.~Balster}
    \affiliation{Department of Astrophysics/IMAPP, Radboud University,
    P.O. Box 9010, 6500 GL, Nijmegen, The Netherlands}

    \author{M.~Bekema}
    \affiliation{NOVA, Netherlands Research School for Astronomy,
    P.O. Box 9513, NL-2300 RA Leiden, The Netherlands}

    \author{E.~Dijkstra}
    \affiliation{NOVA, Netherlands Research School for Astronomy,
    P.O. Box 9513, NL-2300 RA Leiden, The Netherlands}

    \author{P.~Dolron}
    \affiliation{Department of Astrophysics/IMAPP, Radboud University,
    P.O. Box 9010, 6500 GL, Nijmegen, The Netherlands}
    \affiliation{TechnoCenter, Faculty of Science, Radboud University,
    P.O. Box 9010, 6500 GL, Nijmegen, The Netherlands}

    \author{E.~Elswijk}
    \affiliation{NOVA, Netherlands Research School for Astronomy,
    P.O. Box 9513, NL-2300 RA Leiden, The Netherlands}

    \author{A.~van~Elteren}
    \affiliation{Leiden Observatory, Leiden University, P.O. Box 9513,
    NL-2300 RA Leiden, The Netherlands}

    \author{A.~Engels}
    \affiliation{TechnoCenter, Faculty of Science, Radboud University,
    P.O. Box 9010, 6500 GL, Nijmegen, The Netherlands}

    \author{M.~Fokker}
    \affiliation{Department of Astrophysics/IMAPP, Radboud University,
    P.O. Box 9010, 6500 GL, Nijmegen, The Netherlands}

    \author{M.~de~Haan}
    \affiliation{NOVA, Netherlands Research School for Astronomy,
    P.O. Box 9513, NL-2300 RA Leiden, The Netherlands}

    \author{F.~Hahn}
    \affiliation{TechnoCenter, Faculty of Science, Radboud University,
    P.O. Box 9010, 6500 GL, Nijmegen, The Netherlands}

    \author{R.~ter~Horst}
    \affiliation{NOVA, Netherlands Research School for Astronomy,
    P.O. Box 9513, NL-2300 RA Leiden, The Netherlands}

    \author{D.~Lesman}
    \affiliation{NOVA, Netherlands Research School for Astronomy,
    P.O. Box 9513, NL-2300 RA Leiden, The Netherlands}
    
    \author{J.~Kragt}
    \affiliation{NOVA, Netherlands Research School for Astronomy,
    P.O. Box 9513, NL-2300 RA Leiden, The Netherlands}

    \author{J.~Morren}
    \affiliation{Instituut voor Sterrenkunde, KU Leuven,
    Celestijnenlaan 200D, 3001 Leuven, Belgium}

    \author{H. Nillissen}
    \affiliation{Department of Astrophysics/IMAPP, Radboud University,
    P.O. Box 9010, 6500 GL, Nijmegen, The Netherlands}

    \author{W.~Pessemier}
    \affiliation{Instituut voor Sterrenkunde, KU Leuven,
    Celestijnenlaan 200D, 3001 Leuven, Belgium}

    \author[0000-0002-1323-9788]{G.~Raskin}
    \affiliation{Instituut voor Sterrenkunde, KU Leuven,
    Celestijnenlaan 200D, 3001 Leuven, Belgium}
    
    \author{A.~de~Rijke}
    \affiliation{Center for Mathematics and Computer Science, CWI,
    P.O. Box 94079, 1090 GB Amsterdam,The Netherlands}

    \author{L.H.A.~Scheers}
    \affiliation{Center for Mathematics and Computer Science, CWI,
    P.O. Box 94079, 1090 GB Amsterdam,The Netherlands}

    \author{M.~Schuil}
    \affiliation{NOVA, Netherlands Research School for Astronomy,
    P.O. Box 9513, NL-2300 RA Leiden, The Netherlands}

    \author[0000-0003-0223-9368]{S.T.~Timmer}
    \affiliation{Department of Astrophysics/IMAPP, Radboud University,
    P.O. Box 9010, 6500 GL, Nijmegen, The Netherlands}


    \author{L.~Antunes Amaral}
\affiliation{Instituto de F\'{i}sica y Astronom\'{i}a, Universidad de Valpara\'{i}so, Gran Breta\~{n}a 1111, Playa Ancha, Valpara\'iso 2360102, Chile}

 \author{E.~Arancibia-Rojas}
 \affiliation{Instituto de F\'{i}sica y Astronom\'{i}a, Universidad de Valpara\'{i}so, Gran Breta\~{n}a 1111, Playa Ancha, Valpara\'iso 2360102, Chile}

    \author[0000-0001-7090-4898]{I.~Arcavi}
    \affiliation{School of Physics and Astronomy, Tel Aviv University, Tel Aviv 69978, Israel}    

    \author[0000-0003-0901-1606]{N.~Blagorodnova}
    \affiliation{Departament de F\'isica Qu\'antica i Astrof\'isica (FQA), Universitat de Barcelona (UB), Barcelona, Spain}
    \affiliation{Institut de Ci\'encies del Cosmos (ICCUB), Universitat de Barcelona (UB), Barcelona, Spain}
    \affiliation{Institut d'Estudis Espacials de Catalunya (IEEC), Barcelona, Spain}
     \affiliation{Department of Astrophysics/IMAPP, Radboud University,
    P.O. Box 9010, 6500 GL, Nijmegen, The Netherlands}

    \author[0009-0006-7543-1544]{S.~Biswas}
     \affiliation{Department of Astrophysics/IMAPP, Radboud University,
    P.O. Box 9010, 6500 GL, Nijmegen, The Netherlands}

    \author[0000-0001-8522-4983]{R.P.~Breton}
    \affiliation{Department of Physics and Astronomy, The University of Manchester, Oxford Road, Manchester, M13 9PL, United Kingdom}    

    \author[0009-0009-3947-3140]{H.~Dawson}
    \affiliation{Institut f\"ur Physik und Astronomie, Universit\"at Potsdam, Haus 28, Karl-Liebknecht-Str. 24/25, D-14476 Potsdam-Golm, Germany}

    \author[0000-0001-8460-1564]{P.~Dayal}
    \affiliation{Kapteyn Astronomical Institute, University of Groningen, PO Box 800, 9700 AV Groningen, The Netherlands}

    \author[0000-0003-2449-1329]{S.~De Wet}
    \affiliation{Department of Astronomy, University of Cape Town,
    Private Bag X3, Rondebosch, 7701, South Africa}
    \affiliation{ South African Astronomical Observatory, P.O. Box 9,
    Observatory, 7935, South Africa}
    \affiliation{The Inter-University Institute for Data Intensive
    Astronomy, University of Cape Town, Private Bag X3, Rondebosch,
    7701, South Africa}    

    \author[0000-0001-6662-0200]{C.~Duffy}
    \affiliation{Armagh Observatory \& Planetarium, College Hill,
    Armagh, BT61 9DG, Northern Ireland, UK}

    \author[0009-0007-8485-1281]{S.~Faris}
    \affiliation{School of Physics and Astronomy, Tel Aviv University,
    Tel Aviv 69978, Israel}

    \author[0000-0002-9113-7162]{M.~Fausnaugh}
    \affiliation{Department of Physics \& Astronomy, Texas Tech University, Box 41051, Lubbock, TX, USA, 79409-1051}    

    \author[0000-0002-3653-5598]{A.~Gal-Yam}
    \affiliation{Department of Particle Physics and Astrophysics, Weizmann Institute of Science, 234 Herzl St., Rehovot, 76100, Israel}

    \author[0000-0002-9487-6366]{S.~Geier}
    \affiliation{Institut f\"ur Physik und Astronomie, Universit\"at Potsdam, Haus 28, Karl-Liebknecht-Str. 24/25, D-14476 Potsdam-Golm, Germany}

    \author[0000-0002-5936-1156]{A.~Horesh}
    \affiliation{Racah Institute of Physics. The Hebrew University of Jerusalem. Jerusalem 91904, Israel}     

    \author[0000-0002-3054-4135]{C.~Johnston}
    \affiliation{Department of Astrophysics/IMAPP, Radboud University,
    P.O. Box 9010, 6500 GL, Nijmegen, The Netherlands}    
    \affiliation{Max-Planck-Institut f\"ur Astrophysik, Karl-Schwarzschild-Stra{\ss}e 1, 85741 Garching  bei M\"unchen, Germany}

    \author[0009-0009-7528-4347]{G.~Katusiime}
    \affiliation{Departament de Física Quántica i Astrofísica (FQA), Universitat de Barcelona (UB), Barcelona, Spain}
    \affiliation{Institut de Ciéncies del Cosmos (ICCUB), Universitat de Barcelona (UB), Barcelona, Spain}
    \affiliation{Institut d'Estudis Espacials de Catalunya (IEEC), Barcelona, Spain}    

    \author[0000-0003-0416-120X]{C.~Kelley}
    \affiliation{Department of Astrophysics/IMAPP, Radboud University,
    P.O. Box 9010, 6500 GL, Nijmegen, The Netherlands}

    \author[0000-0002-9878-1647]{A.~Kosakowski}
    \affiliation{Department of Physics \& Astronomy, Texas Tech
    University, Box 41051, Lubbock, TX, USA, 79409-1051}

    \author[0000-0002-6540-1484]{T.~Kupfer}  
    \affiliation{Hamburger Sternwarte, University of
    Hamburg, Gojenbergsweg 112, D-21029 Hamburg, Germany}

    \author[0000-0002-8597-0756]{G.~Leloudas}
    \affiliation{DTU Space, National Space Institute,
    Technical University of Denmark, Elektrovej 327, 2800 Kgs. Lyngby,
    Denmark}

    \author[0000-0001-7821-9369]{A.~Levan}
    \affiliation{Department of Astrophysics/IMAPP, Radboud University,
    P.O. Box 9010, 6500 GL, Nijmegen, The Netherlands}

    \author{D.~Modiano}
    \affiliation{Anton Pannekoek Institute for Astronomy, University
    of Amsterdam, P.O. Box 94249, 1090 GE Amsterdam, The Netherlands}

    \author[0000-0002-2261-0382]{O.~Mogawana}
    \affiliation{Department of Astronomy, University of Cape Town,
    Private Bag X3, Rondebosch, 7701, South Africa}
    \affiliation{ South African Astronomical Observatory, P.O. Box 9,
    Observatory, 7935, South Africa}
    \affiliation{The Inter-University Institute for Data Intensive
    Astronomy, University of Cape Town, Private Bag X3, Rondebosch,
    7701, South Africa}        

    \author[0000-0002-1872-5398]{J.~Munday}
    \affiliation{Department of Physics, Gibbet Hill Road, University
    of Warwick, Coventry CV4 7AL, United Kingdom}

    \author[0000-0003-1149-1741]{J.A.~Paice}
    \affiliation{Centre for Extragalactic Astronomy, Ogden Centre for
    Fundamental Physics - West, Department of Physics, Durham
    University, South Road, Durham DH1 3LE}

    \author[0000-0002-0537-3573]{F.~Patat}
    \affiliation{European Southern Observatory, Karl-Schwardschildstra{\ss}e 2, Garching-bei-Munchen, Germany}

    \author[0000-0003-4615-6556]{I.~Pelisoli}
        \affiliation{Department of Physics, Gibbet Hill Road, University
    of Warwick, Coventry CV4 7AL, United Kingdom}

    \author{G.~Ramsay}
    \affiliation{Armagh Observatory \& Planetarium, College Hill,
    Armagh, BT61 9DG, Northern Ireland, UK}

    \author[0000-0001-8470-0220]{P.T.~Ranaivomanana}
    \affiliation{Department of Astrophysics/IMAPP, Radboud University,
    P.O. Box 9010, 6500 GL, Nijmegen, The Netherlands}
    \affiliation{Instituut voor Sterrenkunde, KU Leuven,
    Celestijnenlaan 200D, 3001 Leuven, Belgium}

    \author[0000-0001-7518-1393]{R.~Ruiz-Carmona}
    \affiliation{Racah Institute of Physics. The Hebrew University of Jerusalem. Jerusalem 91904, Israel}         
    \affiliation{Gemini Observatory, NSF’s NOIRLab, Av. J. Cisternas 1500 N, 1720236 La Serena, Chile}            
    \author[0000-0001-6339-6768]{V.~Schaffenroth}
    \affiliation{Institut f\"ur Physik und Astronomie, Universit\"at Potsdam, Haus 28, Karl-Liebknecht-Str. 24/25, D-14476 Potsdam-Golm, Germany}
    \affiliation{Thüringer Landessternwarte Tautenburg, Sternwarte 5, D-07778 Tautenburg, Germany}

    \author[0000-0001-5387-7189]{S.~Scaringi}
    \affiliation{Centre for Extragalactic Astronomy, Ogden Centre for
    Fundamental Physics - West, Department of Physics, Durham
    University, South Road, Durham DH1 3LE}

    \author[0000-0002-3424-8528]{F.~Stoppa}
    \affiliation{Department of Astrophysics/IMAPP, Radboud University,
    P.O. Box 9010, 6500 GL, Nijmegen, The Netherlands}

    \author[0000-0001-6279-0552]{R.~Street}
    \affiliation{Las Cumbres Observatory, 6740 Corotna Drive, Suite 102, Goleta, CA 93117, USA }

    \author[0000-0001-6314-8131]{H. Tranin}
    \affiliation{Departament de Física Quántica i Astrofísica (FQA), Universitat de Barcelona (UB), Barcelona, Spain}
    \affiliation{Institut de Ciéncies del Cosmos (ICCUB), Universitat de Barcelona (UB), Barcelona, Spain}
    \affiliation{Institut d'Estudis Espacials de Catalunya (IEEC), Barcelona, Spain}        

    \author[0000-0002-6603-994X]{M.~Uzundag}
    \affiliation{Instituut voor Sterrenkunde, KU Leuven,
    Celestijnenlaan 200D, 3001 Leuven, Belgium}

    \author{S.~Valenti}
    \affiliation{Department of Physics and Astronomy, University of California, Davis, 1 Shields Avenue, Davis, CA 95616-5270, USA}            

    \author[0000-0002-0146-3096]{M.~Veresvarska}
    \affiliation{Centre for Extragalactic Astronomy, Ogden Centre for
    Fundamental Physics - West, Department of Physics, Durham
    University, South Road, Durham DH1 3LE}

    \author{M. Vu\u{c}kovi\'c}
    \affiliation{Instituto de F\'{i}sica y Astronom\'{i}a, Universidad de Valpara\'{i}so, Gran Breta\~{n}a 1111, Playa Ancha, Valpara\'iso 2360102, Chile}

    \author[0009-0004-1442-619X]{H.C.I.~Wichern}
    \affiliation{DTU Space, National Space Institute,
    Technical University of Denmark, Elektrovej 327, 2800 Kgs. Lyngby,
    Denmark}

    \author[0000-0002-3101-1808]{R.A.M.J.~Wijers}
        \affiliation{Anton Pannekoek Institute for Astronomy, University
    of Amsterdam, P.O. Box 94249, 1090 GE Amsterdam, The Netherlands}
    
    \author[0000-0002-3516-2152]{R.A.D.~Wijnands}
    \affiliation{Anton Pannekoek Institute for Astronomy, University
    of Amsterdam, P.O. Box 94249, 1090 GE Amsterdam, The Netherlands}

    \author[0000-0001-8985-2493]{E. Zimmerman}
    \affiliation{Department of Particle Physics and Astrophysics,
    Weizmann Institute of Science, 234 Herzl St., Rehovot, 76100,
    Israel}

   \date{Received ...; accepted ...}
  \begin{abstract}
      The main science aim of the BlackGEM array is to detect optical counterparts to gravitational wave mergers. Additionally, the array will perform a set of synoptic surveys to detect Local Universe transients and short time-scale variability in stars and binaries, as well as a six-filter all-sky survey down to $\sim$22$^{nd}$~mag.
   The BlackGEM Phase-I array consists of three optical wide-field unit telescopes. Each unit uses an f/5.5 modified Dall-Kirkham (Harmer-Wynne) design with a triplet corrector lens, and a 65~cm primary mirror, coupled with a 110Mpix CCD detector, that provides an instantaneous field-of-view of 2.7~square degrees, sampled at 0.564\arcsec/pixel. The total field-of-view for the array is 8.2 square degrees. Each telescope is equipped with a six-slot filter wheel containing an optimised Sloan set (BG-$u$, BG-$g$, BG-$r$, BG-$i$, BG-$z$) and a wider-band 440-720~nm (BG-$q$) filter. Each unit telescope is independent from the others. Cloud-based data processing is done in real time, and includes a transient-detection routine as well as a full-source optimal-photometry module. 
   BlackGEM has been installed at the ESO La Silla observatory as of October 2019. After a prolonged COVID-19 hiatus, science operations started on April 1, 2023 and will run for five years. Aside from its core scientific program, BlackGEM will give rise to a multitude of additional science cases in multi-color time-domain astronomy, to the benefit of a variety of topics in astrophysics, such as infant supernovae, luminous red novae, asteroseismology of post-main-sequence objects, (ultracompact) binary stars, and the relation between gravitational wave counterparts and other classes of transients.
   \end{abstract}
%

\section{Introduction \label{sec:intro}}
The direct detection of gravitational waves (GW) through the use of laser-interferometry has opened up a completely new window onto the Universe \citep{abbott16}. To fully utilise the scientific promise of this new domain, gravitational wave sources also need to be identified, where possible, in the electromagnetic (EM) window. The detection and subsequent follow-up of the binary neutron star merger GW\,170817 has shown the wealth of (astro)physics that can be determined from a combined GW+EM detection (e.g. \citealt{Abbott17a}, \citealt{Abbott17b}). 
Well before the detection of GW\,170817, it had become clear that sky localisation and sensitivity would be the key issues in a successful EM detection of a GW source, e.g. \cite{Nissanke13}. The localisation capabilities of ground-based laser-interferometers are ultimately limited by the speed of light and the size of the Earth. This results in localisation uncertainties (`error boxes') of tens to thousands of square degrees, depending on the signal-to-noise (SNR) in the GW band, the number of laser interferometers, the orientation of the source location and spin angles with respect to the (plane of) detectors on Earth (e.g. \citealt{Fairhurst11}, \citealt{Nissanke13}, \citealt{Abbott20}, \citealt{Petrov22}). Model analyses of expected EM signatures show that the optical-infrared regime is favorable for an EM detection due to the non-relativistic, semi-isotropic nature of the radiation emitted in a gamma ray burst (GRB) afterglow-like and kilonova-type event, on timescales of hours to days after the merger event itself (e.g. \citealt{Kasen13}, \citealt{Rosswog14a}, \citealt{Grossman14}, \citealt{Setzer23}, \citealt{Kawaguchi20}). Peak apparent magnitudes of kilonova events are expected to be in the range 17--23~mag for binary neutron star mergers up to a GW detection horizon of $d_{\rm hor}\sim<$200~Mpc for the current generation of laser interferometers for a binary neutron star merger, depending on a multitude of, so far poorly constrained, parameters such as the total ejected mass, the exact composition of the material and the observer's viewing angle to the event. Efforts to establish the counterparts to a set of GW events and limit the brightness of kilonova counterparts include \citet{Kasliwal20}, \citet{Andreoni20}, \citet{Gompertz20}, and \citet{DeWet21}.

\subsection{Science requirements}
The BlackGEM array aims to detect optical kilonova signatures of gravitational wave mergers in the kilohertz regime on a timescale of minutes to hours after the merger event. Coupling the expected error box sizes and brightness expectations, the following science requirements were formulated at the start of the project (2012): 
\begin{enumerate}[label={\sl $\bullet$SR}\arabic*:, leftmargin=1cm]
    \item The ability to detect an optical counterpart to a GW merger, down to a brightness of 23$^{rd}$ magnitude in New Moon conditions in a broad band filter; 
    \item within an error box of up to 100 square degrees;
    \item on a timescale of no more than 2 hours; 
    \item with a start of observations no later than 10 minutes after the GW alert.
\end{enumerate}

\subsection{Design requirements}
The science requirements led to the following design requirements: 
\begin{enumerate}[label={\sl $\bullet$DR}\arabic*:,leftmargin=1cm]
    \item A spatial resolution limited by the natural seeing on a high-quality site, nominally taken to be 1\arcsec, to take advantage of the trade-off between depth, aperture and spatial resolution in background-limited observations;
    \item A set of filters spanning the full optical window, including a broad-band filter optimized for throughput;
    \item A data-reduction pipeline capable of processing and analysing the acquired data on time scales initially set to 10 minutes to allow for rapid identification of counterpart candidates;  
    \item A robotic set-up for autonomous follow-up of a GW event.
\end{enumerate}

Requirement {\sl DR1} follows from the required depth of observations. In the sky-background limited regime, the achieved signal-to-noise ratio for point sources scales as the square of the telescope diameter ($D$) over the achieved image quality ("seeing", $\sigma$), i.e. ${SNR} \propto (D/\sigma)^2$, and therefore excellent natural seeing can be used to trade-off against mirror diameter to limit the size (and cost) of a facility. The most restrictive of the science requirements is the combination of wide-field and image quality/depth. An assessment of available facilities and off-the-shelf telescopes showed that no viable option was available in the southern hemisphere that could simultaneously accomplish all science requirements. This gap in facility coverage in the Southern Hemisphere and the long-term membership of the Netherlands in the European Southern Observatory (ESO) determined the choice for a Chilean site, in particular ESO La Silla. Since the start of the project, other Southern Hemisphere coverage gaps have been filled in by the BlackGEM prototype MeerLICHT and ATLAS-Sutherland in South Africa (and the ATLAS station in Chile), as well as the GOTO-South telescopes in Australia, effectively now giving 24/7 coverage of the southern skies (\citealt{Bloemen16}, \citealt{ATLAS18}, \citealt{GOTO22}).

Design requirement {\sl DR2} comes from the unknown/uncertain color of kilonovae, and the expected strong reddening of kilonovae in the first few hours-day, as simulated by e.g. \cite{Kasen13} and demonstrated by GW\,170817 (e.g. \citealt{Arcavi17}). Having a set of filters available will allow BlackGEM to both choose the best filter for an observation as well as determine optical colors for transients found in the error box, where fast reddening is expected to be a strong diagnostic to distinguish kilonovae from, for example, young supernovae and/or Galactic transients, in particular, dwarf novae outbursts (see e.g.  \citealt{jvr19}, \citealt{DeWet21}). 

Design requirement {\sl DR3} follows directly from {\sl SR3} and {\sl DR4} from {\sl SR4}. 
To achieve {\sl SR2}, an instantaneous field-of-view of $\geq$8 square degrees was needed, and this has been achieved by building a 3-unit array. The BlackGEM unit telescopes are labeled BG-2 ("Ruby"), BG-3 ("Opal"), and BG-4 ("Emerald"; see Sect.\ \ref{sec:domes})\footnote{Numbering starts at `2' as the BlackGEM prototype MeerLICHT, installed at SAAO Sutherland, South Africa, is number 1}. 

\section{Optomechanical Design \label{sec:optomechanical}}

The project has opted for an array of smaller unit telescopes with single-detector cameras over a monolithic telescope and/or mosaic detector approach. This was motivated by costs, as a unit telescope could be kept smaller, with less-demanding (slower) and smaller-sized optics, and the added possible benefit of series production, and the difficulty in mosaic cameras to properly align individual chips with each other, as well as to a, generally curved, focal plane, see e.g. the additional measures taken by the Zwicky Transient Facility (ZTF, \citealt{ztf19}) as described in \cite{Dekany20}. The choice was made for an equatorial mount over an alt-az mount to avoid the added complication of a derotator for the camera, as well as the fact that, in alt-az telescopes, the diffraction spikes of bright stars rotate with respect to the sky depending on the parallactic angle, potentially leading to a large number of spurious transients and/or variables \citep{ruxy16}. Smaller optics allow the use of a filter wheel, housing filters of reasonable size ($\sim$10~cm). After the optical design was made, it was realized that even an Atmospheric Dispersion Corrector (ADC) could be accommodated \citep{terhorst16}. 

\subsection{Optical design and manufacturing \label{sec:opticaldesign}}

\begin{figure*}[htb]
\includegraphics[width=14cm]{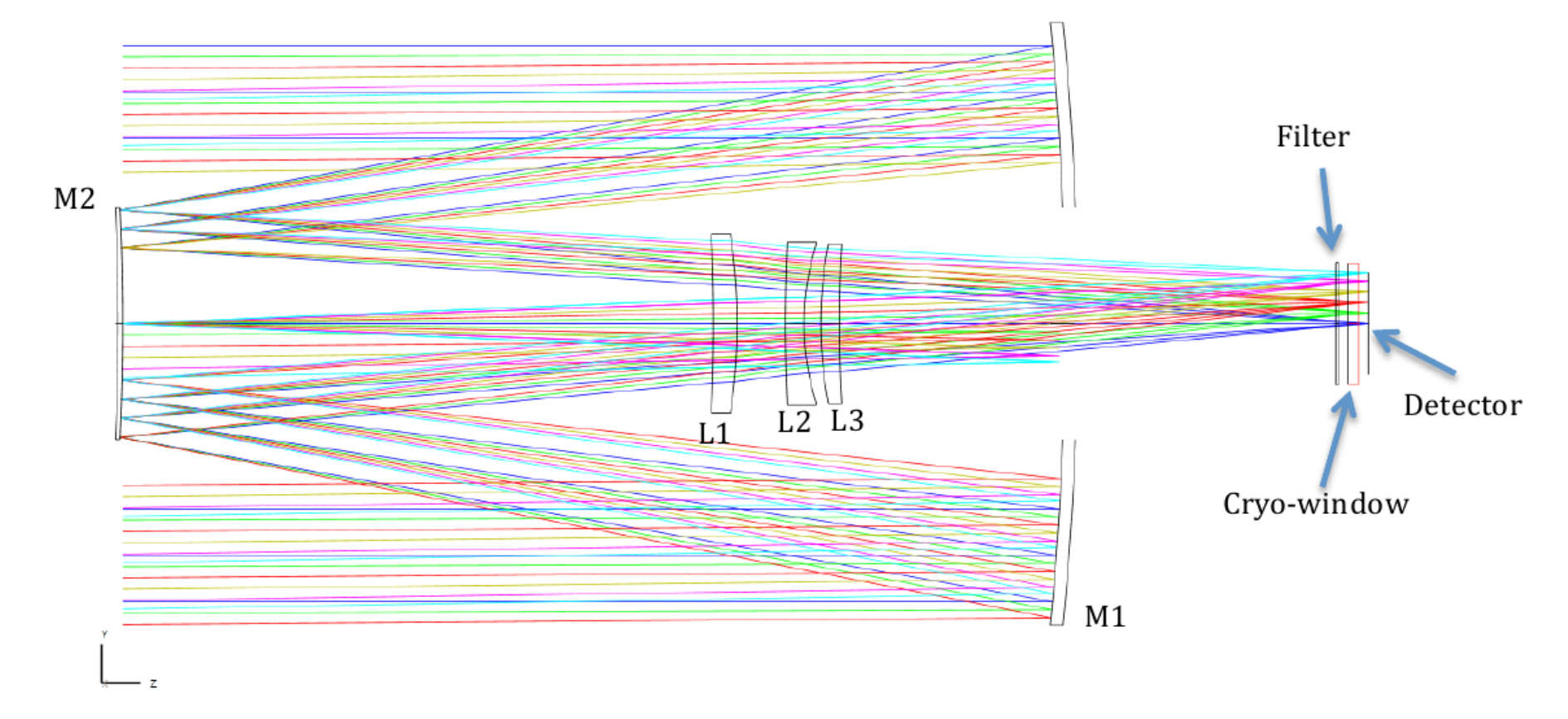}
\caption{Optical layout of the BlackGEM telescopes. Colored lines show principal and off-axis ray paths. Mirrors (`M') and lenses (`L') are indicated.  \label{fig:opticaldesign}}
\end{figure*}

The optical design of the BlackGEM unit telescope is that of a modified Dall-Kirkham (Harmer-Wynne; \citealt{HarmerWynne76}) design consisting of a parabolic primary mirror, a spherical convex secondary mirror and a triplet lens field-corrector (Figure\ \ref{fig:opticaldesign}. After the triplet lens, which doubles as the ADC, light passes through a filter and the entrance window to the cryostat to fall onto the detector. The parameters of the optical elements are given in Table\ \ref{tab:opticaldesign}. The focal ratio of the combined system is F/5.5, resulting in a focal length of $F_z$\,=\,3300~mm, with the $z$-axis taken along the optical path. This results in a plate scale at the focal plane of 16~$\mu$m/arcsecond.  Eighty per cent of the enclosed energy lies within a 9 micron radius even at 1.1$\degr$ away from the optical axis, and the optical design is close to diffraction-limited at all angles away from the optical axis. The focal plane is flat to within 5~$\mu$m over a radius of 150 mm from the optical axis. At the focal plane, the lateral displacement with wavelength is less than 5~$\mu$m.

Chromatic focus variations are within $\Delta z < 10~\mu$m over the full focal-plane, and vignetting is limited to $<10\%$ at the outer radius of the focal plane. The contribution of the optics to the point spread function is $<0.3\arcsec$ over the full focal plane. 

\begin{table*}
\caption{Optical design of the BlackGEM telescopes. Mirrors are referred to as `M', lenses with `L', and surfaces on lenses with `S', followed by a numerical label indicating their order with respect to the (incoming) light path. \label{tab:opticaldesign}}
\begin{minipage}{16cm}
\begin{tabular}{lrrll}
 & Radius & Thickness/{\sl distance\,}\footnote{distances from the reference plane are given in italics} & Material & Diameter\\
  & (mm) & (mm) & & (mm)\\
  \hline
  M1 & --3411.50 & {\sl --1010.90} & Fused Silica & 650 \\
  M2 & --2956.00 & {\sl 634.00} & Fused Silica & 260\\
  L1,S1 & --2201.48 & 25.00 & BK7 & 200\\
  L1,S2 & --678.55 & {\sl 52.47} & - & - \\
  L2,S1 & 1999.58 & 20.00 & BK7 & 190\\
  L2,S2 & 302.68 & {\sl 17.67} & - & - \\
  L3,S1 & 466.73 & 20.00 & BK7 & 180 \\
  L3,S2 (ADC) & 1669.17 & {\sl 532.38} & - & - \\
  Filter, S1 & infinity & 3 & Fused Silica & 119x119\\
  Filter, S2 & infinity & {\sl 10} & - & -\\
  Cryo-window, S1 & infinity & 12 & Fused Silica & 160x160\\
  Cryo-window, S2 & infinity & {\sl 10} & - & -\\
  Detector & infinity & - & - & 100x100\\
  \hline
\end{tabular}
\end{minipage}
\end{table*}

The fused-silica parabolic primary mirror (M1) is 65~cm in diameter, oversized with respect to the 60~mm optical aperture, to correct for the fact that the pupil is located on the secondary mirror, M2. The primary mirror was shaped by Lockwood (Illinois, USA). The primary mirrors show a peak-to-valley (PV) rms of $\lambda$/5 (BG-2), $\lambda$/2 (BG-3), and $\lambda$/3 (BG-4)  at a reference wavelength of $\lambda$\,=\,650~nm.  The lower accuracy value of BG3 is due to residual astigmatism that proved to be present in the primary mirror shape after production. This has been partially corrected by a tensioning clamp. 

The fused-silica spherical secondary mirror was made at Schott, Switzerland, and then polished in-house at the Netherlands Research School for Astronomy (NOVA\footnote{www.astronomie.nl}) Optical-Infrared group. Performance tests show a root-mean-square variation of 58~nm,  $\lambda$/9 wavefront error, at the reference wavelength of $\lambda$\,=\,500~nm.

The triplet lens field corrector was manufactured at FineOptix, Germany. Anti-reflection coatings are put on all transmission optics to suppress ghosting, required by the combination of the desire to detect faint (23$^{rd}$ mag) transients and a wide field-of-view which will always contain stars of $<$10$^{th}$ magnitude that can cause a significant reflected-light background.

To obtain maximum throughput over the full wavelength range (0.3 - 1.0~$\mu$m), both the primary and secondary mirrors were initially coated with UV-enhanced protected silver coatings with very high reflectivity ($>90\%$) at $\lambda>350$~nm. 
However, during commissioning, this coating very quickly tarnished on all mirrors. In collaboration with NOIRLab\footnote{www.noirlab.edu} and ESO, the original protected silver coatings on both M1 and M2 were replaced with a protected silver coating on M1 and a bare aluminium coating on M2.  


The BlackGEM telescope uses an 
ADC as part of the field-flattening optics \citep{terhorst16}. The third lens in the triplet lens field-corrector can be moved in the plane perpendicular to the light path by two rotating notch actuators. Due to its plano-concave shape, this lens thereby acts as a prism when it is off-centred, which, at the right position, gives a full-field first-order atmospheric dispersion correction. This allows the BlackGEM telescopes to remain seeing-limited in blue and broad-band filter observations up to zenith distances of 60\degr. 

\subsection{Filters \label{sec:filters}}

The BlackGEM filter set is similar to the Sloan set ($u,g,r,i,z$; \citealt{Doi10}) with the addition of a wider-band, 440-720~nm, $q$-band filter. The filter response curves were optimized for the overall throughput and steepness of the flanks. The change-over from 1\%--99\% transmission occurs within 10~\AA\ for all filters on both sides of their wavelength ranges. This sharp change-over was chosen to minimise the effect of back-reflection of light off of the filters, thereby suppressing the halos of saturated stars as well as the contribution to the background by scattered light. With the BlackGEM field-of-view, it is unavoidable to have bright stars ($g<10$) in the field, whose scattered light can significantly contribute to the background level if the optics are not properly designed and manufactured. The $g_{BG}$ and $r_{BG}$ filters were purposefully separated by 13~nm, such that the strong sky emission line of [O{\sc i}]~{557.7~nm} falls in neither of the two filters. 

\begin{table}
\caption{Passband of the BlackGEM filters. Start and end wavelengths are defined as the wavelengths where 1\% transmission is crossed.  \label{tab:filterssum}}
\begin{tabular}{lrr}
Filter & Start Wavelength & End Wavelength \\
 & (nm) & (nm)\\ \hline
$u_{BG}$ & 346 & 410\\
$g_{BG}$ & 409 & 551 \\
$q_{BG}$ & 437 & 722 \\
$r_{BG}$ & 559 & 690 \\
$i_{BG}$ & 687 & 843 \\
$z_{BG}$& 837 & 994 \\
\end{tabular}
\end{table}

The BlackGEM filter set was designed and made at Astrodon Inc (USA), now part of Optical Structures Inc. 
An overview of the measured transmission curves is given in
Table\ \ref{tab:filterssum}, Figure\ \ref{fig:filtertrans} and the
5~\AA-step throughput values are given in Table\ \ref{tab:filters} in the Appendix\ \ref{app:filterthroughput}. 

\begin{figure}[htb]
\includegraphics[width=8cm]{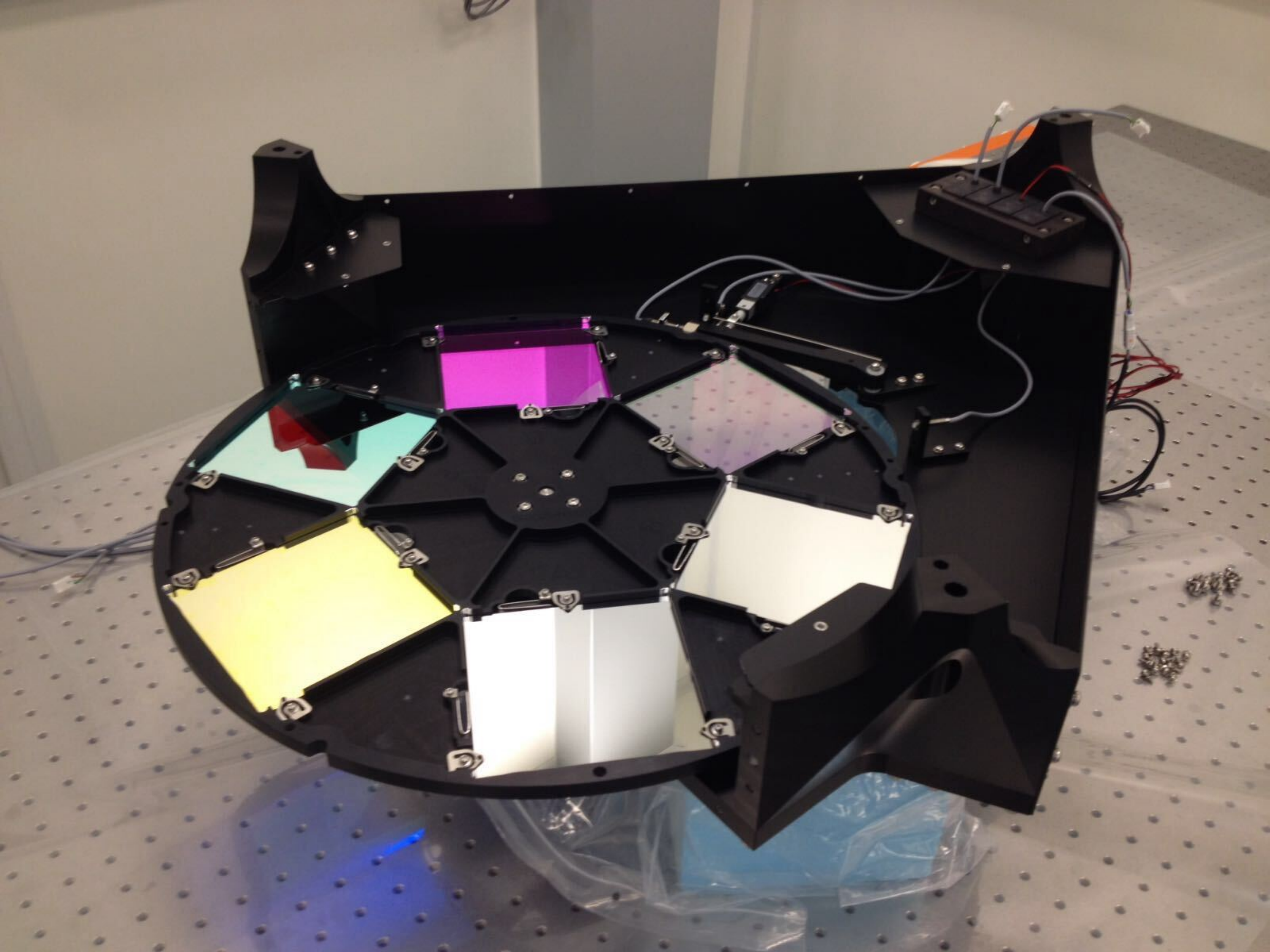}
\caption{Photo of the opened filter wheel during assembly.\label{fig:filterwheel}}
\end{figure}

The filters are housed in a six-slot filter wheel, which can be rotated in both directions and is located in the camera house, which is between the lens barrel and the entrance window to the cryostat (Figure\ \ref{fig:filterwheel}). Filter changes take up to $\sim$3~seconds between the extreme positions. 

\begin{figure}[htb]
\includegraphics[width=10cm]{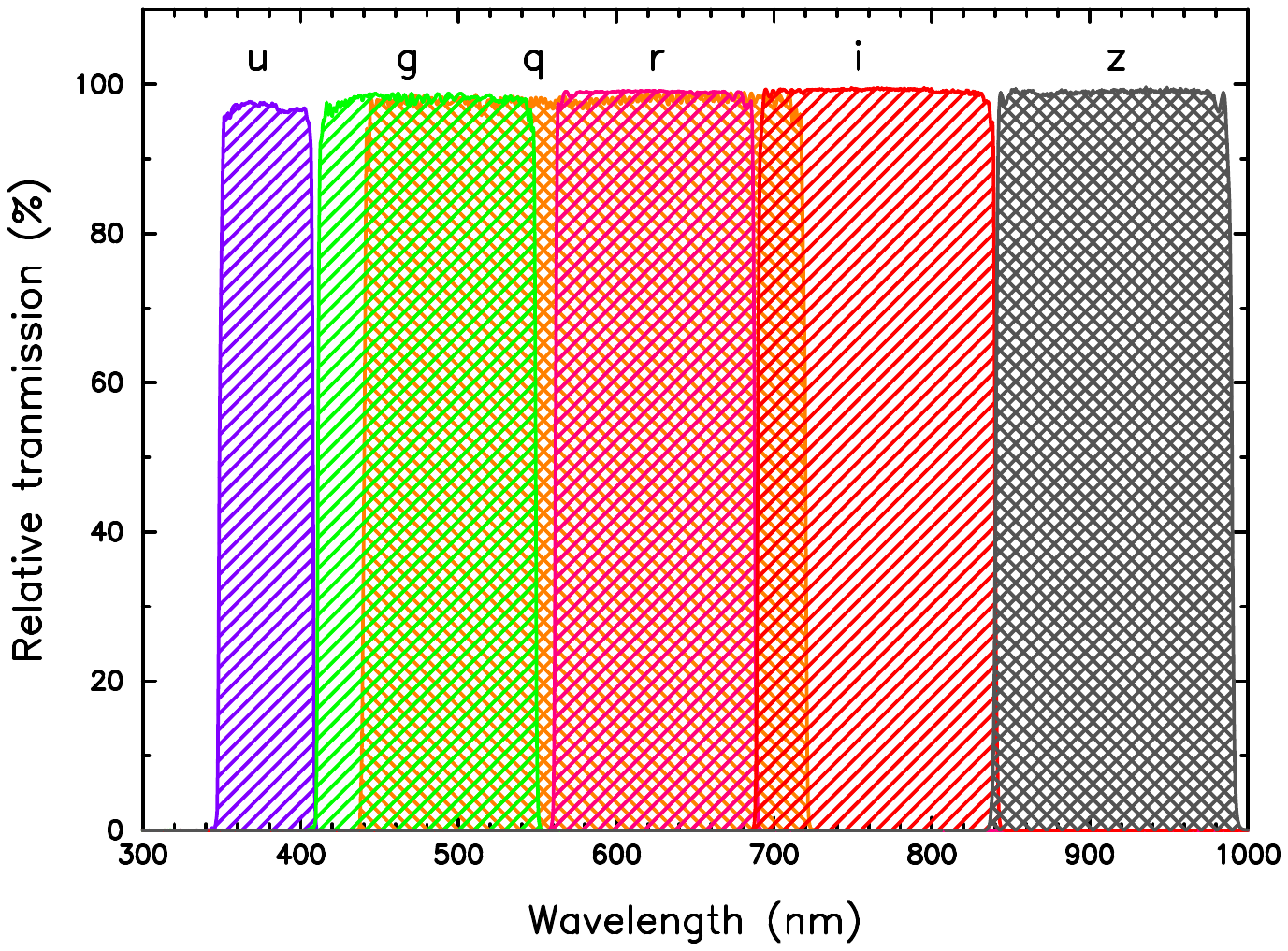}
\caption[]{Transmission curves of the BlackGEM filters.  Purple is the $u$-band, green the $g$-band, orange the $q$-band, red the $r$-band, light-red the $i$-band and in grey the $z$-band. \label{fig:filtertrans}}
\end{figure}

\subsection{Detector \& Cryostat}
\begin{figure}[htb]
\includegraphics[width=9cm]{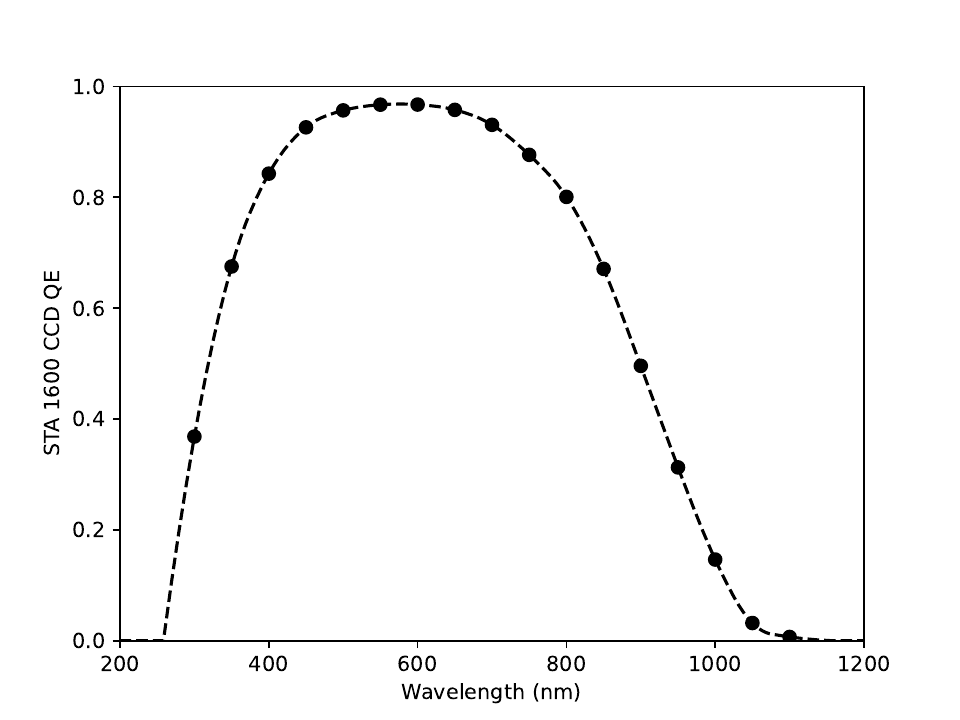}
\caption{Quantum efficiency as a function of wavelength for the STA\,1600 detector used by the BlackGEM array. Black dots indicate measured wavelengths, and the dashed line our interpolation. \label{fig:ccdqe}}
\end{figure}

Each BlackGEM unit telescope uses a single STA1600, 10\,560 $\times$10\,560 pixel, CCD detector with a square pixel size of 9~$\mu$m $\times$ 9~$\mu$m. The detectors were produced at Semiconductor Technology Associates  (USA) and are equipped with a broadband coating. The quantum efficiency curve is shown in Figure\ \ref{fig:ccdqe}. All BlackGEM detectors have a very low number of bad pixels and columns, $\ll1\%$, and a few minor blemishes with lower quantum efficiency (QE). 

Each detector is read out through an STA Archon controller via 16 ports over four video boards at 1~MHz read-out speed. The set-up results in a 7s read-out time and a read-out-noise of 7--8~electrons. Some cross-talk is present over the channels, which is corrected for in the data reduction process. 

\begin{figure}[htb]
\includegraphics[width=8cm]{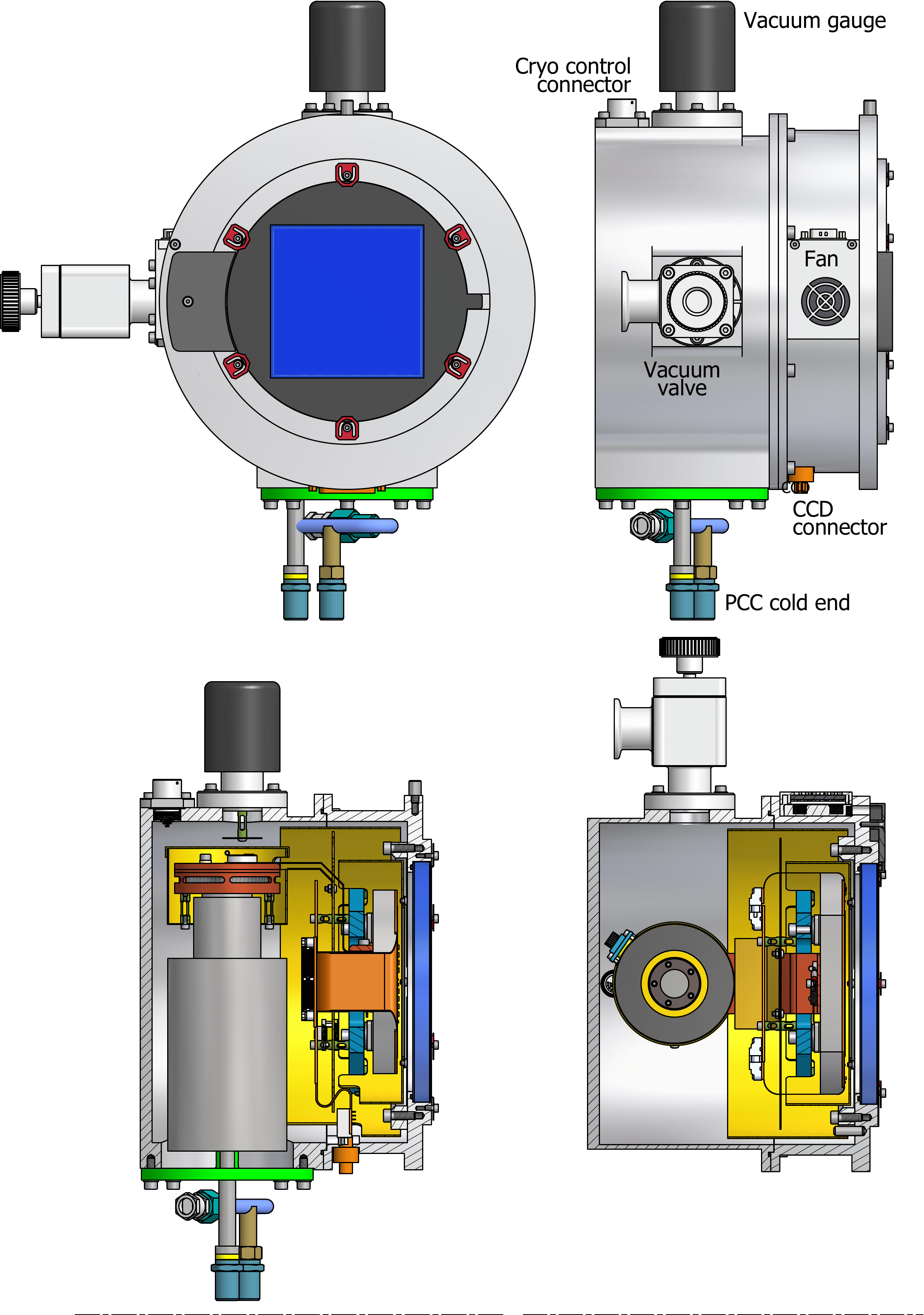}
\caption{Cross-cut through the design of the BlackGEM cryostat for the STA\,1600 CCD detector. \label{fig:cryo}}
\end{figure}

\begin{figure}[htb]
\includegraphics[width=9cm]{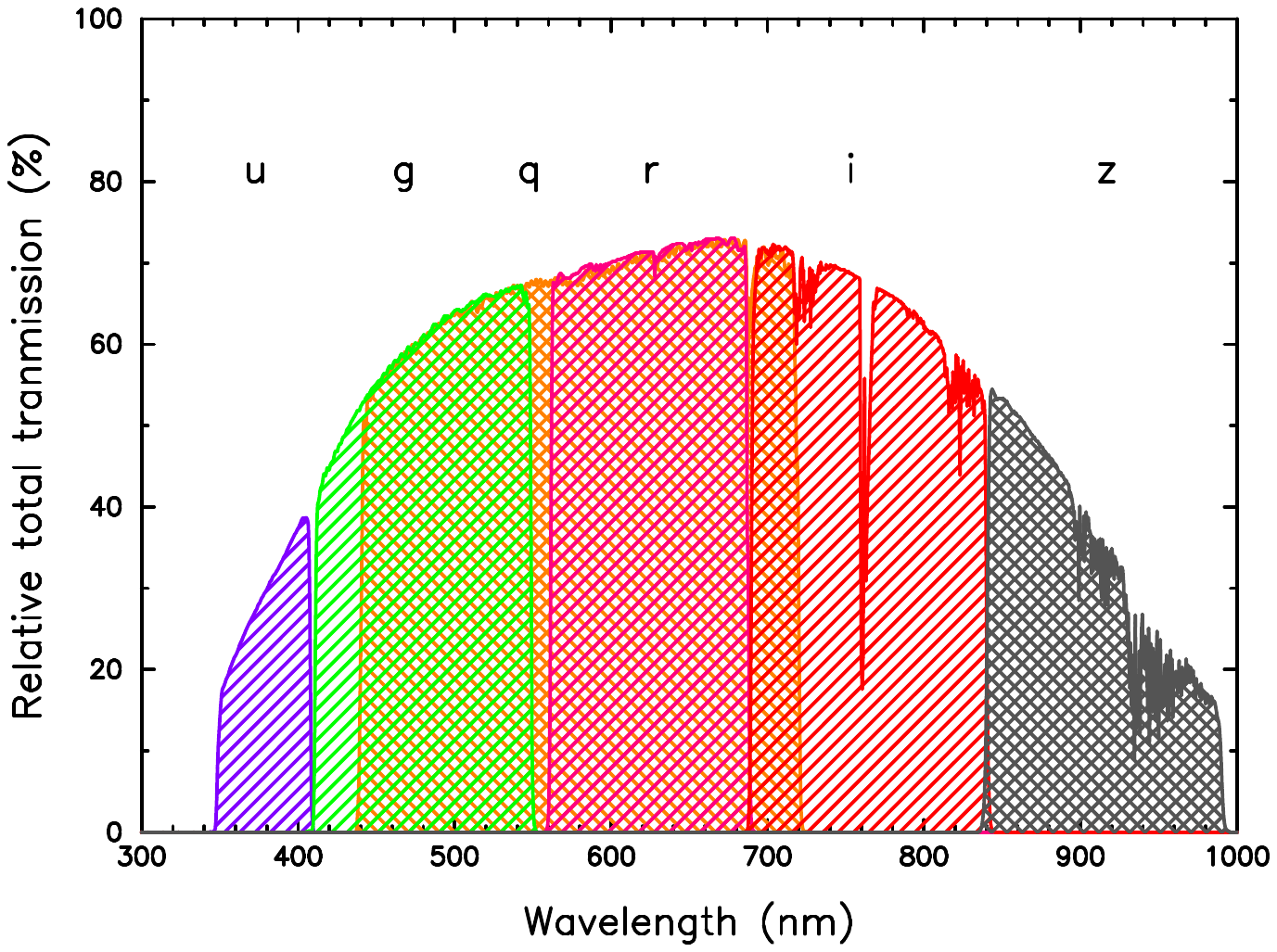}
\caption{Overall relative transmission of the BlackGEM telescope system, from the top of the atmosphere to the detector, as a function of wavelength. Color coding as in Figure\ \ref{fig:filtertrans} \label{fig:reflectivity}}
\end{figure}

The detector's cryostat was designed and manufactured at the Institute
for Astronomy at the KU Leuven \citep{raskin16}. The cryostat design
was inspired by that of the cryostats of the MAIA instrument at the
Mercator telescope on La Palma \citep{raskin13}. The detector is
cooled by a Joule-Thomson cooler (Polycold Compact Cooler) using PT16
gas. A heater inside the cryostat stabilizes the detector temperature
at 160~K to within 0.1~K. The cryostat is controlled by a PLC
controller, custom-built at the  KU Leuven \citep{raskin16}. An
overview of the cryostat design is given in Figure\ \ref{fig:cryo} and
the subsequent overall system transmissio of the BlackGEM telescopes
in Figure\ \ref{fig:reflectivity} and Table\ \ref{tab:transmission}. 

\subsection{Mechanical design \label{sec:mechanicaldesign}}

\begin{figure*}[htb]
\includegraphics[width=15cm]{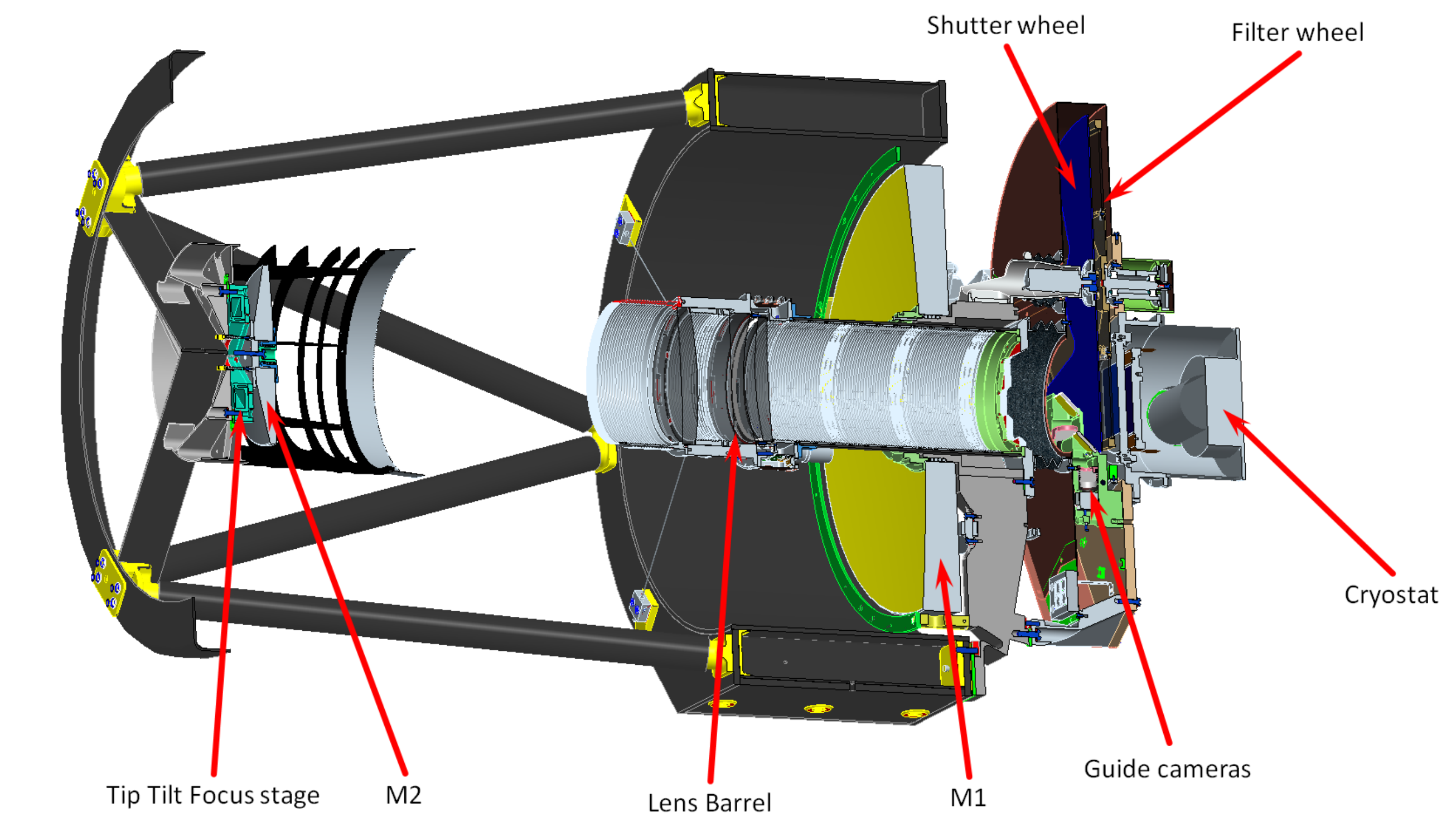}
\caption[]{Annotated design drawing of the BlackGEM telescopes. \label{fig:mech-xcut}}
\end{figure*}

\begin{figure}[htb]
\begin{center}
\includegraphics[width=7cm]{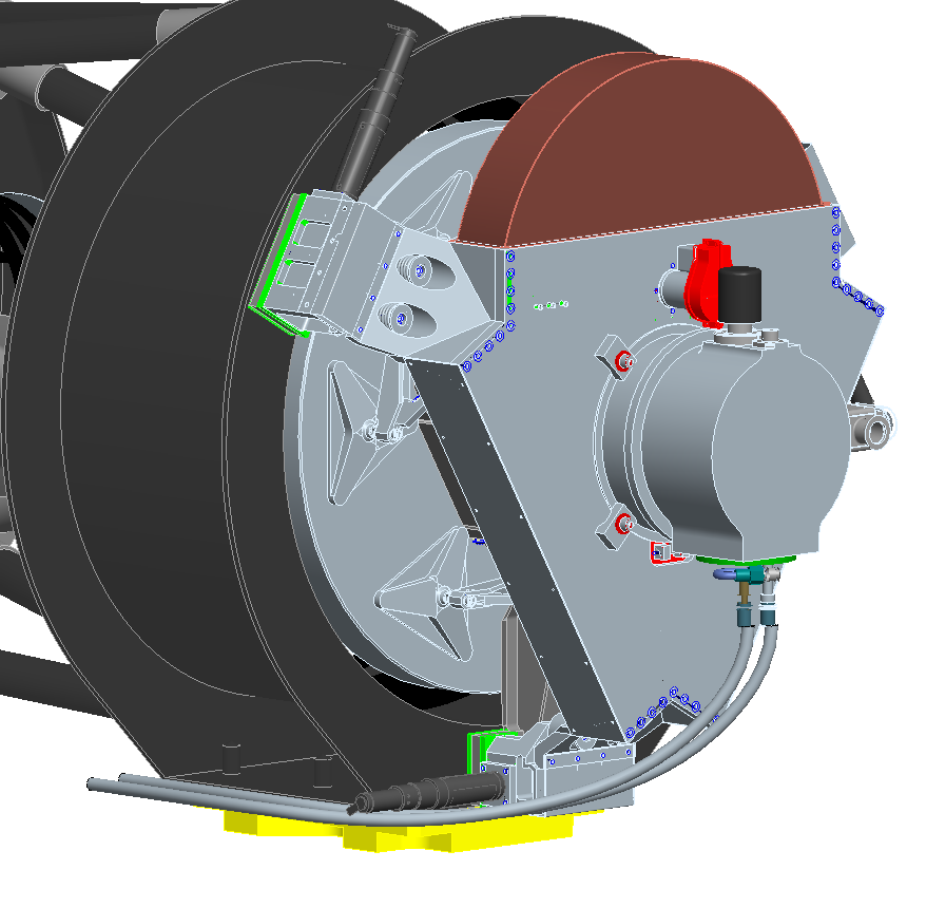}
\caption{Backside view of the mechanical design of the BlackGEM telescope. At the far left, the carbon-fibre mirror cell (black) can be seen with the main mirror (grey) supported on an 18-point, 6-pad, 3-arm whiffle tree. The shutter/filterwheel housing and the cryostat (brown/grey) are attached to the carbon fibre mirror cell on three spring-blade clamping points (green). The attachment to the mount is via an interface plate, which is part of the carbon-fibre mirror cell (yellow). 
\label{fig:mechback}}
\end{center}
\end{figure}

The mechanical design of the optical telescope assembly (OTA) was made at the NOVA Optical Infrared Group and needed to comply with the requirement that optical performance remains seeing limited (i.e. optical distortions $\leq0.3$\arcsec) at 60\degr zenith distance over an integration time of 60~seconds. Also, the complete structure needs to be able to withstand wind loads up to 15~m s$^{-1}$ in an open clam-shell dome. This resulted in stringent requirements on the stiffness of the OTA, requiring an all carbon-fiber design (Figs.\ \ref{fig:mech-xcut} and \ref{fig:mechback}). This was designed with, and manufactured by, Airborne BV (The Netherlands). The OTA consists of a main ring, which attaches to the mount through an interface plate; trusses that connect to the top ring; and the top ring, which holds the M2 unit, connected to the OTA by a spider consisting of four blade-springs. The main ring is open at the back, leaving M1 exposed to the air in order to improve the thermal balance between the optics and the environment. M1 is supported by an 18-pad whiffle tree, allowing precise positioning and position-independent support of the mirror. Thereby, the shape of the mirror (which is not actively controlled) remains within specifications up to 70\degr zenith distance. 
Connected to the main ring by blade connectors is the camera unit, holding the filter and shutter wheel in a closed compartment and the cryostat/detector assembly. The M2 unit holds the secondary mirror which is itself mounted on a piezo stage that allows for tip-tilt control, used for guiding, and lateral motion along the optical axis, used for focussing. 

The triplet corrector lens in its barrel is mounted to the whiffle tree assembly which also supports M1. The barrel is positioned through the central hole of the primary. The primary mirror maintains its lateral position by the lens barrel location. 

\begin{figure}[htb]
\begin{center}
\includegraphics[width=6cm]{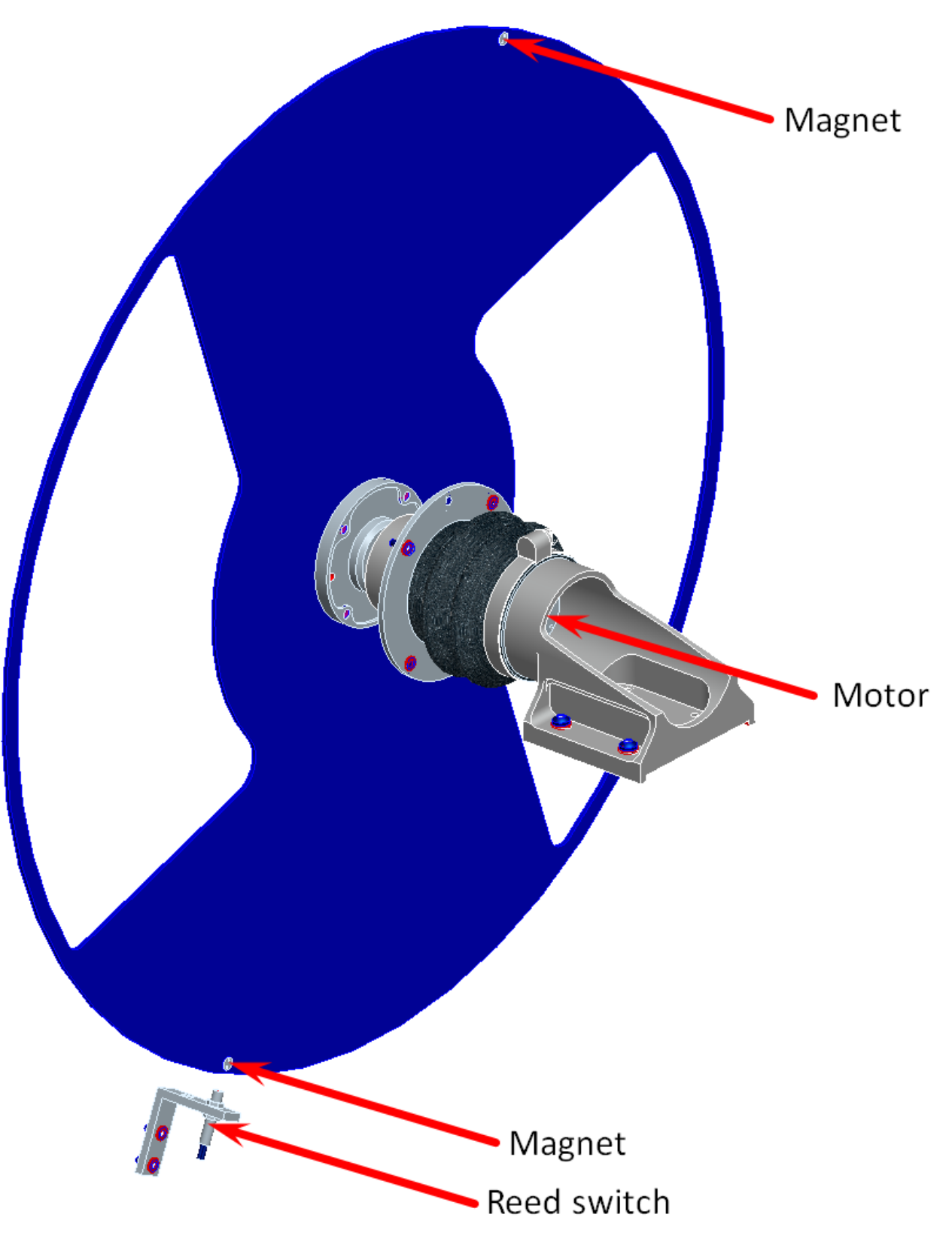}
\caption[]{Butterfly design of the BlackGEM shutter, showing the curved sides that ensure equal illumination per pixel during opening and closing of the shutter. \label{fig:shutter}}
\end{center}
\end{figure}

The camera house with the six-slot filter wheel also contains the shutter, which is a butterfly-type shutter. Two apertures are cut such that they provide uniform illumination per pixel over the focal plane during the length of an exposure (Figure\ \ref{fig:shutter}). The shutter takes $\sim$0.8~second to cross the focal plane. This limits the accuracy of the BlackGEM timing. More accurate timing requires a `per-pixel' illumination model. 
Timing for BlackGEM observations is provided by a trigger from the shutter controller to a GPS-based Meinberg LANTIME M3000 clock. Both `shutter open' as well as `shutter closed' GPS times are recorded, and the mid-time of each observation is derived from these two triggers. 

\subsection{Focus and guiding \label{sec:focus}}

After an initial positioning and alignment of the main optical elements, focus is controlled through a PI P-528.TCD piezo-stage on which M2 is mounted. The stage can be moved over a 200~$\mu m$ range along the optical axis. Collimation is done manually by shifting and tilting M1. 

The lateral movement of the piezo stage allows for a $\pm$200\arcsec\ offset from the nominal position of the optical axis. This range is used for autoguiding through three guide cameras mounted at the sides of the main field-of-view next to the science CCD. The guide cameras are UI-3060CP-M-GL Rev.2 from IDS Imaging and are operated through a USB-3.0 connection to the M2 control computer located in the counterweight. 

The guide cameras contain a 1936$\times$1216 pixel CMOS detector. With a plate scale of 5.86~$\mu$m/pixel (0.36\arcsec/pixel), the field of view of each camera is 11.8\arcmin$\times$7.4\arcmin. 
In standard operations, a position offset is calculated from the set of guide stars detected on the three cameras at a rate of 1~Hz. Guiding is started on the `shutter open' signal generated by the CCD controller and stopped by the `shutter close' signal. As the guide camera pick-off mirrors are located before the shutter/filter wheel, guiding is done in white light and is independent of the chosen filter.    

\subsection{Mount and Cable wraps \label{sec:mount}}
Each BlackGEM telescope is mounted on its own second generation Fornax-200 mount, a German equatorial mount which was co-designed by the BlackGEM project and the Fornax company (Hungary). The mount consists of two identical drive systems, one for declination and one for hour angle. Each drive consists of a worm wheel of 605~mm in diameter with 250~cogs, in a worm - worm wheel ratio of 1:250. The wheels are made of bronze for stiffness and durability. A set of Renishaw absolute encoders are used for pointing and allow a precision $\leq$\,0.1\arcsec. In practice the pointing accuracy is limited by physical effects such as flexure of the structure, play between gears etc, to about 1\arcmin, which is acceptable given the large field-of-view. A second generation mount controller fully using the absolute encoders is under development. 

Some of the telescope's electronic subsystems (filter-wheel/shutter controller, piezo controller, ADC controller, M2 control computer, DC power supplies) are located in the counterweight to serve as part of the counterweight. Only the CCD controller is placed as close as possible to the cryostat to prevent electronic pick-up noise. The counterweight, the CCD controller, and the guide cameras are water/glycol-cooled to prevent heat-dumping near the telescope.

\begin{figure}[htb]
\begin{center}
\includegraphics[width=8cm]{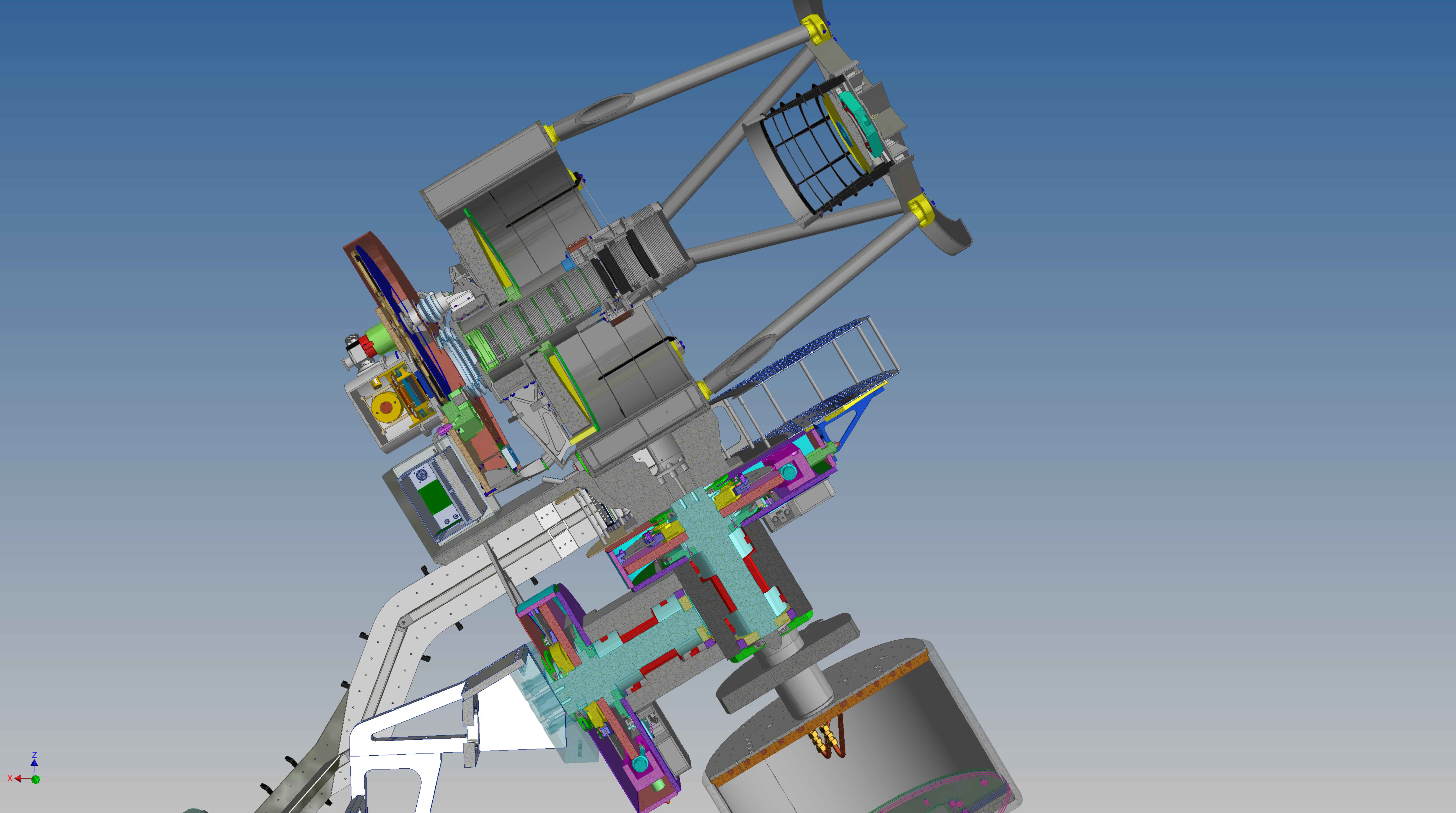}
\caption[]{Cut-open view of the BlackGEM telescope on the Fornax 200 mount, also showing the drum-shaped counter-weight and the fly-over cable wrap\label{fig:mount}}
\end{center}
\end{figure}

Cabling from the telescope down to the computer rack goes via a cable wrap on the declination axis first and then through a second wrap behind the hour angle axis. The cable wraps are made of a caterpillar type commonly used in astronomical telescopes where a flexible tray allows freedom of rotation in excess of one full rotation (see Figure\ \ref{fig:mount}).

\subsection{Dome, housing and site \label{sec:domes}}

\begin{figure}[htb]
\begin{center}
\includegraphics[width=8cm]{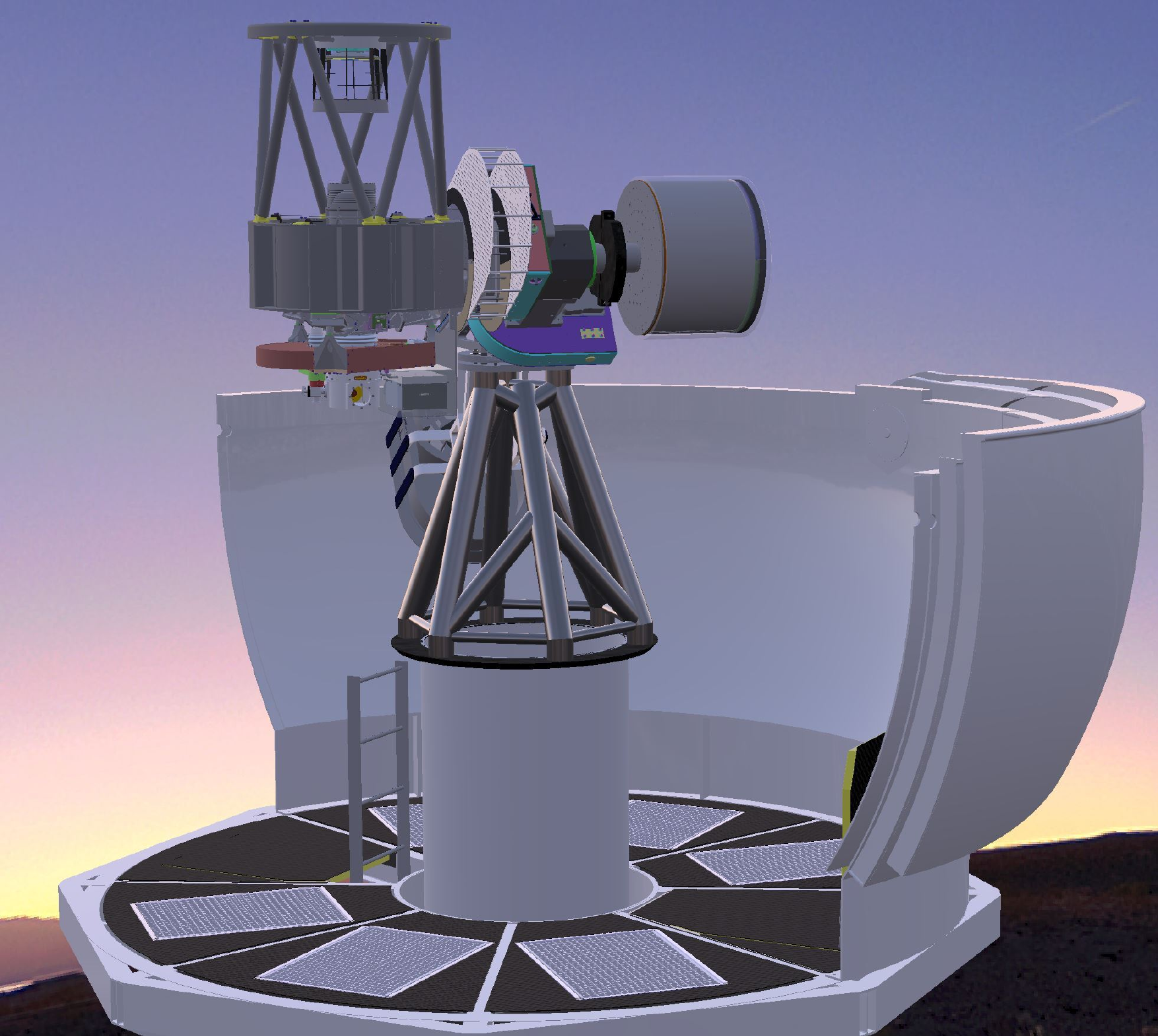}
\caption[]{Design overview of a BlackGEM unit telescope and counterweight attached to the Fornax mount (purple/green) and the pyramidal pier and steel cylinder base. A cut-open version of the clam shell dome in the open state is shown. \label{fig:pier}}
\end{center}
\end{figure}

Each telescope is mounted on a triangulated pyramidal pier consisting of hollow steel pipes to maximize stiffness (Figure\ \ref{fig:pier}). The pyramidal pier itself is bolted onto on a steel 1m diameter hollow cylinder, 5m in height. The telescope pier is nested inside a second steel pier, 1.3m in diameter, which holds the observing floor and an Astroshell 5m diameter clam-shell dome. The telescope and dome pier are separately anchored in concrete feet and only connect to each other through the La Silla bedrock. The rotation centre of the mount is located 8m above the ground and 3m above the floor of the dome. Due to the position of the telescope on the west side of the pier, observations towards the east are taken with the telescope high up, above the mount, whereas observations towards the west have the telescope lower to the dome floor. This also limits the visibility to the West to +3hr in hour angle. 

\begin{figure*}[htb]
\begin{center}
\includegraphics[width=14cm]{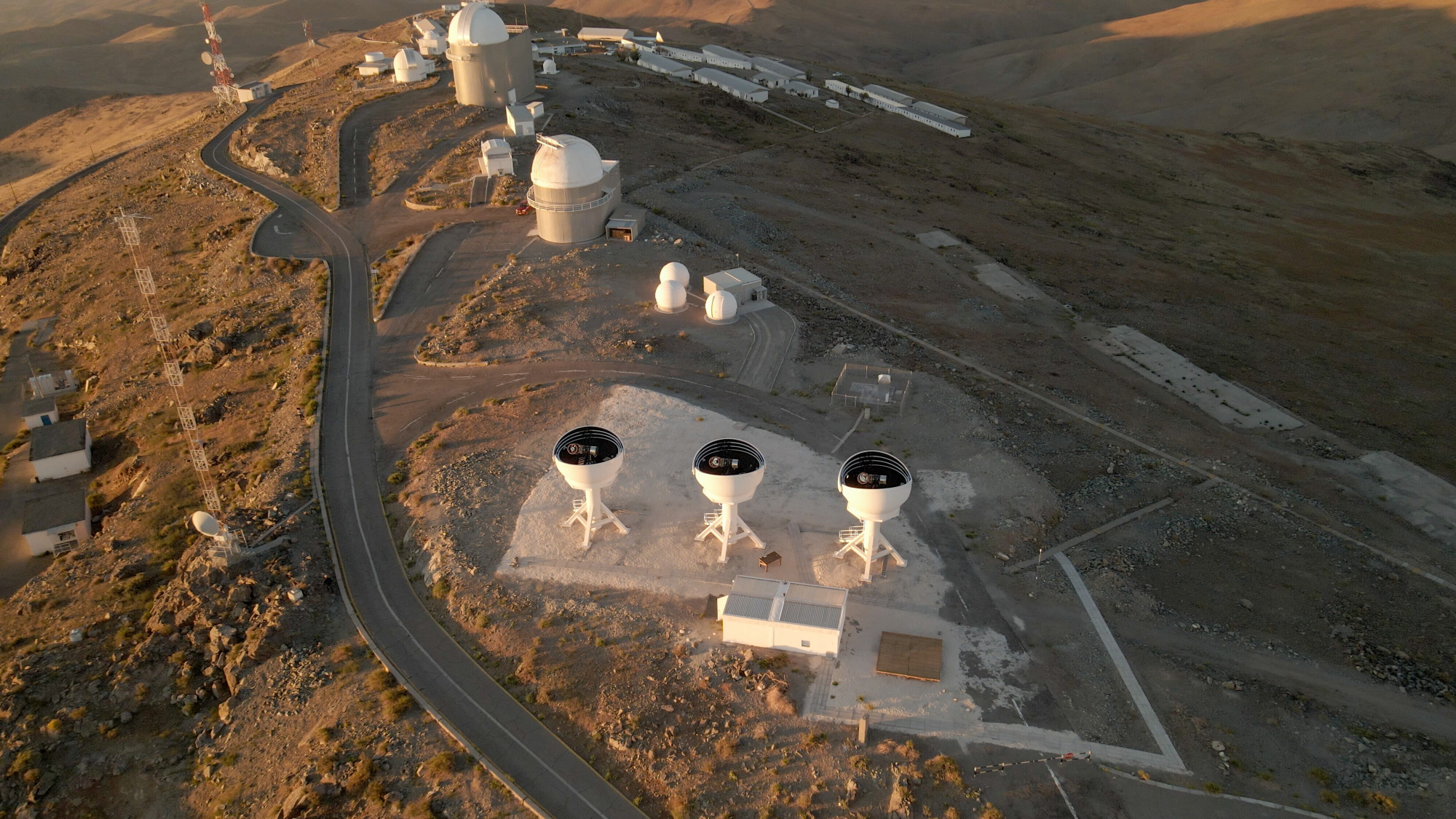}
\caption[]{Aerial view of the BlackGEM array site at ESO La Silla, seen from the south-east. The BlackGEM control building, formerly belonging to the Marseille telescope, can be seen in front of BG-2 and BG-3. The ExTra, ESO 1m, REM and ESO 1.5m telescopes can be seen to the North of BlackGEM. \label{fig:site}}
\end{center}
\end{figure*}

At La Silla, the BlackGEM array is located on the site of the former GPO/Marly and Marseille telescopes, at the top of the `ridge'. The former Marseille building now functions as the BlackGEM control room and warehouse building, also housing the cryocoolers, control servers, power and networking systems. Internet- and power-wise the BlackGEM array is completely integrated into the ESO La Silla infrastructure and is one of the hosted projects. An overview photo is shown in Figure\ \ref{fig:site}.  The lay-out of the site has been made such that a maximum of 15~BlackGEM telescopes can be accommodated on the former GPO `island'. 

The geographic location of each of the three BlackGEM telescopes is given in Table\ \ref{tab:location}. All 
telescopes are at the same altitude of 2382~m with respect to the Earth's ellipsoid \citep{McCarthyPetit2004}, and aligned along an East-West line, with BG2 being the eastern-most telescope and BG4 the western-most telescope. 

\begin{table}
\caption{Geographic location of the BlackGEM telescopes \label{tab:location}}
\begin{minipage}{8cm}
\begin{tabular}{lrr}
 Telescope & Longitude & Latitude \\ 
 &(d)&(d)\\
  \hline
BG-2 (Ruby)    & --70.737850 & --29.257469\\
BG-3 (Opal)    & --70.737969 & --29.257469\\
BG-4 (Emerald) & --70.738067 & --29.257469\\
  \hline
\end{tabular}
\end{minipage}
\end{table}

\section{Telescope Control system \label{sec:abot}}
BlackGEM is controlled through the ABOT system developed by Sybilla Technologies (Poland). The ABOT system is derived from the Solaris project \citep{Kozlowski17}. A web-interface and cloud-based observing queues are used to control the telescope remotely and/or robotically. Dedicated procedures have been defined with Sybilla to robotise the taking of flatfields, biases, and regular science data. 

The PLC-based ObservatoryWatch system supplied by Cilium Technologies (Poland) monitors weather sensors and supplies this information to the ABOT system. It also supplies a direct trigger to the dome in case of adverse conditions, which ensures that the dome is closed even if ABOT is not operational. The ABOT telescope control system interfaces with the Microsoft Azure cloud platform, which allows access to the system from anywhere in the world. 

\section{Data processing \label{sec:dataprocessing}}
All raw data are transported in real-time from La Silla to Santiago de Chile, where they are ingested into the Google Cloud environment. Final data products (calibrated images, source catalogues, and transient catalogues) are available within (currently) 15~minutes after the end of the exposure. A full description of the BlackGEM data processing and data products will be described in Vreeswijk et al. (in prep.), hence only a short summary is given here. 

All BlackGEM data are debiased and (twilight-sky) flatfielded. No defringing is required for the redder bands due to the very low levels of fringing in the STA 1600 detector. 
Astrometric calibrations are made using positions from the Gaia Data Release (DR) 3 astrometric catalogue, and all BlackGEM positions are in the ICRS frame with epoch J2016.0 \citep{GaiaEDR3_Astrometry21,GaiaDR3_23}.
Astrometric solutions generally show root-mean-square residuals $<$50~$\mu$as in each coordinate. 

Photometric calibrations are performed using the Gaia DR3 low-resolution spectra of selected stars in the field-of-view. A set of 70~million Gaia DR3 low-resolution spectra were convolved with the BlackGEM throughput curves, taking the La Silla atmospheric extinction into account. Synthetic BlackGEM-system apparent magnitudes were generated for all stars in the reference catalogue. For selected stars in each frame the zeropoint is deduced by comparing the measured instrumental magnitudes with the reference magnitude. From the resulting histogram of zero-points a median value, as well as a zeropoint accuracy, is derived. A quality control step is performed where, based on a number of performance statistics, including e.g. the zeropoint, the limiting magnitude, the derived image quality, the elongation of stars etc., the frame is classified in one of four ranges, running from `green', to `yellow', `orange' and `red'. Everything but red-classified frames can be used for science. 

Using a reference frame, image differencing is performed using the ZOGY formalism \citep{ZOGY16}. From the resulting corrected statistics (S{\sc corr}) map, transients and variables are identified when their S{\sc corr} value is larger than 6, corresponding to a $>6\sigma$ deviation from the mean noise pattern in the difference image. All objects thus identified are saved in a `transient catalogue'. These include positive as well as negative transients and variable objects. A real-bogus machine-learnt algorithm is then applied to each detection to produce a real-bogus score between 0 (bogus) and 1 (real) \citep{MeerCrab21}.  

In addition, a match is made to all possible asteroids within 20\arcsec, as known to the Minor Planet Center (MPC) (Pieterse et al., in prep.). The current MPC database is downloaded on a daily basis at 17:00 local time. 

On each reduced frame, a forced optimal-photometry measurement \citep{Horne86} is made at the position of all Gaia DR3 objects within the field of view, where proper motion has been taken into account. The output from the forced optimal-photometry routine is saved in the `full-source catalogue'. The transient database is incorporated into a query-able MonetDB database provided by DataSpex BV in the Netherlands. The full-source database is a Google BigQuery database, implemented into the GoogleCloud environment by the Dutch branch of Cloud Technology Solutions.

\section{Operations \label{sec:operations}}

The BlackGEM telescopes are robotically operating at the ESO La Silla Observatory. The daily routine consists of: \begin{itemize}
\item Biases: a set of five biases taken each afternoon before sunset and each morning before sunrise in a closed-up dome; 
\item Flatfields are taken each twilight for each of the six filters, with 5~exposures per filter using dynamic exposure times to obtain, on average, 13\,000~counts per pixel in each flatfield. The filters are ordered as $u$,$z$,$r$,$g$,$i$,$q$ during the evening twilight and the reverse for the morning twilight. This order is set by the combined effects of system throughput in a given filter and the darkening (brightening) evening (morning) sky. Flatfields are taken at a telescope position of Dec\,=$-29$\degr, HA\,=\,1hr away from the direction to the Sun. A 10\arcsec\ dither in both RA and Dec is applied between each flatfield exposure in a filter sequence; 
\item Survey program: 50~minutes after sunset, the night program starts. During the day, a survey plan will have been generated with a priority (from 1, lowest to 5, highest). During normal operation, this survey plan will be executed, unless it is overridden by an automatic trigger, in particular for the GW program. 
\end{itemize}

Each BlackGEM telescope is mounted on the west side of the pier, to allow for a maximum eastern hour angle in detecting GW afterglows events at the earliest possible time. No meridian flip can be performed due to the fixed cable wraps. Operating limits for the BlackGEM telescopes are: 
\begin{itemize}
\item a pointing limit of $>20\degr$ above the horizon in elevation;
\item a pointing limit of $HA>-7$~hrs on the east side of the pier (as long as it does not conflict with the horizon limit); 
\item a pointing limit of $HA<+3$~hrs on the west side of the pier due to vignetting by the bottom ring of the dome;
\item a hardware limit of $Dec<+30\degr$, due to a hardware stop on the declination axis. 
\end{itemize}

\section{Science Program}

The BlackGEM array is designed and built for the identification and characterisation of gravitational wave counterparts. To enable this search, and in-between GW triggers, a fixed set of survey programs will be executed. Full details on the survey program of BlackGEM will be given in an accompanying paper (Van Roestel et al., in prep) and are therefore omitted here. We provide a concise description here, to highlight the broad potential of science cases to be covered with BlackGEM.

\subsection{Sky grid \label{sec:skygrid}}
All BlackGEM observations are executed on a predefined sky grid. If specific targets are requested they will be executed on a `snap-to-grid' basis. The existence of the predefined grid strongly simplifies data reduction  procedures, e.g. with respect to reference frames, as well as database structures. 

The predefined grid consists of 15\,946 grid points that tile the sky from the celestial South Pole up to Dec\,=\,+62\degr. In addition to this regular grid, 140 `special fields' have been defined that cover the baryonic mass overdensities in the nearby Universe, as well as special regions of interest in our own Milky Way Galaxy, e.g. the Galactic Bulge region, selected globular and open clusters. The special fields are labeled 16\,000--16\,139. The grid definition is available from the BlackGEM website\footnote{ \url{www.blackgem.org}}.  

\subsection{Gravitational wave follow-up}
The top priority of the BlackGEM science program is the follow-up of gravitational wave (GW) triggers generated by the second generation laser-interferometer detectors Advanced LIGO, Advanced Virgo, KAGRA and, in the future, LIGO-India. Simulations, e.g. \citep{Nissanke13} and real-world performances of LIGO-Virgo \citep{GWTC3} show that the expected size of error boxes in the coming phases of these GW detectors should result in sky areas of $\sim$100--200~square degrees and detection horizon limits of $\sim$150--200~Mpc for binary neutron star mergers and neutron star - black hole mergers when three detectors are operational. Assuming an optical brightness similar to that of GW\,170817, the BlackGEM array can detect counterparts up to $\sim$200~Mpc \citep{Chase22}. Kilonova signals are expected to display a rapid reddening (\citealt{Kasen13}), setting them apart from other types of rapid transients, with blue components (possibly) visible at early times ($\sim$hrs). For this reason, the BlackGEM telescopes have been equipped with a broad set of filters, and, if possible, a multi-color strategy will be used. 

The main requirement in the GW follow-up with BlackGEM is a cadence over the full part of the (observable) error box within 2~hrs. Scanning will be done in as many filters as can be fit into the 2-hour cadence on the observable error box. If possible, all six filters will be used. Alternatively, the $u$, $q$ and $i$ filters will be used, and in the worst case (largest error boxes), only the $q$-band will be used. An example of the strategy used by BlackGEM is given in \cite{DeWet21} for the follow-up of GW\,190814 with MeerLICHT and \cite{DeWet23} for GRB\,220627A. 

For scanning the error box, a sky grid (Sect.\ \ref{sec:skygrid}) will be used in conjunction with the ranked-tiling method defined in \citet{Ghosh16}. The LVK-provided localisation map is convolved with the sky grid to provide a probability per grid tile, after which the visible tiles are ordered on descending probability. Observations are made per set of ten tiles, which are then immediately repeated to veto against asteroids. Processing of data is done as outlined in Sect.\ \ref{sec:dataprocessing}, and candidate counterparts are sent to the BlackGEM GW team for a last-step human vetting and further processing. All validated candidates are reported to the General Coordinates Network (GCN\footnote{\url{gcn.nasa.gov}}) and the Transient Name Server (TNS\footnote{\url{wis-tns.org}}).

As part of the GW program we will also survey a `GW footprint' area consisting of $\sim$2700 square degrees once a week in one filter to be able to identify recent interlopers to GW candidates, such as supernovae, tidal disruption events, dwarf nova outburts (\citealt{VanRoestel19}). The sky area is targeting the majority of stellar mass within 150 Mpc. 

\subsection{Southern All Sky Survey}
Transients can only be detected when a high-quality, deep reference frame is available. In addition, the southern skies have so far seen a limited number of (full-sky) multi-color surveys, in particular, DELVE DR2 (\citealt{DELVEDR2}), VPHAS+ \citep{Drew14, VPHASDR2}, SkyMapper (\citealt{SM_DR4}) and the southern part of Pan-STARRS (\citealt{PS1_DR2}). This limits the ability to construct a pre-operations reference frame from archival observations. In agreement with ESO, the BlackGEM project will therefore perform a six-band, 30\,000~square degree ($\delta<+30$\degr) survey of the southern sky down to a depth of 22$^{nd}$~mag. This BlackGEM Southern All-Sky Survey (BG-SASS) will become available through the ESO Archive. As a starting point for transient detections, the BlackGEM array will use reference images obtained with the MeerLICHT prototype telescope \citep{Bloemen16}. 

\subsection{Local Transient Survey}
The aimed-for detection horizon to binary neutron star systems of the second generation GW detectors is, at full sensitivity, $>$200~Mpc. To rule out existing nearby transients as GW sources and obtain high quality data on the nearest (and brightest) transients, it is imperative to obtain a thorough and extensive set of observations of the major mass concentrations in the nearby Universe, such as the Fornax, Norma, Puppis and Eridanus galaxy clusters. In addition, within our very local Universe ($<$15\,Mpc), a sizeable portion of nearby galaxies  have been observed by the Hubble Space Telescope (HST), and contain valuable information for the identification of the progenitor stars (and their associated precursor emission) for supernova explosions \citep{Smartt2015PASA,Qin2023arXiv}, luminous red novae \citep{Blagorodnova2017,Blagorodnova2021} and luminous blue variables \citep{Smith2011MNRAS} among others.

The Local Transient Survey of BlackGEM will include fields with major mass concentrations as well as $<$15\,Mpc galaxies with deep HST coverage ($>$\,300s). The observations will be done in three bands ($u,q,i$) with a cadence of 3~hrs. These fields will enter the watch list depending on the available slots and the visibility of each field. These fields will be kept on the list until they become unobservable due to seasonal shifts. 

\subsection{Fast Synoptic Survey}
In the last decade, the optical time-domain Universe at cadences below one hour
has become available at precisions of $\mu$mag. The past {\sl Kepler\/} space telescope and its refurbished K2 version, as well as the ongoing Transiting 
Exoplanet Space Survey (TESS) are instrumental in opening up this window of high-precision uninterrupted space photometry. This has given rise to a flood of novel science cases and discoveries aside from their prime aim of exoplanet hunting, notably asteroseismology \citep{Aerts21}. Periodic variables with shorter time scales of the order of a few minutes were monitored as well, with the {\it Kepler}/K2 and TESS short-cadence modes of about 1 minute and 20\,s respectively. This delivered vastly new modelling potential for asteroseismology of compact pulsators such as white dwarfs \citep[e.g.,][]{Hermes17,Giammichele18,Corsico19} and subdwarfs \citep[e.g.][]{Uzundag21}. As any instrument on the ground, BlackGEM cannot deliver $\mu$mag precision nor uninterrupted light curves. However, it will be able to aid in asteroseismology of faint compact pulsators at mmag level by delivering good candidate pulsators for follow-up, from their dominant frequency accessible by BlackGEM combined with their positions in the color-magnitude diagram, as already demonstrated from MeerLICHT photometry \citep{Princy23}.  

{\it Kepler\/}/K2 and TESS also led to breakthrough science of 
fast accretion phenomena in various compact binaries including black holes \citep{Scaringi15} and white dwarfs \citep{Scaringi22}.
At the level of mmag-precision ground-based photometry, the
{\sl RATS} and {\sl OmegaWhite} surveys covered $>$400~square degrees of Galactic Plane fields at a 5~minute cadence in a single filter (\citealt{RATS1}, \citealt{RATS2}, \citealt{Macfarlane15}). The ZTF High Cadence survey (\citealt{Kupfer21}) is hugely successful in finding, among others, ultracompact binaries (e.g. \citealt{Burdge19}, \citealt{Burdge20}, \citealt{JvR22}, \citealt{Burdge23}). These binaries are also strong sources of gravitational wave emission, but in the lower frequency ranges to be covered by the LISA mission (\citealt{Ramsay18}, \citealt{Kupfer23}). 

To discover and study the population of ultracompact binaries as well as short-period single and close binary pulsators, the BlackGEM array will perform a Fast Synoptic Survey, which will use one telescope continuously. The observing strategy will consist of a 3-4 hr monitor of a single field continuously with 60s exposures while cycling through the filters in a sequence of ($q,u,q,i,q,u,q,i...$). Hereby, the $q$-band filter gets a higher cadence than $u$ and $i$, allowing for shorter duration signals to be resolved (e.g. ingress/egress features of short-period eclipsing systems). Two fields will be observed per night, and each will be repeated the next night for an identical set of observations. In addition, each observed field will be targeted once per night in the $q$-band with an additional 60s exposure for 14 nights before and after the 2-day high-cadence observations. 

\section{Science operations, preliminary performance and early science results \label{sec:installation}}

The BlackGEM array was installed at the ESO La Silla Observatory over the period August 2019 - February 2020. All installation and commissioning activities had to be suspended due to the COVID-19 pandemic in March 2020 and could only be resumed in March 2022: a full two-year hiatus due to suspended travel between Europe and South America, closure of ESO sites, and strict travel restrictions on either the European or Chilean sides. Scientific operations started on April 1, 2023 and will run for a five-year period. This first overview paper will be followed by a survey plan paper, a data reduction and data-basing paper, and the commissioning results. 

\begin{figure}[htb]
\begin{center}
\includegraphics[width=8cm]{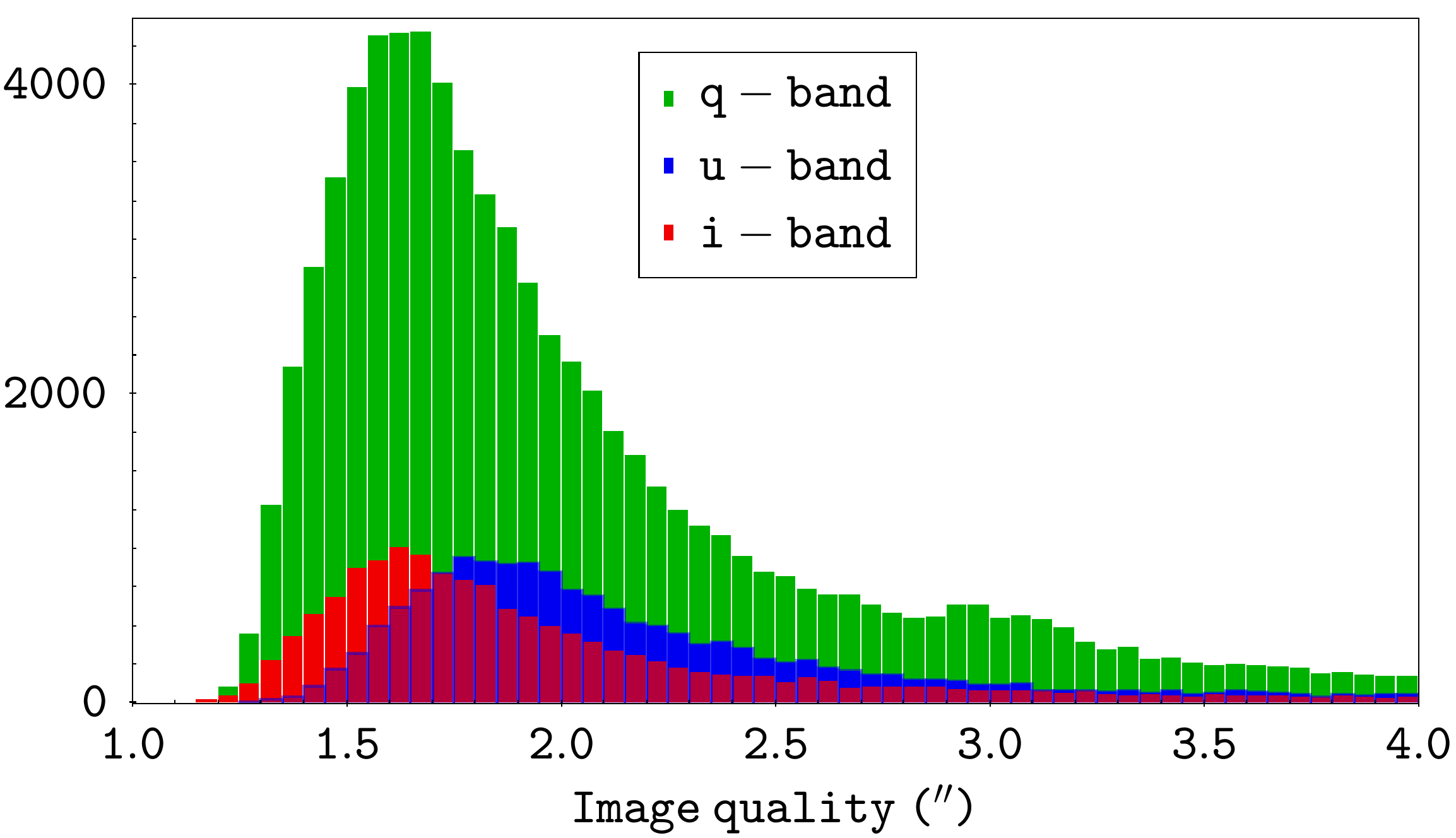}
\caption[]{Distribution of measured image quality in the $u$ (blue),$q$ (green) and $i$-bands (red) on BG4 over the period April 1, 2023 to October 20, 2024.  \label{fig:seeing}}
\end{center}
\end{figure}

\begin{figure}[htb]
\begin{center}
\includegraphics[width=8cm]{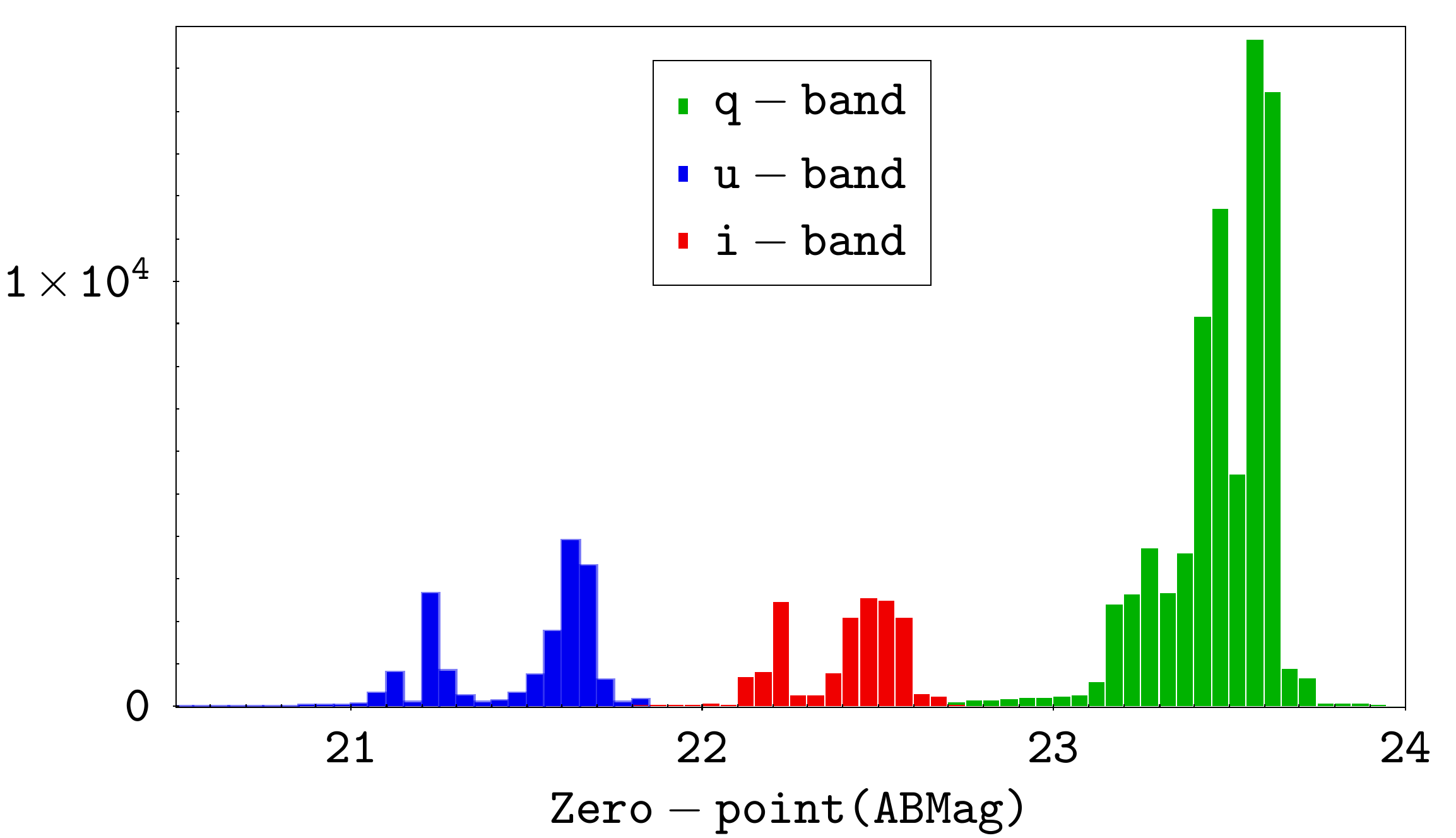}
\caption[]{Distribution of measured zero-points in the $u$ (blue),$q$ (green) and $i$-bands (red) on BG4 over the period April 1, 2023 to October 20, 2024.  \label{fig:zeropoint}}
\end{center}
\end{figure}

\begin{figure}[htb]
\begin{center}
\includegraphics[width=8cm]{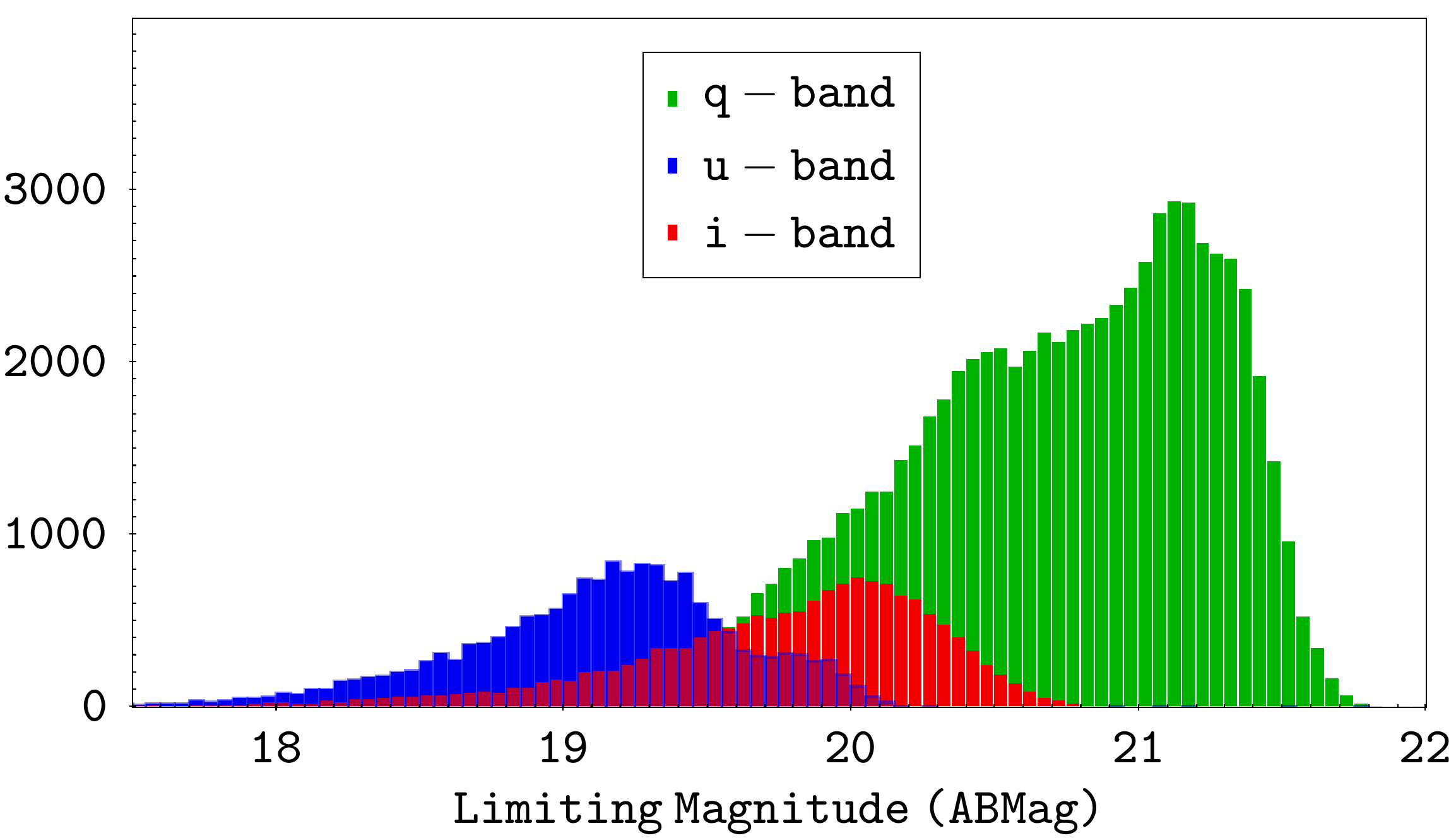}
\caption[]{Distribution of measured seeing values in the $u$ (blue),$q$ (green) and $i$-bands (red) on BG4 over the period April 1, 2023 to October 20, 2024.  \label{fig:limmag}}
\end{center}
\end{figure}

In Figs.\ \ref{fig:seeing}, \ref{fig:zeropoint} and \ref{fig:limmag} we show the performance of BG4-Emerald over the period April 1, 2023 - October 21, 2024. The performances of BG2-Ruby and BG3-Opal are similar. It can be seen in Figure\ \ref{fig:seeing} that the aim of 1\arcsec\, median image quality is not achieved yet. Here image quality is defined as the sigma-clipped full-width-at-half-maximum determined by SExtractor of the profile of the brightest 33\% of non-saturated, non-flagged stars on the inner 90\% of each image.  
For the $q$-band and $i$-band the median image quality is $\sim$1.65\arcsec\ and for the $u$-band $\sim$1.9\arcsec. Additionally a floor at around 1.2\arcsec\ can be seen, as well as a long tail towards high seeing values. Preliminary analysis shows that two factors are at play: high-frequency vibrations and non-uniform tracking by the mount cause the higher-than-desired median values and the floor to the seeing, and wind-shake causes the high-value tail. Remediate action is currently being taken by better wind isolation of the M2 piezo-stage as well as a higher-quality mount controller that uses the absolute encoder on the worm-wheel, to be installed in the next six months. 

Figs.\ \ref{fig:zeropoint} and \ref{fig:limmag} show the deduced zero-point of BG4 over the same period of time in the $u$, $q$, and $i$-bands. Zero-point is defined as the AB magnitude that would give a source-integrated electron flux of 1 e$^{-}$ s$^{-1}$, assuming no atmospheric extinction, which is corrected for separately in the photometric calibration. The zero-point distribution for each filter shows a relatively narrow spread with several sub-peaks. This is due to thin cirrus clouds which cause the system to be less sensitive, but still acceptable to the quality control steps, as well as a gradual degradation of the mirror and lens reflectivities due to dust accumulation. The BG4 primary mirror as well as the entry lens to the lens barrel were cleaned in August 2024, which caused an 0.5 magnitude gain in the zero-point. 

The limiting magnitude in the three bands is displayed in Figure\ \ref{fig:limmag}. These are the 5$\sigma$ point-source sensitivities in a standard 60s exposure. The broad distribution is caused by the variation in zero-point, seeing and sky background levels. The $q$-band is, on average, about 1.2 magnitude more sensitive than the $i$-band and almost 2 magnitudes more sensitive than the $u$-band. For this reason, the discovery of faint transients is best performed in the $q$-band. With a limiting magnitude q$_{\rm AB} >$ 21 mag, BlackGEM is currently the most sensitive dedicated synoptic survey based in the Southern Hemisphere. Compared to other projects such as ATLAS (\citep{ATLAS18}) and GOTO (\citep{GOTO22}), BlackGEM has chosen for a strategy that goes deeper, but covers less area per night ($\sim$1\,000 square degrees per telescope), and with a multi-band coverage.

Early science results from BlackGEM include the detection of over 550 new transients in the period January 1, 2024 - October 20, 2024, as reported to the Transient Name Server\footnote{https://wis-tns.org}. Of these 227 were reported by the BlackGEM Consortium first. The main aim of detecting a gravitational wave counterpart is still to be achieved, primarily due to a lower than expected number of binary neutron star and neutron star - black hole mergers in the nearby Universe during the currently running O4 LIGO-Virgo-KAGRA observing run. These non-detections have now pushed down the expected merger rate in the local universe for binary neutron stars to $<1.7\times$10$^3$ Gpc$^{-3}$ yr$^{-1}$ \citep{GWTC23}, which translates to $<$ 10 events per year within 200 Mpc. This is still based on a single event gravitational wave event, but in line with the rate of short gamma-ray bursts, see \cite{Mandel22}.

The transients detected by BlackGEM sofar include rare events such as the brightest-ever gamma-ray burst GRB\,221009A ("the BOAT", \citealt{Dichiara22}), which was the first-ever transient detection with BlackGEM (\citealt{Groot22}); the extragalactic nova AT2024pid / BGEM\,J004734.42--251926.3 in NGC\,253 \citep{Tranin2024, Duarte24}; and the lowest-luminosity-ever supernovae, of Type Iax, SN2024vjm / BGEM\,J190925.80--635001.7 in NGC\,6744 \citep{Groot24a, Asquini24, Srivastav24}. In particular these last two intrinsically low-luminosity events are testimony to the type of transient events that are now within reach with the deeper grasp of BlackGEM. 

\section{Summary \label{sec:summary}}

The BlackGEM array consists of three 65cm wide-field unit telescopes installed at the ESO La Silla Observatory in Chile. Together they provide an 8.2 square degree field-of-view. Standard survey operations consist of 60s exposures, in one of six filters ($u,g,q,r,i,z$) that cover the optical wavelength range. The science focus is on time-domain astronomy, in particular counterparts to gravitational wave mergers, in combination with transients in Local Universe (binary) stellar populations and ultracompact binaries in our Milky Way Galaxy. BlackGEM will offer a high-cadence ($<$1 day), multi-color view of transients and variables in the Local Universe. 

\begin{acknowledgements}
BlackGEM has been made possible with financial aid from Radboud University, the Netherlands Research School for Astronomy (NOVA), the KU Leuven Department of Physics and Astronomy, the Netherlands Organisation for Scientific Research (NWO), 
the Research Foundation Flanders (FWO) under grant agreement G0A2917N (BlackGEM, with PIs C. Aerts \& G. Raskin and personnel C. Johnston and P. Ranaivomanana), the European Research Council, F. Stoppa is funded by the NOVA Research School for Astronomy in the Netherlands.
P.J.G. is partly supported by the National Research Foundation of South Africa, through SARChI grant 111692. HD and VS were supported by the Deutsche Forschungsgemeinschaft (DFG) through grants GE2506/17-1 and GE2506/12-1, respectively.
J.v.R. acknowledges funding by the VENI grant VI.Veni.212.201 by the Dutch Research Council (NWO)   )
M. U. acknowledges funding from the Research Foundation Flanders (FWO) by means of a junior postdoctoral fellowship (grant agreement No. 1247624N).
S.S. is supported by the Science and Technology Facilities Council (STFC) grant ST/X001075/1.
M.V. acknowledges the support of the Science and Technology Facilities Council (STFC) studentship ST/W507428/1.
I.A. acknowledges support from the European Research Council (ERC) under the European Union’s Horizon 2020 research and innovation program (grant agreement number 852097), from the Israel Science Foundation (grant number 2752/19), from the United States - Israel Binational Science Foundation (BSF; grant number 2018166), and from the Pazy foundation (grant number 216312).
G.L. was supported by a research grant (19054) from VILLUM FONDEN.
N. B. acknowledges to be funded by the European Union (ERC, CET-3PO, 101042610). Views and opinions expressed are however those of the author(s) only and do not necessarily reflect those of the European Union or the European Research Council Executive Agency. Neither the European Union nor the granting authority can be held responsible for them.
RPB acknowledges support from the European Research Council (ERC) under the European Union’s Horizon 2020 research and innovation program (grant agreement No. 715051; Spiders)
\end{acknowledgements}


\bibliographystyle{aasjournal}


\begin{thebibliography}{}
\expandafter\ifx\csname natexlab\endcsname\relax\def\natexlab#1{#1}\fi
\providecommand{\url}[1]{\href{#1}{#1}}
\providecommand{\dodoi}[1]{doi:~\href{http://doi.org/#1}{\nolinkurl{#1}}}
\providecommand{\doeprint}[1]{\href{http://ascl.net/#1}{\nolinkurl{http://ascl.net/#1}}}
\providecommand{\doarXiv}[1]{\href{https://arxiv.org/abs/#1}{\nolinkurl{https://arxiv.org/abs/#1}}}

\bibitem[{{Abbott} {et~al.}(2016){Abbott}, {Abbott}, {Abbott}, {Abernathy}, {Acernese}, {Ackley}, {Adams}, {Adams}, {Addesso}, {Adhikari}, {Adya}, {Affeldt}, {Agathos}, {Agatsuma}, {Aggarwal}, {Aguiar}, {Aiello}, {Ain}, {Ajith}, {Allen}, {Allocca}, {Altin}, {Anderson}, {Anderson}, {Arai}, {Arain}, {Araya}, {Arceneaux}, {Areeda}, {Arnaud}, {Arun}, {Ascenzi}, {Ashton}, {Ast}, {Aston}, {Astone}, {Aufmuth}, {Aulbert}, {Babak}, {Bacon}, {Bader}, {Baker}, {Baldaccini}, {Ballardin}, {Ballmer}, {Barayoga}, {Barclay}, {Barish}, {Barker}, {Barone}, {Barr}, {Barsotti}, {Barsuglia}, {Barta}, {Bartlett}, {Barton}, {Bartos}, {Bassiri}, {Basti}, {Batch}, {Baune}, {Bavigadda}, {Bazzan}, {Behnke}, {Bejger}, {Belczynski}, {Bell}, {Bell}, {Berger}, {Bergman}, {Bergmann}, {Berry}, {Bersanetti}, {Bertolini}, {Betzwieser}, {Bhagwat}, {Bhandare}, {Bilenko}, {Billingsley}, {Birch}, {Birney}, {Birnholtz}, {Biscans}, {Bisht}, {Bitossi}, {Biwer}, {Bizouard}, {Blackburn}, {Blair}, {Blair}, {Blair}, {Bloemen}, {Bock}, {Bodiya}, {Boer},
  {Bogaert}, {Bogan}, {Bohe}, {Bojtos}, {Bond}, {Bondu}, {Bonnand}, {Boom}, {Bork}, {Boschi}, {Bose}, {Bouffanais}, {Bozzi}, {Bradaschia}, {Brady}, {Braginsky}, {Branchesi}, {Brau}, {Briant}, {Brillet}, {Brinkmann}, {Brisson}, {Brockill}, {Brooks}, {Brown}, {Brown}, {Brown}, {Buchanan}, {Buikema}, {Bulik}, {Bulten}, {Buonanno}, {Buskulic}, {Buy}, {Byer}, {Cabero}, {Cadonati}, {Cagnoli}, {Cahillane}, {Bustillo}, {Callister}, {Calloni}, {Camp}, {Cannon}, {Cao}, {Capano}, {Capocasa}, {Carbognani}, {Caride}, {Casanueva Diaz}, {Casentini}, {Caudill}, {Cavagli{\`a}}, {Cavalier}, {Cavalieri}, {Cella}, {Cepeda}, {Baiardi}, {Cerretani}, {Cesarini}, {Chakraborty}, {Chalermsongsak}, {Chamberlin}, {Chan}, {Chao}, {Charlton}, {Chassande-Mottin}, {Chen}, {Chen}, {Cheng}, {Chincarini}, {Chiummo}, {Cho}, {Cho}, {Chow}, {Christensen}, {Chu}, {Chua}, {Chung}, {Ciani}, {Clara}, {Clark}, {Cleva}, {Coccia}, {Cohadon}, {Colla}, {Collette}, {Cominsky}, {Constancio}, {Conte}, {Conti}, {Cook}, {Corbitt}, {Cornish}, {Corsi},
  {Cortese}, {Costa}, {Coughlin}, {Coughlin}, {Coulon}, {Countryman}, {Couvares}, {Cowan}, {Coward}, {Cowart}, {Coyne}, {Coyne}, {Craig}, {Creighton}, {Creighton}, {Cripe}, {Crowder}, {Cruise}, {Cumming}, {Cunningham}, {Cuoco}, {Dal Canton}, {Danilishin}, {D'Antonio}, {Danzmann}, {Darman}, {Da Silva Costa}, {Dattilo}, {Dave}, {Daveloza}, {Davier}, {Davies}, {Daw}, {Day}, {De}, {DeBra}, {Debreczeni}, {Degallaix}, {De Laurentis}, {Del{\'e}glise}, {Del Pozzo}, {Denker}, {Dent}, {Dereli}, {Dergachev}, {DeRosa}, {De Rosa}, {DeSalvo}, {Dhurandhar}, {D{\'\i}az}, {Di Fiore}, {Di Giovanni}, {Di Lieto}, {Di Pace}, {Di Palma}, {Di Virgilio}, {Dojcinoski}, {Dolique}, {Donovan}, {Dooley}, {Doravari}, {Douglas}, {Downes}, {Drago}, {Drever}, {Driggers}, {Du}, {Ducrot}, {Dwyer}, {Edo}, {Edwards}, {Effler}, {Eggenstein}, {Ehrens}, {Eichholz}, {Eikenberry}, {Engels}, {Essick}, {Etzel}, {Evans}, {Evans}, {Everett}, {Factourovich}, {Fafone}, {Fair}, {Fairhurst}, {Fan}, {Fang}, {Farinon}, {Farr}, {Farr}, {Favata}, {Fays},
  {Fehrmann}, {Fejer}, {Feldbaum}, {Ferrante}, {Ferreira}, {Ferrini}, {Fidecaro}, {Finn}, {Fiori}, {Fiorucci}, {Fisher}, {Flaminio}, {Fletcher}, {Fong}, {Fournier}, {Franco}, {Frasca}, {Frasconi}, {Frede}, {Frei}, {Freise}, {Frey}, {Frey}, {Fricke}, {Fritschel}, {Frolov}, {Fulda}, {Fyffe}, {Gabbard}, {Gair}, {Gammaitoni}, {Gaonkar}, {Garufi}, {Gatto}, {Gaur}, {Gehrels}, {Gemme}, {Gendre}, {Genin}, {Gennai}, {George}, {Gergely}, {Germain}, {Ghosh}, {Ghosh}, {Ghosh}, {Giaime}, {Giardina}, {Giazotto}, {Gill}, {Glaefke}, {Gleason}, {Goetz}, {Goetz}, {Gondan}, {Gonz{\'a}lez}, {Castro}, {Gopakumar}, {Gordon}, {Gorodetsky}, {Gossan}, {Gosselin}, {Gouaty}, {Graef}, {Graff}, {Granata}, {Grant}, {Gras}, {Gray}, {Greco}, {Green}, {Greenhalgh}, {Groot}, {Grote}, {Grunewald}, {Guidi}, {Guo}, {Gupta}, {Gupta}, {Gushwa}, {Gustafson}, {Gustafson}, {Hacker}, {Hall}, {Hall}, {Hammond}, {Haney}, {Hanke}, {Hanks}, {Hanna}, {Hannam}, {Hanson}, {Hardwick}, {Harms}, {Harry}, {Harry}, {Hart}, {Hartman}, {Haster}, {Haughian},
  {Healy}, {Heefner}, {Heidmann}, {Heintze}, {Heinzel}, {Heitmann}, {Hello}, {Hemming}, {Hendry}, {Heng}, {Hennig}, {Heptonstall}, {Heurs}, {Hild}, {Hoak}, {Hodge}, {Hofman}, {Hollitt}, {Holt}, {Holz}, {Hopkins}, {Hosken}, {Hough}, {Houston}, {Howell}, {Hu}, {Huang}, {Huerta}, {Huet}, {Hughey}, {Husa}, {Huttner}, {Huynh-Dinh}, {Idrisy}, {Indik}, {Ingram}, {Inta}, {Isa}, {Isac}, {Isi}, {Islas}, {Isogai}, {Iyer}, {Izumi}, {Jacobson}, {Jacqmin}, {Jang}, {Jani}, {Jaranowski}, {Jawahar}, {Jim{\'e}nez-Forteza}, {Johnson}, {Johnson-McDaniel}, {Jones}, {Jones}, {Jonker}, {Ju}, {Haris}, {Kalaghatgi}, {Kalogera}, {Kandhasamy}, {Kang}, {Kanner}, {Karki}, {Kasprzack}, {Katsavounidis}, {Katzman}, {Kaufer}, {Kaur}, {Kawabe}, {Kawazoe}, {K{\'e}f{\'e}lian}, {Kehl}, {Keitel}, {Kelley}, {Kells}, {Kennedy}, {Keppel}, {Key}, {Khalaidovski}, {Khalili}, {Khan}, {Khan}, {Khan}, {Khazanov}, {Kijbunchoo}, {Kim}, {Kim}, {Kim}, {Kim}, {Kim}, {Kim}, {King}, {King}, {Kinzel}, {Kissel}, {Kleybolte}, {Klimenko}, {Koehlenbeck}, {Kokeyama},
  {Koley}, {Kondrashov}, {Kontos}, {Koranda}, {Korobko}, {Korth}, {Kowalska}, {Kozak}, {Kringel}, {Krishnan}, {Kr{\'o}lak}, {Krueger}, {Kuehn}, {Kumar}, {Kumar}, {Kuo}, {Kutynia}, {Kwee}, {Lackey}, {Landry}, {Lange}, {Lantz}, {Lasky}, {Lazzarini}, {Lazzaro}, {Leaci}, {Leavey}, {Lebigot}, {Lee}, {Lee}, {Lee}, {Lee}, {Lenon}, {Leonardi}, {Leong}, {Leroy}, {Letendre}, {Levin}, {Levine}, {Li}, {Libson}, {Littenberg}, {Lockerbie}, {Logue}, {Lombardi}, {London}, {Lord}, {Lorenzini}, {Loriette}, {Lormand}, {Losurdo}, {Lough}, {Lousto}, {Lovelace}, {L{\"u}ck}, {Lundgren}, {Luo}, {Lynch}, {Ma}, {MacDonald}, {Machenschalk}, {MacInnis}, {Macleod}, {Maga{\~n}a-Sandoval}, {Magee}, {Mageswaran}, {Majorana}, {Maksimovic}, {Malvezzi}, {Man}, {Mandel}, {Mandic}, {Mangano}, {Mansell}, {Manske}, {Mantovani}, {Marchesoni}, {Marion}, {M{\'a}rka}, {M{\'a}rka}, {Markosyan}, {Maros}, {Martelli}, {Martellini}, {Martin}, {Martin}, {Martynov}, {Marx}, {Mason}, {Masserot}, {Massinger}, {Masso-Reid}, {Matichard}, {Matone}, {Mavalvala},
  {Mazumder}, {Mazzolo}, {McCarthy}, {McClelland}, {McCormick}, {McGuire}, {McIntyre}, {McIver}, {McManus}, {McWilliams}, {Meacher}, {Meadors}, {Meidam}, {Melatos}, {Mendell}, {Mendoza-Gandara}, {Mercer}, {Merilh}, {Merzougui}, {Meshkov}, {Messenger}, {Messick}, {Meyers}, {Mezzani}, {Miao}, {Michel}, {Middleton}, {Mikhailov}, {Milano}, {Miller}, {Millhouse}, {Minenkov}, {Ming}, {Mirshekari}, {Mishra}, {Mitra}, {Mitrofanov}, {Mitselmakher}, {Mittleman}, {Moggi}, {Mohan}, {Mohapatra}, {Montani}, {Moore}, {Moore}, {Moraru}, {Moreno}, {Morriss}, {Mossavi}, {Mours}, {Mow-Lowry}, {Mueller}, {Mueller}, {Muir}, {Mukherjee}, {Mukherjee}, {Mukherjee}, {Mukund}, {Mullavey}, {Munch}, {Murphy}, {Murray}, {Mytidis}, {Nardecchia}, {Naticchioni}, {Nayak}, {Necula}, {Nedkova}, {Nelemans}, {Neri}, {Neunzert}, {Newton}, {Nguyen}, {Nielsen}, {Nissanke}, {Nitz}, {Nocera}, {Nolting}, {Normandin}, {Nuttall}, {Oberling}, {Ochsner}, {O'Dell}, {Oelker}, {Ogin}, {Oh}, {Oh}, {Ohme}, {Oliver}, {Oppermann}, {Oram}, {O'Reilly},
  {O'Shaughnessy}, {Ott}, {Ottaway}, {Ottens}, {Overmier}, {Owen}, {Pai}, {Pai}, {Palamos}, {Palashov}, {Palomba}, {Pal-Singh}, {Pan}, {Pan}, {Pankow}, {Pannarale}, {Pant}, {Paoletti}, {Paoli}, {Papa}, {Paris}, {Parker}, {Pascucci}, {Pasqualetti}, {Passaquieti}, {Passuello}, {Patricelli}, {Patrick}, {Pearlstone}, {Pedraza}, {Pedurand}, {Pekowsky}, {Pele}, {Penn}, {Perreca}, {Pfeiffer}, {Phelps}, {Piccinni}, {Pichot}, {Pickenpack}, {Piergiovanni}, {Pierro}, {Pillant}, {Pinard}, {Pinto}, {Pitkin}, {Poeld}, {Poggiani}, {Popolizio}, {Post}, {Powell}, {Prasad}, {Predoi}, {Premachandra}, {Prestegard}, {Price}, {Prijatelj}, {Principe}, {Privitera}, {Prix}, {Prodi}, {Prokhorov}, {Puncken}, {Punturo}, {Puppo}, {P{\"u}rrer}, {Qi}, {Qin}, {Quetschke}, {Quintero}, {Quitzow-James}, {Raab}, {Rabeling}, {Radkins}, {Raffai}, {Raja}, {Rakhmanov}, {Ramet}, {Rapagnani}, {Raymond}, {Razzano}, {Re}, {Read}, {Reed}, {Regimbau}, {Rei}, {Reid}, {Reitze}, {Rew}, {Reyes}, {Ricci}, {Riles}, {Robertson}, {Robie}, {Robinet}, {Rocchi},
  {Rolland}, {Rollins}, {Roma}, {Romano}, {Romano}, {Romanov}, {Romie}, {Rosi{\'n}ska}, {Rowan}, {R{\"u}diger}, {Ruggi}, {Ryan}, {Sachdev}, {Sadecki}, {Sadeghian}, {Salconi}, {Saleem}, {Salemi}, {Samajdar}, {Sammut}, {Sampson}, {Sanchez}, {Sandberg}, {Sandeen}, {Sanders}, {Sanders}, {Sassolas}, {Sathyaprakash}, {Saulson}, {Sauter}, {Savage}, {Sawadsky}, {Schale}, {Schilling}, {Schmidt}, {Schmidt}, {Schnabel}, {Schofield}, {Sch{\"o}nbeck}, {Schreiber}, {Schuette}, {Schutz}, {Scott}, {Scott}, {Sellers}, {Sengupta}, {Sentenac}, {Sequino}, {Sergeev}, {Serna}, {Setyawati}, {Sevigny}, {Shaddock}, {Shaffer}, {Shah}, {Shahriar}, {Shaltev}, {Shao}, {Shapiro}, {Shawhan}, {Sheperd}, {Shoemaker}, {Shoemaker}, {Siellez}, {Siemens}, {Sigg}, {Silva}, {Simakov}, {Singer}, {Singer}, {Singh}, {Singh}, {Singhal}, {Sintes}, {Slagmolen}, {Smith}, {Smith}, {Smith}, {Smith}, {Son}, {Sorazu}, {Sorrentino}, {Souradeep}, {Srivastava}, {Staley}, {Steinke}, {Steinlechner}, {Steinlechner}, {Steinmeyer}, {Stephens}, {Stevenson}, {Stone},
  {Strain}, {Straniero}, {Stratta}, {Strauss}, {Strigin}, {Sturani}, {Stuver}, {Summerscales}, {Sun}, {Sutton}, {Swinkels}, {Szczepa{\'n}czyk}, {Tacca}, {Talukder}, {Tanner}, {T{\'a}pai}, {Tarabrin}, {Taracchini}, {Taylor}, {Theeg}, {Thirugnanasambandam}, {Thomas}, {Thomas}, {Thomas}, {Thorne}, {Thorne}, {Thrane}, {Tiwari}, {Tiwari}, {Tokmakov}, {Tomlinson}, {Tonelli}, {Torres}, {Torrie}, {T{\"o}yr{\"a}}, {Travasso}, {Traylor}, {Trifir{\`o}}, {Tringali}, {Trozzo}, {Tse}, {Turconi}, {Tuyenbayev}, {Ugolini}, {Unnikrishnan}, {Urban}, {Usman}, {Vahlbruch}, {Vajente}, {Valdes}, {Vallisneri}, {van Bakel}, {van Beuzekom}, {van den Brand}, {Van Den Broeck}, {Vander-Hyde}, {van der Schaaf}, {van Heijningen}, {van Veggel}, {Vardaro}, {Vass}, {Vas{\'u}th}, {Vaulin}, {Vecchio}, {Vedovato}, {Veitch}, {Veitch}, {Venkateswara}, {Verkindt}, {Vetrano}, {Vicer{\'e}}, {Vinciguerra}, {Vine}, {Vinet}, {Vitale}, {Vo}, {Vocca}, {Vorvick}, {Voss}, {Vousden}, {Vyatchanin}, {Wade}, {Wade}, {Wade}, {Waldman}, {Walker}, {Wallace},
  {Walsh}, {Wang}, {Wang}, {Wang}, {Wang}, {Wang}, {Ward}, {Ward}, {Warner}, {Was}, {Weaver}, {Wei}, {Weinert}, {Weinstein}, {Weiss}, {Welborn}, {Wen}, {We{\ss}els}, {Westphal}, {Wette}, {Whelan}, {Whitcomb}, {White}, {Whiting}, {Wiesner}, {Wilkinson}, {Willems}, {Williams}, {Williams}, {Williamson}, {Willis}, {Willke}, {Wimmer}, {Winkelmann}, {Winkler}, {Wipf}, {Wiseman}, {Wittel}, {Woan}, {Worden}, {Wright}, {Wu}, {Yablon}, {Yakushin}, {Yam}, {Yamamoto}, {Yancey}, {Yap}, {Yu}, {Yvert}, {Zadro{\.Z}ny}, {Zangrando}, {Zanolin}, {Zendri}, {Zevin}, {Zhang}, {Zhang}, {Zhang}, {Zhang}, {Zhao}, {Zhou}, {Zhou}, {Zhu}, {Zucker}, {Zuraw}, {Zweizig}, {LIGO Scientific Collaboration}, \& {Virgo Collaboration}}]{abbott16}
{Abbott}, B.~P., {Abbott}, R., {Abbott}, T.~D., {et~al.} 2016, \prl, 116, 061102, \dodoi{10.1103/PhysRevLett.116.061102}

\bibitem[{{Abbott} {et~al.}(2017{\natexlab{a}}){Abbott}, {Abbott}, {Abbott}, {Acernese}, {Ackley}, {Adams}, {Adams}, {Addesso}, {Adhikari}, {Adya}, {Affeldt}, {Afrough}, {Agarwal}, {Agathos}, {Agatsuma}, {Aggarwal}, {Aguiar}, {Aiello}, {Ain}, {Ajith}, {Allen}, {Allen}, {Allocca}, {Altin}, {Amato}, {Ananyeva}, {Anderson}, {Anderson}, {Angelova}, {Antier}, {Appert}, {Arai}, {Araya}, {Areeda}, {Arnaud}, {Arun}, {Ascenzi}, {Ashton}, {Ast}, {Aston}, {Astone}, {Atallah}, {Aufmuth}, {Aulbert}, {AultONeal}, {Austin}, {Avila-Alvarez}, {Babak}, {Bacon}, {Bader}, {Bae}, {Bailes}, {Baker}, {Baldaccini}, {Ballardin}, {Ballmer}, {Banagiri}, {Barayoga}, {Barclay}, {Barish}, {Barker}, {Barkett}, {Barone}, {Barr}, {Barsotti}, {Barsuglia}, {Barta}, {Barthelmy}, {Bartlett}, {Bartos}, {Bassiri}, {Basti}, {Batch}, {Bawaj}, {Bayley}, {Bazzan}, {B{\'e}csy}, {Beer}, {Bejger}, {Belahcene}, {Bell}, {Berger}, {Bergmann}, {Bernuzzi}, {Bero}, {Berry}, {Bersanetti}, {Bertolini}, {Betzwieser}, {Bhagwat}, {Bhandare}, {Bilenko},
  {Billingsley}, {Billman}, {Birch}, {Birney}, {Birnholtz}, {Biscans}, {Biscoveanu}, {Bisht}, {Bitossi}, {Biwer}, {Bizouard}, {Blackburn}, {Blackman}, {Blair}, {Blair}, {Blair}, {Bloemen}, {Bock}, {Bode}, {Boer}, {Bogaert}, {Bohe}, {Bondu}, {Bonilla}, {Bonnand}, {Boom}, {Bork}, {Boschi}, {Bose}, {Bossie}, {Bouffanais}, {Bozzi}, {Bradaschia}, {Brady}, {Branchesi}, {Brau}, {Briant}, {Brillet}, {Brinkmann}, {Brisson}, {Brockill}, {Broida}, {Brooks}, {Brown}, {Brown}, {Brunett}, {Buchanan}, {Buikema}, {Bulik}, {Bulten}, {Buonanno}, {Buskulic}, {Buy}, {Byer}, {Cabero}, {Cadonati}, {Cagnoli}, {Cahillane}, {Calder{\'o}n Bustillo}, {Callister}, {Calloni}, {Camp}, {Canepa}, {Canizares}, {Cannon}, {Cao}, {Cao}, {Capano}, {Capocasa}, {Carbognani}, {Caride}, {Carney}, {Carullo}, {Casanueva Diaz}, {Casentini}, {Caudill}, {Cavagli{\`a}}, {Cavalier}, {Cavalieri}, {Cella}, {Cepeda}, {Cerd{\'a}-Dur{\'a}n}, {Cerretani}, {Cesarini}, {Chamberlin}, {Chan}, {Chao}, {Charlton}, {Chase}, {Chassande-Mottin}, {Chatterjee},
  {Chatziioannou}, {Cheeseboro}, {Chen}, {Chen}, {Chen}, {Cheng}, {Chia}, {Chincarini}, {Chiummo}, {Chmiel}, {Cho}, {Cho}, {Chow}, {Christensen}, {Chu}, {Chua}, {Chua}, {Chung}, {Chung}, {Ciani}, {Ciolfi}, {Cirelli}, {Cirone}, {Clara}, {Clark}, {Clearwater}, {Cleva}, {Cocchieri}, {Coccia}, {Cohadon}, {Cohen}, {Colla}, {Collette}, {Cominsky}, {Constancio}, {Conti}, {Cooper}, {Corban}, {Corbitt}, {Cordero-Carri{\'o}n}, {Corley}, {Cornish}, {Corsi}, {Cortese}, {Costa}, {Coughlin}, {Coughlin}, {Coulon}, {Countryman}, {Couvares}, {Covas}, {Cowan}, {Coward}, {Cowart}, {Coyne}, {Coyne}, {Creighton}, {Creighton}, {Cripe}, {Crowder}, {Cullen}, {Cumming}, {Cunningham}, {Cuoco}, {Dal Canton}, {D{\'a}lya}, {Danilishin}, {D'Antonio}, {Danzmann}, {Dasgupta}, {Da Silva Costa}, {Dattilo}, {Dave}, {Davier}, {Davis}, {Daw}, {Day}, {De}, {DeBra}, {Degallaix}, {De Laurentis}, {Del{\'e}glise}, {Del Pozzo}, {Demos}, {Denker}, {Dent}, {De Pietri}, {Dergachev}, {De Rosa}, {DeRosa}, {De Rossi}, {DeSalvo}, {de Varona}, {Devenson},
  {Dhurandhar}, {D{\'\i}az}, {Dietrich}, {Di Fiore}, {Di Giovanni}, {Di Girolamo}, {Di Lieto}, {Di Pace}, {Di Palma}, {Di Renzo}, {Doctor}, {Dolique}, {Donovan}, {Dooley}, {Doravari}, {Dorrington}, {Douglas}, {Dovale {\'A}lvarez}, {Downes}, {Drago}, {Dreissigacker}, {Driggers}, {Du}, {Ducrot}, {Dudi}, {Dupej}, {Dwyer}, {Edo}, {Edwards}, {Effler}, {Eggenstein}, {Ehrens}, {Eichholz}, {Eikenberry}, {Eisenstein}, {Essick}, {Estevez}, {Etienne}, {Etzel}, {Evans}, {Evans}, {Factourovich}, {Fafone}, {Fair}, {Fairhurst}, {Fan}, {Farinon}, {Farr}, {Farr}, {Fauchon-Jones}, {Favata}, {Fays}, {Fee}, {Fehrmann}, {Feicht}, {Fejer}, {Fernandez-Galiana}, {Ferrante}, {Ferreira}, {Ferrini}, {Fidecaro}, {Finstad}, {Fiori}, {Fiorucci}, {Fishbach}, {Fisher}, {Fitz-Axen}, {Flaminio}, {Fletcher}, {Fong}, {Font}, {Forsyth}, {Forsyth}, {Fournier}, {Frasca}, {Frasconi}, {Frei}, {Freise}, {Frey}, {Frey}, {Fries}, {Fritschel}, {Frolov}, {Fulda}, {Fyffe}, {Gabbard}, {Gadre}, {Gaebel}, {Gair}, {Gammaitoni}, {Ganija}, {Gaonkar},
  {Garcia-Quiros}, {Garufi}, {Gateley}, {Gaudio}, {Gaur}, {Gayathri}, {Gehrels}, {Gemme}, {Genin}, {Gennai}, {George}, {George}, {Gergely}, {Germain}, {Ghonge}, {Ghosh}, {Ghosh}, {Ghosh}, {Giaime}, {Giardina}, {Giazotto}, {Gill}, {Glover}, {Goetz}, {Goetz}, {Gomes}, {Goncharov}, {Gonz{\'a}lez}, {Gonzalez Castro}, {Gopakumar}, {Gorodetsky}, {Gossan}, {Gosselin}, {Gouaty}, {Grado}, {Graef}, {Granata}, {Grant}, {Gras}, {Gray}, {Greco}, {Green}, {Gretarsson}, {Groot}, {Grote}, {Grunewald}, {Gruning}, {Guidi}, {Guo}, {Gupta}, {Gupta}, {Gushwa}, {Gustafson}, {Gustafson}, {Halim}, {Hall}, {Hall}, {Hamilton}, {Hammond}, {Haney}, {Hanke}, {Hanks}, {Hanna}, {Hannam}, {Hannuksela}, {Hanson}, {Hardwick}, {Harms}, {Harry}, {Harry}, {Hart}, {Haster}, {Haughian}, {Healy}, {Heidmann}, {Heintze}, {Heitmann}, {Hello}, {Hemming}, {Hendry}, {Heng}, {Hennig}, {Heptonstall}, {Heurs}, {Hild}, {Hinderer}, {Ho}, {Hoak}, {Hofman}, {Holt}, {Holz}, {Hopkins}, {Horst}, {Hough}, {Houston}, {Howell}, {Hreibi}, {Hu}, {Huerta}, {Huet},
  {Hughey}, {Husa}, {Huttner}, {Huynh-Dinh}, {Indik}, {Inta}, {Intini}, {Isa}, {Isac}, {Isi}, {Iyer}, {Izumi}, {Jacqmin}, {Jani}, {Jaranowski}, {Jawahar}, {Jim{\'e}nez-Forteza}, {Johnson}, {Johnson-McDaniel}, {Jones}, {Jones}, {Jonker}, {Ju}, {Junker}, {Kalaghatgi}, {Kalogera}, {Kamai}, {Kandhasamy}, {Kang}, {Kanner}, {Kapadia}, {Karki}, {Karvinen}, {Kasprzack}, {Kastaun}, {Katolik}, {Katsavounidis}, {Katzman}, {Kaufer}, {Kawabe}, {K{\'e}f{\'e}lian}, {Keitel}, {Kemball}, {Kennedy}, {Kent}, {Key}, {Khalili}, {Khan}, {Khan}, {Khan}, {Khazanov}, {Kijbunchoo}, {Kim}, {Kim}, {Kim}, {Kim}, {Kim}, {Kim}, {Kimbrell}, {King}, {King}, {Kinley-Hanlon}, {Kirchhoff}, {Kissel}, {Kleybolte}, {Klimenko}, {Knowles}, {Koch}, {Koehlenbeck}, {Koley}, {Kondrashov}, {Kontos}, {Korobko}, {Korth}, {Kowalska}, {Kozak}, {Kr{\"a}mer}, {Kringel}, {Krishnan}, {Kr{\'o}lak}, {Kuehn}, {Kumar}, {Kumar}, {Kumar}, {Kuo}, {Kutynia}, {Kwang}, {Lackey}, {Lai}, {Landry}, {Lang}, {Lange}, {Lantz}, {Lanza}, {Larson}, {Lartaux-Vollard}, {Lasky},
  {Laxen}, {Lazzarini}, {Lazzaro}, {Leaci}, {Leavey}, {Lee}, {Lee}, {Lee}, {Lee}, {Lee}, {Lehmann}, {Lenon}, {Leon}, {Leonardi}, {Leroy}, {Letendre}, {Levin}, {Li}, {Linker}, {Littenberg}, {Liu}, {Liu}, {Lo}, {Lockerbie}, {London}, {Lord}, {Lorenzini}, {Loriette}, {Lormand}, {Losurdo}, {Lough}, {Lousto}, {Lovelace}, {L{\"u}ck}, {Lumaca}, {Lundgren}, {Lynch}, {Ma}, {Macas}, {Macfoy}, {Machenschalk}, {MacInnis}, {Macleod}, {Maga{\~n}a Hernandez}, {Maga{\~n}a-Sandoval}, {Maga{\~n}a Zertuche}, {Magee}, {Majorana}, {Maksimovic}, {Man}, {Mandic}, {Mangano}, {Mansell}, {Manske}, {Mantovani}, {Marchesoni}, {Marion}, {M{\'a}rka}, {M{\'a}rka}, {Markakis}, {Markosyan}, {Markowitz}, {Maros}, {Marquina}, {Marsh}, {Martelli}, {Martellini}, {Martin}, {Martin}, {Martynov}, {Marx}, {Mason}, {Massera}, {Masserot}, {Massinger}, {Masso-Reid}, {Mastrogiovanni}, {Matas}, {Matichard}, {Matone}, {Mavalvala}, {Mazumder}, {McCarthy}, {McClelland}, {McCormick}, {McCuller}, {McGuire}, {McIntyre}, {McIver}, {McManus}, {McNeill}, {McRae},
  {McWilliams}, {Meacher}, {Meadors}, {Mehmet}, {Meidam}, {Mejuto-Villa}, {Melatos}, {Mendell}, {Mercer}, {Merilh}, {Merzougui}, {Meshkov}, {Messenger}, {Messick}, {Metzdorff}, {Meyers}, {Miao}, {Michel}, {Middleton}, {Mikhailov}, {Milano}, {Miller}, {Miller}, {Miller}, {Millhouse}, {Milovich-Goff}, {Minazzoli}, {Minenkov}, {Ming}, {Mishra}, {Mitra}, {Mitrofanov}, {Mitselmakher}, {Mittleman}, {Moffa}, {Moggi}, {Mogushi}, {Mohan}, {Mohapatra}, {Molina}, {Montani}, {Moore}, {Moraru}, {Moreno}, {Morisaki}, {Morriss}, {Mours}, {Mow-Lowry}, {Mueller}, {Muir}, {Mukherjee}, {Mukherjee}, {Mukherjee}, {Mukund}, {Mullavey}, {Munch}, {Mu{\~n}iz}, {Muratore}, {Murray}, {Nagar}, {Napier}, {Nardecchia}, {Naticchioni}, {Nayak}, {Neilson}, {Nelemans}, {Nelson}, {Nery}, {Neunzert}, {Nevin}, {Newport}, {Newton}, {Ng}, {Nguyen}, {Nguyen}, {Nichols}, {Nielsen}, {Nissanke}, {Nitz}, {Noack}, {Nocera}, {Nolting}, {North}, {Nuttall}, {Oberling}, {O'Dea}, {Ogin}, {Oh}, {Oh}, {Ohme}, {Okada}, {Oliver}, {Oppermann}, {Oram}, {O'Reilly},
  {Ormiston}, {Ortega}, {O'Shaughnessy}, {Ossokine}, {Ottaway}, {Overmier}, {Owen}, {Pace}, {Page}, {Page}, {Pai}, {Pai}, {Palamos}, {Palashov}, {Palomba}, {Pal-Singh}, {Pan}, {Pan}, {Pang}, {Pang}, {Pankow}, {Pannarale}, {Pant}, {Paoletti}, {Paoli}, {Papa}, {Parida}, {Parker}, {Pascucci}, {Pasqualetti}, {Passaquieti}, {Passuello}, {Patil}, {Patricelli}, {Pearlstone}, {Pedraza}, {Pedurand}, {Pekowsky}, {Pele}, {Penn}, {Perez}, {Perreca}, {Perri}, {Pfeiffer}, {Phelps}, {Piccinni}, {Pichot}, {Piergiovanni}, {Pierro}, {Pillant}, {Pinard}, {Pinto}, {Pirello}, {Pitkin}, {Poe}, {Poggiani}, {Popolizio}, {Porter}, {Post}, {Powell}, {Prasad}, {Pratt}, {Pratten}, {Predoi}, {Prestegard}, {Prijatelj}, {Principe}, {Privitera}, {Prix}, {Prodi}, {Prokhorov}, {Puncken}, {Punturo}, {Puppo}, {P{\"u}rrer}, {Qi}, {Quetschke}, {Quintero}, {Quitzow-James}, {Raab}, {Rabeling}, {Radkins}, {Raffai}, {Raja}, {Rajan}, {Rajbhandari}, {Rakhmanov}, {Ramirez}, {Ramos-Buades}, {Rapagnani}, {Raymond}, {Razzano}, {Read}, {Regimbau}, {Rei},
  {Reid}, {Reitze}, {Ren}, {Reyes}, {Ricci}, {Ricker}, {Rieger}, {Riles}, {Rizzo}, {Robertson}, {Robie}, {Robinet}, {Rocchi}, {Rolland}, {Rollins}, {Roma}, {Romano}, {Romano}, {Romel}, {Romie}, {Rosi{\'n}ska}, {Ross}, {Rowan}, {R{\"u}diger}, {Ruggi}, {Rutins}, {Ryan}, {Sachdev}, {Sadecki}, {Sadeghian}, {Sakellariadou}, {Salconi}, {Saleem}, {Salemi}, {Samajdar}, {Sammut}, {Sampson}, {Sanchez}, {Sanchez}, {Sanchis-Gual}, {Sandberg}, {Sanders}, {Sassolas}, {Sathyaprakash}, {Saulson}, {Sauter}, {Savage}, {Sawadsky}, {Schale}, {Scheel}, {Scheuer}, {Schmidt}, {Schmidt}, {Schnabel}, {Schofield}, {Sch{\"o}nbeck}, {Schreiber}, {Schuette}, {Schulte}, {Schutz}, {Schwalbe}, {Scott}, {Scott}, {Seidel}, {Sellers}, {Sengupta}, {Sentenac}, {Sequino}, {Sergeev}, {Shaddock}, {Shaffer}, {Shah}, {Shahriar}, {Shaner}, {Shao}, {Shapiro}, {Shawhan}, {Sheperd}, {Shoemaker}, {Shoemaker}, {Siellez}, {Siemens}, {Sieniawska}, {Sigg}, {Silva}, {Singer}, {Singh}, {Singhal}, {Sintes}, {Slagmolen}, {Smith}, {Smith}, {Smith}, {Somala},
  {Son}, {Sonnenberg}, {Sorazu}, {Sorrentino}, {Souradeep}, {Spencer}, {Srivastava}, {Staats}, {Staley}, {Steinke}, {Steinlechner}, {Steinlechner}, {Steinmeyer}, {Stevenson}, {Stone}, {Stops}, {Strain}, {Stratta}, {Strigin}, {Strunk}, {Sturani}, {Stuver}, {Summerscales}, {Sun}, {Sunil}, {Suresh}, {Sutton}, {Swinkels}, {Szczepa{\'n}czyk}, {Tacca}, {Tait}, {Talbot}, {Talukder}, {Tanner}, {T{\'a}pai}, {Taracchini}, {Tasson}, {Taylor}, {Taylor}, {Tewari}, {Theeg}, {Thies}, {Thomas}, {Thomas}, {Thomas}, {Thorne}, {Thorne}, {Thrane}, {Tiwari}, {Tiwari}, {Tokmakov}, {Toland}, {Tonelli}, {Tornasi}, {Torres-Forn{\'e}}, {Torrie}, {T{\"o}yr{\"a}}, {Travasso}, {Traylor}, {Trinastic}, {Tringali}, {Trozzo}, {Tsang}, {Tse}, {Tso}, {Tsukada}, {Tsuna}, {Tuyenbayev}, {Ueno}, {Ugolini}, {Unnikrishnan}, {Urban}, {Usman}, {Vahlbruch}, {Vajente}, {Valdes}, {Vallisneri}, {van Bakel}, {van Beuzekom}, {van den Brand}, {Van Den Broeck}, {Vander-Hyde}, {van der Schaaf}, {van Heijningen}, {van Veggel}, {Vardaro}, {Varma}, {Vass},
  {Vas{\'u}th}, {Vecchio}, {Vedovato}, {Veitch}, {Veitch}, {Venkateswara}, {Venugopalan}, {Verkindt}, {Vetrano}, {Vicer{\'e}}, {Viets}, {Vinciguerra}, {Vine}, {Vinet}, {Vitale}, {Vo}, {Vocca}, {Vorvick}, {Vyatchanin}, {Wade}, {Wade}, {Wade}, {Walet}, {Walker}, {Wallace}, {Walsh}, {Wang}, {Wang}, {Wang}, {Wang}, {Wang}, {Ward}, {Warner}, {Was}, {Watchi}, {Weaver}, {Wei}, {Weinert}, {Weinstein}, {Weiss}, {Wen}, {Wessel}, {We{\ss}els}, {Westerweck}, {Westphal}, {Wette}, {Whelan}, {Whitcomb}, {Whiting}, {Whittle}, {Wilken}, {Williams}, {Williams}, {Williamson}, {Willis}, {Willke}, {Wimmer}, {Winkler}, {Wipf}, {Wittel}, {Woan}, {Woehler}, {Wofford}, {Wong}, {Worden}, {Wright}, {Wu}, {Wysocki}, {Xiao}, {Yamamoto}, {Yancey}, {Yang}, {Yap}, {Yazback}, {Yu}, {Yu}, {Yvert}, {Zadro{\.Z}ny}, {Zanolin}, {Zelenova}, {Zendri}, {Zevin}, {Zhang}, {Zhang}, {Zhang}, {Zhang}, {Zhao}, {Zhou}, {Zhou}, {Zhu}, {Zhu}, {Zimmerman}, {Zucker}, {Zweizig}, {LIGO Scientific Collaboration}, \& {Virgo Collaboration}}]{Abbott17a}
---. 2017{\natexlab{a}}, \prl, 119, 161101, \dodoi{10.1103/PhysRevLett.119.161101}

\bibitem[{{Abbott} {et~al.}(2017{\natexlab{b}}){Abbott}, {Abbott}, {Abbott}, {Acernese}, {Ackley}, {Adams}, {Adams}, {Addesso}, {Adhikari}, {Adya}, {Affeldt}, {Afrough}, {Agarwal}, {Agathos}, {Agatsuma}, {Aggarwal}, {Aguiar}, {Aiello}, {Ain}, {Ajith}, {Allen}, {Allen}, {Allocca}, {Altin}, {Amato}, {Ananyeva}, {Anderson}, {Anderson}, {Angelova}, {Antier}, {Appert}, {Arai}, {Araya}, {Areeda}, {Arnaud}, {Arun}, {Ascenzi}, {Ashton}, {Ast}, {Aston}, {Astone}, {Atallah}, {Aufmuth}, {Aulbert}, {AultONeal}, {Austin}, {Avila-Alvarez}, {Babak}, {Bacon}, {Bader}, {Bae}, {Baker}, {Baldaccini}, {Ballardin}, {Ballmer}, {Banagiri}, {Barayoga}, {Barclay}, {Barish}, {Barker}, {Barkett}, {Barone}, {Barr}, {Barsotti}, {Barsuglia}, {Barta}, {Barthelmy}, {Bartlett}, {Bartos}, {Bassiri}, {Basti}, {Batch}, {Bawaj}, {Bayley}, {Bazzan}, {B{\'e}csy}, {Beer}, {Bejger}, {Belahcene}, {Bell}, {Berger}, {Bergmann}, {Bero}, {Berry}, {Bersanetti}, {Bertolini}, {Betzwieser}, {Bhagwat}, {Bhandare}, {Bilenko}, {Billingsley}, {Billman}, {Birch},
  {Birney}, {Birnholtz}, {Biscans}, {Biscoveanu}, {Bisht}, {Bitossi}, {Biwer}, {Bizouard}, {Blackburn}, {Blackman}, {Blair}, {Blair}, {Blair}, {Bloemen}, {Bock}, {Bode}, {Boer}, {Bogaert}, {Bohe}, {Bondu}, {Bonilla}, {Bonnand}, {Boom}, {Bork}, {Boschi}, {Bose}, {Bossie}, {Bouffanais}, {Bozzi}, {Bradaschia}, {Brady}, {Branchesi}, {Brau}, {Briant}, {Brillet}, {Brinkmann}, {Brisson}, {Brockill}, {Broida}, {Brooks}, {Brown}, {Brown}, {Brunett}, {Buchanan}, {Buikema}, {Bulik}, {Bulten}, {Buonanno}, {Buskulic}, {Buy}, {Byer}, {Cabero}, {Cadonati}, {Cagnoli}, {Cahillane}, {Calder{\'o}n Bustillo}, {Callister}, {Calloni}, {Camp}, {Canepa}, {Canizares}, {Cannon}, {Cao}, {Cao}, {Capano}, {Capocasa}, {Carbognani}, {Caride}, {Carney}, {Casanueva Diaz}, {Casentini}, {Caudill}, {Cavagli{\`a}}, {Cavalier}, {Cavalieri}, {Cella}, {Cepeda}, {Cerd{\'a}-Dur{\'a}n}, {Cerretani}, {Cesarini}, {Chamberlin}, {Chan}, {Chao}, {Charlton}, {Chase}, {Chassande-Mottin}, {Chatterjee}, {Chatziioannou}, {Cheeseboro}, {Chen}, {Chen}, {Chen},
  {Cheng}, {Chia}, {Chincarini}, {Chiummo}, {Chmiel}, {Cho}, {Cho}, {Chow}, {Christensen}, {Chu}, {Chua}, {Chua}, {Chung}, {Chung}, {Ciani}, {Ciolfi}, {Cirelli}, {Cirone}, {Clara}, {Clark}, {Clearwater}, {Cleva}, {Cocchieri}, {Coccia}, {Cohadon}, {Cohen}, {Colla}, {Collette}, {Cominsky}, {Constancio}, {Conti}, {Cooper}, {Corban}, {Corbitt}, {Cordero-Carri{\'o}n}, {Corley}, {Cornish}, {Corsi}, {Cortese}, {Costa}, {Coughlin}, {Coughlin}, {Coulon}, {Countryman}, {Couvares}, {Covas}, {Cowan}, {Coward}, {Cowart}, {Coyne}, {Coyne}, {Creighton}, {Creighton}, {Cripe}, {Crowder}, {Cullen}, {Cumming}, {Cunningham}, {Cuoco}, {Dal Canton}, {D{\'a}lya}, {Danilishin}, {D'Antonio}, {Danzmann}, {Dasgupta}, {Da Silva Costa}, {Dattilo}, {Dave}, {Davier}, {Davis}, {Daw}, {Day}, {De}, {DeBra}, {Degallaix}, {De Laurentis}, {Del{\'e}glise}, {Del Pozzo}, {Demos}, {Denker}, {Dent}, {De Pietri}, {Dergachev}, {De Rosa}, {DeRosa}, {De Rossi}, {DeSalvo}, {de Varona}, {Devenson}, {Dhurandhar}, {D{\'\i}az}, {Di Fiore}, {Di Giovanni}, {Di
  Girolamo}, {Di Lieto}, {Di Pace}, {Di Palma}, {Di Renzo}, {Doctor}, {Dolique}, {Donovan}, {Dooley}, {Doravari}, {Dorrington}, {Douglas}, {Dovale {\'A}lvarez}, {Downes}, {Drago}, {Dreissigacker}, {Driggers}, {Du}, {Ducrot}, {Dupej}, {Dwyer}, {Edo}, {Edwards}, {Effler}, {Ehrens}, {Eichholz}, {Eikenberry}, {Eisenstein}, {Essick}, {Estevez}, {Etienne}, {Etzel}, {Evans}, {Evans}, {Factourovich}, {Fafone}, {Fair}, {Fairhurst}, {Fan}, {Farinon}, {Farr}, {Farr}, {Fauchon-Jones}, {Favata}, {Fays}, {Fee}, {Fehrmann}, {Feicht}, {Fejer}, {Fernandez-Galiana}, {Ferrante}, {Ferreira}, {Ferrini}, {Fidecaro}, {Finstad}, {Fiori}, {Fiorucci}, {Fishbach}, {Fisher}, {Fitz-Axen}, {Flaminio}, {Fletcher}, {Fong}, {Font}, {Forsyth}, {Forsyth}, {Fournier}, {Frasca}, {Frasconi}, {Frei}, {Freise}, {Frey}, {Frey}, {Fries}, {Fritschel}, {Frolov}, {Fulda}, {Fyffe}, {Gabbard}, {Gadre}, {Gaebel}, {Gair}, {Gammaitoni}, {Ganija}, {Gaonkar}, {Garcia-Quiros}, {Garufi}, {Gateley}, {Gaudio}, {Gaur}, {Gayathri}, {Gehrels}, {Gemme}, {Genin},
  {Gennai}, {George}, {George}, {Gergely}, {Germain}, {Ghonge}, {Ghosh}, {Ghosh}, {Ghosh}, {Giaime}, {Giardina}, {Giazotto}, {Gill}, {Glover}, {Goetz}, {Goetz}, {Gomes}, {Goncharov}, {Gonz{\'a}lez}, {Gonzalez Castro}, {Gopakumar}, {Gorodetsky}, {Gossan}, {Gosselin}, {Gouaty}, {Grado}, {Graef}, {Granata}, {Grant}, {Gras}, {Gray}, {Greco}, {Green}, {Gretarsson}, {Griswold}, {Groot}, {Grote}, {Grunewald}, {Gruning}, {Guidi}, {Guo}, {Gupta}, {Gupta}, {Gushwa}, {Gustafson}, {Gustafson}, {Halim}, {Hall}, {Hall}, {Hamilton}, {Hammond}, {Haney}, {Hanke}, {Hanks}, {Hanna}, {Hannam}, {Hannuksela}, {Hanson}, {Hardwick}, {Harms}, {Harry}, {Harry}, {Hart}, {Haster}, {Haughian}, {Healy}, {Heidmann}, {Heintze}, {Heitmann}, {Hello}, {Hemming}, {Hendry}, {Heng}, {Hennig}, {Heptonstall}, {Heurs}, {Hild}, {Hinderer}, {Hoak}, {Hofman}, {Holt}, {Holz}, {Hopkins}, {Horst}, {Hough}, {Houston}, {Howell}, {Hreibi}, {Hu}, {Huerta}, {Huet}, {Hughey}, {Husa}, {Huttner}, {Huynh-Dinh}, {Indik}, {Inta}, {Intini}, {Isa}, {Isac}, {Isi},
  {Iyer}, {Izumi}, {Jacqmin}, {Jani}, {Jaranowski}, {Jawahar}, {Jim{\'e}nez-Forteza}, {Johnson}, {Jones}, {Jones}, {Jonker}, {Ju}, {Junker}, {Kalaghatgi}, {Kalogera}, {Kamai}, {Kandhasamy}, {Kang}, {Kanner}, {Kapadia}, {Karki}, {Karvinen}, {Kasprzack}, {Katolik}, {Katsavounidis}, {Katzman}, {Kaufer}, {Kawabe}, {K{\'e}f{\'e}lian}, {Keitel}, {Kemball}, {Kennedy}, {Kent}, {Key}, {Khalili}, {Khan}, {Khan}, {Khan}, {Khazanov}, {Kijbunchoo}, {Kim}, {Kim}, {Kim}, {Kim}, {Kim}, {Kim}, {Kimbrell}, {King}, {King}, {Kinley-Hanlon}, {Kirchhoff}, {Kissel}, {Kleybolte}, {Klimenko}, {Knowles}, {Koch}, {Koehlenbeck}, {Koley}, {Kondrashov}, {Kontos}, {Korobko}, {Korth}, {Kowalska}, {Kozak}, {Kr{\"a}mer}, {Kringel}, {Krishnan}, {Kr{\'o}lak}, {Kuehn}, {Kumar}, {Kumar}, {Kumar}, {Kuo}, {Kutynia}, {Kwang}, {Lackey}, {Lai}, {Landry}, {Lang}, {Lange}, {Lantz}, {Lanza}, {Larson}, {Lartaux-Vollard}, {Lasky}, {Laxen}, {Lazzarini}, {Lazzaro}, {Leaci}, {Leavey}, {Lee}, {Lee}, {Lee}, {Lee}, {Lee}, {Lehmann}, {Lenon}, {Leonardi}, {Leroy},
  {Letendre}, {Levin}, {Li}, {Linker}, {Littenberg}, {Liu}, {Lo}, {Lockerbie}, {London}, {Lord}, {Lorenzini}, {Loriette}, {Lormand}, {Losurdo}, {Lough}, {Lousto}, {Lovelace}, {L{\"u}ck}, {Lumaca}, {Lundgren}, {Lynch}, {Ma}, {Macas}, {Macfoy}, {Machenschalk}, {MacInnis}, {Macleod}, {Maga{\~n}a Hernandez}, {Maga{\~n}a-Sandoval}, {Maga{\~n}a Zertuche}, {Magee}, {Majorana}, {Maksimovic}, {Man}, {Mandic}, {Mangano}, {Mansell}, {Manske}, {Mantovani}, {Marchesoni}, {Marion}, {M{\'a}rka}, {M{\'a}rka}, {Markakis}, {Markosyan}, {Markowitz}, {Maros}, {Marquina}, {Marsh}, {Martelli}, {Martellini}, {Martin}, {Martin}, {Martynov}, {Mason}, {Massera}, {Masserot}, {Massinger}, {Masso-Reid}, {Mastrogiovanni}, {Matas}, {Matichard}, {Matone}, {Mavalvala}, {Mazumder}, {McCarthy}, {McClelland}, {McCormick}, {McCuller}, {McGuire}, {McIntyre}, {McIver}, {McManus}, {McNeill}, {McRae}, {McWilliams}, {Meacher}, {Meadors}, {Mehmet}, {Meidam}, {Mejuto-Villa}, {Melatos}, {Mendell}, {Mercer}, {Merilh}, {Merzougui}, {Meshkov}, {Messenger},
  {Messick}, {Metzdorff}, {Meyers}, {Miao}, {Michel}, {Middleton}, {Mikhailov}, {Milano}, {Miller}, {Miller}, {Miller}, {Millhouse}, {Milovich-Goff}, {Minazzoli}, {Minenkov}, {Ming}, {Mishra}, {Mitra}, {Mitrofanov}, {Mitselmakher}, {Mittleman}, {Moffa}, {Moggi}, {Mogushi}, {Mohan}, {Mohapatra}, {Montani}, {Moore}, {Moraru}, {Moreno}, {Morriss}, {Mours}, {Mow-Lowry}, {Mueller}, {Muir}, {Mukherjee}, {Mukherjee}, {Mukherjee}, {Mukund}, {Mullavey}, {Munch}, {Mu{\~n}iz}, {Muratore}, {Murray}, {Napier}, {Nardecchia}, {Naticchioni}, {Nayak}, {Neilson}, {Nelemans}, {Nelson}, {Nery}, {Neunzert}, {Nevin}, {Newport}, {Newton}, {Ng}, {Nguyen}, {Nguyen}, {Nichols}, {Nielsen}, {Nissanke}, {Nitz}, {Noack}, {Nocera}, {Nolting}, {North}, {Nuttall}, {Oberling}, {O'Dea}, {Ogin}, {Oh}, {Oh}, {Ohme}, {Okada}, {Oliver}, {Oppermann}, {Oram}, {O'Reilly}, {Ormiston}, {Ortega}, {O'Shaughnessy}, {Ossokine}, {Ottaway}, {Overmier}, {Owen}, {Pace}, {Page}, {Page}, {Pai}, {Pai}, {Palamos}, {Palashov}, {Palomba}, {Pal-Singh}, {Pan}, {Pan},
  {Pang}, {Pang}, {Pankow}, {Pannarale}, {Pant}, {Paoletti}, {Paoli}, {Papa}, {Parida}, {Parker}, {Pascucci}, {Pasqualetti}, {Passaquieti}, {Passuello}, {Patil}, {Patricelli}, {Pearlstone}, {Pedraza}, {Pedurand}, {Pekowsky}, {Pele}, {Penn}, {Perez}, {Perreca}, {Perri}, {Pfeiffer}, {Phelps}, {Piccinni}, {Pichot}, {Piergiovanni}, {Pierro}, {Pillant}, {Pinard}, {Pinto}, {Pirello}, {Pitkin}, {Poe}, {Poggiani}, {Popolizio}, {Porter}, {Post}, {Powell}, {Prasad}, {Pratt}, {Pratten}, {Predoi}, {Prestegard}, {Price}, {Prijatelj}, {Principe}, {Privitera}, {Prodi}, {Prokhorov}, {Puncken}, {Punturo}, {Puppo}, {P{\"u}rrer}, {Qi}, {Quetschke}, {Quintero}, {Quitzow-James}, {Raab}, {Rabeling}, {Radkins}, {Raffai}, {Raja}, {Rajan}, {Rajbhandari}, {Rakhmanov}, {Ramirez}, {Ramos-Buades}, {Rapagnani}, {Raymond}, {Razzano}, {Read}, {Regimbau}, {Rei}, {Reid}, {Reitze}, {Ren}, {Reyes}, {Ricci}, {Ricker}, {Rieger}, {Riles}, {Rizzo}, {Robertson}, {Robie}, {Robinet}, {Rocchi}, {Rolland}, {Rollins}, {Roma}, {Romano}, {Romel}, {Romie},
  {Rosi{\'n}ska}, {Ross}, {Rowan}, {R{\"u}diger}, {Ruggi}, {Rutins}, {Ryan}, {Sachdev}, {Sadecki}, {Sadeghian}, {Sakellariadou}, {Salconi}, {Saleem}, {Salemi}, {Samajdar}, {Sammut}, {Sampson}, {Sanchez}, {Sanchez}, {Sanchis-Gual}, {Sandberg}, {Sanders}, {Sassolas}, {Sathyaprakash}, {Saulson}, {Sauter}, {Savage}, {Sawadsky}, {Schale}, {Scheel}, {Scheuer}, {Schmidt}, {Schmidt}, {Schnabel}, {Schofield}, {Sch{\"o}nbeck}, {Schreiber}, {Schuette}, {Schulte}, {Schutz}, {Schwalbe}, {Scott}, {Scott}, {Seidel}, {Sellers}, {Sengupta}, {Sentenac}, {Sequino}, {Sergeev}, {Shaddock}, {Shaffer}, {Shah}, {Shahriar}, {Shaner}, {Shao}, {Shapiro}, {Shawhan}, {Sheperd}, {Shoemaker}, {Shoemaker}, {Siellez}, {Siemens}, {Sieniawska}, {Sigg}, {Silva}, {Singer}, {Singh}, {Singhal}, {Sintes}, {Slagmolen}, {Smith}, {Smith}, {Smith}, {Somala}, {Son}, {Sonnenberg}, {Sorazu}, {Sorrentino}, {Souradeep}, {Spencer}, {Srivastava}, {Staats}, {Staley}, {Steinke}, {Steinlechner}, {Steinlechner}, {Steinmeyer}, {Stevenson}, {Stone}, {Stops},
  {Strain}, {Stratta}, {Strigin}, {Strunk}, {Sturani}, {Stuver}, {Summerscales}, {Sun}, {Sunil}, {Suresh}, {Sutton}, {Swinkels}, {Szczepa{\'n}czyk}, {Tacca}, {Tait}, {Talbot}, {Talukder}, {Tanner}, {T{\'a}pai}, {Taracchini}, {Tasson}, {Taylor}, {Taylor}, {Tewari}, {Theeg}, {Thies}, {Thomas}, {Thomas}, {Thomas}, {Thorne}, {Thorne}, {Thrane}, {Tiwari}, {Tiwari}, {Tokmakov}, {Toland}, {Tonelli}, {Tornasi}, {Torres-Forn{\'e}}, {Torrie}, {T{\"o}yr{\"a}}, {Travasso}, {Traylor}, {Trinastic}, {Tringali}, {Trozzo}, {Tsang}, {Tse}, {Tso}, {Tsukada}, {Tsuna}, {Tuyenbayev}, {Ueno}, {Ugolini}, {Unnikrishnan}, {Urban}, {Usman}, {Vahlbruch}, {Vajente}, {Valdes}, {van Bakel}, {van Beuzekom}, {van den Brand}, {Van Den Broeck}, {Vander-Hyde}, {van der Schaaf}, {van Heijningen}, {van Veggel}, {Vardaro}, {Varma}, {Vass}, {Vas{\'u}th}, {Vecchio}, {Vedovato}, {Veitch}, {Veitch}, {Venkateswara}, {Venugopalan}, {Verkindt}, {Vetrano}, {Vicer{\'e}}, {Viets}, {Vinciguerra}, {Vine}, {Vinet}, {Vitale}, {Vo}, {Vocca}, {Vorvick},
  {Vyatchanin}, {Wade}, {Wade}, {Wade}, {Walet}, {Walker}, {Wallace}, {Walsh}, {Wang}, {Wang}, {Wang}, {Wang}, {Wang}, {Ward}, {Warner}, {Was}, {Watchi}, {Weaver}, {Wei}, {Weinert}, {Weinstein}, {Weiss}, {Wen}, {Wessel}, {Wessels}, {Westerweck}, {Westphal}, {Wette}, {Whelan}, {Whitcomb}, {Whiting}, {Whittle}, {Wilken}, {Williams}, {Williams}, {Williamson}, {Willis}, {Willke}, {Wimmer}, {Winkler}, {Wipf}, {Wittel}, {Woan}, {Woehler}, {Wofford}, {Wong}, {Worden}, {Wright}, {Wu}, {Wysocki}, {Xiao}, {Yamamoto}, {Yancey}, {Yang}, {Yap}, {Yazback}, {Yu}, {Yu}, {Yvert}, {Zadro{\.z}ny}, {Zanolin}, {Zelenova}, {Zendri}, {Zevin}, {Zhang}, {Zhang}, {Zhang}, {Zhang}, {Zhao}, {Zhou}, {Zhou}, {Zhu}, {Zhu}, {Zimmerman}, {Zucker}, {Zweizig}, {LIGO Scientific Collaboration}, {Virgo Collaboration}, {Wilson-Hodge}, {Bissaldi}, {Blackburn}, {Briggs}, {Burns}, {Cleveland}, {Connaughton}, {Gibby}, {Giles}, {Goldstein}, {Hamburg}, {Jenke}, {Hui}, {Kippen}, {Kocevski}, {McBreen}, {Meegan}, {Paciesas}, {Poolakkil}, {Preece},
  {Racusin}, {Roberts}, {Stanbro}, {Veres}, {von Kienlin}, {GBM}, {Savchenko}, {Ferrigno}, {Kuulkers}, {Bazzano}, {Bozzo}, {Brandt}, {Chenevez}, {Courvoisier}, {Diehl}, {Domingo}, {Hanlon}, {Jourdain}, {Laurent}, {Lebrun}, {Lutovinov}, {Martin-Carrillo}, {Mereghetti}, {Natalucci}, {Rodi}, {Roques}, {Sunyaev}, {Ubertini}, {INTEGRAL}, {Aartsen}, {Ackermann}, {Adams}, {Aguilar}, {Ahlers}, {Ahrens}, {Samarai}, {Altmann}, {Andeen}, {Anderson}, {Ansseau}, {Anton}, {Arg{\"u}elles}, {Auffenberg}, {Axani}, {Bagherpour}, {Bai}, {Barron}, {Barwick}, {Baum}, {Bay}, {Beatty}, {Becker Tjus}, {Bernardini}, {Besson}, {Binder}, {Bindig}, {Blaufuss}, {Blot}, {Bohm}, {B{\"o}rner}, {Bos}, {Bose}, {B{\"o}ser}, {Botner}, {Bourbeau}, {Bourbeau}, {Bradascio}, {Braun}, {Brayeur}, {Brenzke}, {Bretz}, {Bron}, {Brostean-Kaiser}, {Burgman}, {Carver}, {Casey}, {Casier}, {Cheung}, {Chirkin}, {Christov}, {Clark}, {Classen}, {Coenders}, {Collin}, {Conrad}, {Cowen}, {Cross}, {Day}, {de Andr{\'e}}, {De Clercq}, {DeLaunay}, {Dembinski}, {De
  Ridder}, {Desiati}, {de Vries}, {de Wasseige}, {de With}, {DeYoung}, {D{\'\i}az-V{\'e}lez}, {di Lorenzo}, {Dujmovic}, {Dumm}, {Dunkman}, {Dvorak}, {Eberhardt}, {Ehrhardt}, {Eichmann}, {Eller}, {Evenson}, {Fahey}, {Fazely}, {Felde}, {Filimonov}, {Finley}, {Flis}, {Franckowiak}, {Friedman}, {Fuchs}, {Gaisser}, {Gallagher}, {Gerhardt}, {Ghorbani}, {Giang}, {Glauch}, {Gl{\"u}senkamp}, {Goldschmidt}, {Gonzalez}, {Grant}, {Griffith}, {Haack}, {Hallgren}, {Halzen}, {Hanson}, {Hebecker}, {Heereman}, {Helbing}, {Hellauer}, {Hickford}, {Hignight}, {Hill}, {Hoffman}, {Hoffmann}, {Hokanson-Fasig}, {Hoshina}, {Huang}, {Huber}, {Hultqvist}, {H{\"u}nnefeld}, {In}, {Ishihara}, {Jacobi}, {Japaridze}, {Jeong}, {Jero}, {Jones}, {Kalaczynski}, {Kang}, {Kappes}, {Karg}, {Karle}, {Kauer}, {Keivani}, {Kelley}, {Kheirandish}, {Kim}, {Kim}, {Kintscher}, {Kiryluk}, {Kittler}, {Klein}, {Kohnen}, {Koirala}, {Kolanoski}, {K{\"o}pke}, {Kopper}, {Kopper}, {Koschinsky}, {Koskinen}, {Kowalski}, {Krings}, {Kroll}, {Kr{\"u}ckl}, {Kunnen},
  {Kunwar}, {Kurahashi}, {Kuwabara}, {Kyriacou}, {Labare}, {Lanfranchi}, {Larson}, {Lauber}, {Lesiak-Bzdak}, {Leuermann}, {Liu}, {Lu}, {L{\"u}nemann}, {Luszczak}, {Madsen}, {Maggi}, {Mahn}, {Mancina}, {Maruyama}, {Mase}, {Maunu}, {McNally}, {Meagher}, {Medici}, {Meier}, {Menne}, {Merino}, {Meures}, {Miarecki}, {Micallef}, {Moment{\'e}}, {Montaruli}, {Moore}, {Moulai}, {Nahnhauer}, {Nakarmi}, {Naumann}, {Neer}, {Niederhausen}, {Nowicki}, {Nygren}, {Obertacke Pollmann}, {Olivas}, {O'Murchadha}, {Palczewski}, {Pandya}, {Pankova}, {Peiffer}, {Pepper}, {P{\'e}rez de los Heros}, {Pieloth}, {Pinat}, {Price}, {Przybylski}, {Raab}, {R{\"a}del}, {Rameez}, {Rawlins}, {Rea}, {Reimann}, {Relethford}, {Relich}, {Resconi}, {Rhode}, {Richman}, {Robertson}, {Rongen}, {Rott}, {Ruhe}, {Ryckbosch}, {Rysewyk}, {S{\"a}lzer}, {Sanchez Herrera}, {Sandrock}, {Sandroos}, {Santander}, {Sarkar}, {Sarkar}, {Satalecka}, {Schlunder}, {Schmidt}, {Schneider}, {Schoenen}, {Sch{\"o}neberg}, {Schumacher}, {Seckel}, {Seunarine}, {Soedingrekso},
  {Soldin}, {Song}, {Spiczak}, {Spiering}, {Stachurska}, {Stamatikos}, {Stanev}, {Stasik}, {Stettner}, {Steuer}, {Stezelberger}, {Stokstad}, {St{\"o}ssl}, {Strotjohann}, {Stuttard}, {Sullivan}, {Sutherland}, {Taboada}, {Tatar}, {Tenholt}, {Ter-Antonyan}, {Terliuk}, {Te{\v{s}}i{\'c}}, {Tilav}, {Toale}, {Tobin}, {Toscano}, {Tosi}, {Tselengidou}, {Tung}, {Turcati}, {Turley}, {Ty}, {Unger}, {Usner}, {Vandenbroucke}, {Van Driessche}, {van Eijndhoven}, {Vanheule}, {van Santen}, {Vehring}, {Vogel}, {Vraeghe}, {Walck}, {Wallace}, {Wallraff}, {Wandler}, {Wandkowsky}, {Waza}, {Weaver}, {Weiss}, {Wendt}, {Werthebach}, {Whelan}, {Wiebe}, {Wiebusch}, {Wille}, {Williams}, {Wills}, {Wolf}, {Wood}, {Woolsey}, {Woschnagg}, {Xu}, {Xu}, {Xu}, {Yanez}, {Yodh}, {Yoshida}, {Yuan}, {Zoll}, {IceCube Collaboration}, {Balasubramanian}, {Mate}, {Bhalerao}, {Bhattacharya}, {Vibhute}, {Dewangan}, {Rao}, {Vadawale}, {AstroSat Cadmium Zinc Telluride Imager Team}, {Svinkin}, {Hurley}, {Aptekar}, {Frederiks}, {Golenetskii}, {Kozlova},
  {Lysenko}, {Oleynik}, {Tsvetkova}, {Ulanov}, {Cline}, {IPN Collaboration}, {Li}, {Xiong}, {Zhang}, {Lu}, {Song}, {Cao}, {Chang}, {Chen}, {Chen}, {Chen}, {Chen}, {Chen}, {Chen}, {Cui}, {Cui}, {Deng}, {Dong}, {Du}, {Fu}, {Gao}, {Gao}, {Gao}, {Ge}, {Gu}, {Guan}, {Guo}, {Han}, {Hu}, {Huang}, {Huo}, {Jia}, {Jiang}, {Jiang}, {Jin}, {Jin}, {Li}, {Li}, {Li}, {Li}, {Li}, {Li}, {Li}, {Li}, {Li}, {Li}, {Li}, {Liang}, {Liao}, {Liu}, {Liu}, {Liu}, {Liu}, {Liu}, {Liu}, {Liu}, {Lu}, {Lu}, {Luo}, {Ma}, {Meng}, {Nang}, {Nie}, {Ou}, {Qu}, {Sai}, {Sun}, {Tan}, {Tao}, {Tao}, {Tuo}, {Wang}, {Wang}, {Wang}, {Wang}, {Wang}, {Wen}, {Wu}, {Wu}, {Xiao}, {Xu}, {Xu}, {Yan}, {Yang}, {Yang}, {Yang}, {Zhang}, {Zhang}, {Zhang}, {Zhang}, {Zhang}, {Zhang}, {Zhang}, {Zhang}, {Zhang}, {Zhang}, {Zhang}, {Zhang}, {Zhang}, {Zhang}, {Zhang}, {Zhang}, {Zhang}, {Zhang}, {Zhao}, {Zhao}, {Zhao}, {Zheng}, {Zhu}, {Zhu}, {Zou}, {Insight-HXMT Collaboration}, {Albert}, {Andr{\'e}}, {Anghinolfi}, {Ardid}, {Aubert}, {Aublin}, {Avgitas}, {Baret},
  {Barrios-Mart{\'\i}}, {Basa}, {Belhorma}, {Bertin}, {Biagi}, {Bormuth}, {Bourret}, {Bouwhuis}, {Br{\^a}nza{\c{s}}}, {Bruijn}, {Brunner}, {Busto}, {Capone}, {Caramete}, {Carr}, {Celli}, {Cherkaoui El Moursli}, {Chiarusi}, {Circella}, {Coelho}, {Coleiro}, {Coniglione}, {Costantini}, {Coyle}, {Creusot}, {D{\'\i}az}, {Deschamps}, {De Bonis}, {Distefano}, {Di Palma}, {Domi}, {Donzaud}, {Dornic}, {Drouhin}, {Eberl}, {El Bojaddaini}, {El Khayati}, {Els{\"a}sser}, {Enzenh{\"o}fer}, {Ettahiri}, {Fassi}, {Felis}, {Fusco}, {Gay}, {Giordano}, {Glotin}, {Gr{\'e}goire}, {Ruiz}, {Graf}, {Hallmann}, {van Haren}, {Heijboer}, {Hello}, {Hern{\'a}ndez-Rey}, {H{\"o}ssl}, {Hofest{\"a}dt}, {Hugon}, {Illuminati}, {James}, {de Jong}, {Jongen}, {Kadler}, {Kalekin}, {Katz}, {Kiessling}, {Kouchner}, {Kreter}, {Kreykenbohm}, {Kulikovskiy}, {Lachaud}, {Lahmann}, {Lef{\`e}vre}, {Leonora}, {Lotze}, {Loucatos}, {Marcelin}, {Margiotta}, {Marinelli}, {Mart{\'\i}nez-Mora}, {Mele}, {Melis}, {Michael}, {Migliozzi}, {Moussa}, {Navas}, {Nezri},
  {Organokov}, {P{\u{a}}v{\u{a}}la{\c{s}}}, {Pellegrino}, {Perrina}, {Piattelli}, {Popa}, {Pradier}, {Quinn}, {Racca}, {Riccobene}, {S{\'a}nchez-Losa}, {Salda{\~n}a}, {Salvadori}, {Samtleben}, {Sanguineti}, {Sapienza}, {Sieger}, {Spurio}, {Stolarczyk}, {Taiuti}, {Tayalati}, {Trovato}, {Turpin}, {T{\"o}nnis}, {Vallage}, {Van Elewyck}, {Versari}, {Vivolo}, {Vizzoca}, {Wilms}, {Zornoza}, {Z{\'u}{\~n}iga}, {ANTARES Collaboration}, {Beardmore}, {Breeveld}, {Burrows}, {Cenko}, {Cusumano}, {D'A{\`\i}}, {de Pasquale}, {Emery}, {Evans}, {Giommi}, {Gronwall}, {Kennea}, {Krimm}, {Kuin}, {Lien}, {Marshall}, {Melandri}, {Nousek}, {Oates}, {Osborne}, {Pagani}, {Page}, {Palmer}, {Perri}, {Siegel}, {Sbarufatti}, {Tagliaferri}, {Tohuvavohu}, {Swift Collaboration}, {Tavani}, {Verrecchia}, {Bulgarelli}, {Evangelista}, {Pacciani}, {Feroci}, {Pittori}, {Giuliani}, {Del Monte}, {Donnarumma}, {Argan}, {Trois}, {Ursi}, {Cardillo}, {Piano}, {Longo}, {Lucarelli}, {Munar-Adrover}, {Fuschino}, {Labanti}, {Marisaldi}, {Minervini},
  {Fioretti}, {Parmiggiani}, {Gianotti}, {Trifoglio}, {Di Persio}, {Antonelli}, {Barbiellini}, {Caraveo}, {Cattaneo}, {Costa}, {Colafrancesco}, {D'Amico}, {Ferrari}, {Morselli}, {Paoletti}, {Picozza}, {Pilia}, {Rappoldi}, {Soffitta}, {Vercellone}, {AGILE Team}, {Foley}, {Coulter}, {Kilpatrick}, {Drout}, {Piro}, {Shappee}, {Siebert}, {Simon}, {Ulloa}, {Kasen}, {Madore}, {Murguia-Berthier}, {Pan}, {Prochaska}, {Ramirez-Ruiz}, {Rest}, {Rojas-Bravo}, {1M2H Team}, {Berger}, {Soares-Santos}, {Annis}, {Alexander}, {Allam}, {Balbinot}, {Blanchard}, {Brout}, {Butler}, {Chornock}, {Cook}, {Cowperthwaite}, {Diehl}, {Drlica-Wagner}, {Drout}, {Durret}, {Eftekhari}, {Finley}, {Fong}, {Frieman}, {Fryer}, {Garc{\'\i}a-Bellido}, {Gruendl}, {Hartley}, {Herner}, {Kessler}, {Lin}, {Lopes}, {Louren{\c{c}}o}, {Margutti}, {Marshall}, {Matheson}, {Medina}, {Metzger}, {Mu{\~n}oz}, {Muir}, {Nicholl}, {Nugent}, {Palmese}, {Paz-Chinch{\'o}n}, {Quataert}, {Sako}, {Sauseda}, {Schlegel}, {Scolnic}, {Secco}, {Smith}, {Sobreira}, {Villar},
  {Vivas}, {Wester}, {Williams}, {Yanny}, {Zenteno}, {Zhang}, {Abbott}, {Banerji}, {Bechtol}, {Benoit-L{\'e}vy}, {Bertin}, {Brooks}, {Buckley-Geer}, {Burke}, {Capozzi}, {Carnero Rosell}, {Carrasco Kind}, {Castander}, {Crocce}, {Cunha}, {D'Andrea}, {da Costa}, {Davis}, {DePoy}, {Desai}, {Dietrich}, {Eifler}, {Fernandez}, {Flaugher}, {Fosalba}, {Gaztanaga}, {Gerdes}, {Giannantonio}, {Goldstein}, {Gruen}, {Gschwend}, {Gutierrez}, {Honscheid}, {James}, {Jeltema}, {Johnson}, {Johnson}, {Kent}, {Krause}, {Kron}, {Kuehn}, {Lahav}, {Lima}, {Maia}, {March}, {Martini}, {McMahon}, {Menanteau}, {Miller}, {Miquel}, {Mohr}, {Nichol}, {Ogando}, {Plazas}, {Romer}, {Roodman}, {Rykoff}, {Sanchez}, {Scarpine}, {Schindler}, {Schubnell}, {Sevilla-Noarbe}, {Sheldon}, {Smith}, {Smith}, {Stebbins}, {Suchyta}, {Swanson}, {Tarle}, {Thomas}, {Troxel}, {Tucker}, {Vikram}, {Walker}, {Wechsler}, {Weller}, {Carlin}, {Gill}, {Li}, {Marriner}, {Neilsen}, {Dark Energy Camera GW-EM Collaboration}, {DES Collaboration}, {Haislip}, {Kouprianov},
  {Reichart}, {Sand}, {Tartaglia}, {Valenti}, {Yang}, {DLT40 Collaboration}, {Benetti}, {Brocato}, {Campana}, {Cappellaro}, {Covino}, {D'Avanzo}, {D'Elia}, {Getman}, {Ghirlanda}, {Ghisellini}, {Limatola}, {Nicastro}, {Palazzi}, {Pian}, {Piranomonte}, {Possenti}, {Rossi}, {Salafia}, {Tomasella}, {Amati}, {Antonelli}, {Bernardini}, {Bufano}, {Capaccioli}, {Casella}, {Dadina}, {De Cesare}, {Di Paola}, {Giuffrida}, {Giunta}, {Israel}, {Lisi}, {Maiorano}, {Mapelli}, {Masetti}, {Pescalli}, {Pulone}, {Salvaterra}, {Schipani}, {Spera}, {Stamerra}, {Stella}, {Testa}, {Turatto}, {Vergani}, {Aresu}, {Bachetti}, {Buffa}, {Burgay}, {Buttu}, {Caria}, {Carretti}, {Casasola}, {Castangia}, {Carboni}, {Casu}, {Concu}, {Corongiu}, {Deiana}, {Egron}, {Fara}, {Gaudiomonte}, {Gusai}, {Ladu}, {Loru}, {Leurini}, {Marongiu}, {Melis}, {Melis}, {Migoni}, {Milia}, {Navarrini}, {Orlati}, {Ortu}, {Palmas}, {Pellizzoni}, {Perrodin}, {Pisanu}, {Poppi}, {Righini}, {Saba}, {Serra}, {Serrau}, {Stagni}, {Surcis}, {Vacca}, {Vargiu}, {Hunt},
  {Jin}, {Klose}, {Kouveliotou}, {Mazzali}, {M{\o}ller}, {Nava}, {Piran}, {Selsing}, {Vergani}, {Wiersema}, {Toma}, {Higgins}, {Mundell}, {di Serego Alighieri}, {G{\'o}tz}, {Gao}, {Gomboc}, {Kaper}, {Kobayashi}, {Kopac}, {Mao}, {Starling}, {Steele}, {van der Horst}, {GRAWITA: GRAvitational Wave Inaf TeAm}, {Acero}, {Atwood}, {Baldini}, {Barbiellini}, {Bastieri}, {Berenji}, {Bellazzini}, {Bissaldi}, {Blandford}, {Bloom}, {Bonino}, {Bottacini}, {Bregeon}, {Buehler}, {Buson}, {Cameron}, {Caputo}, {Caraveo}, {Cavazzuti}, {Chekhtman}, {Cheung}, {Chiang}, {Ciprini}, {Cohen-Tanugi}, {Cominsky}, {Costantin}, {Cuoco}, {D'Ammando}, {de Palma}, {Digel}, {Di Lalla}, {Di Mauro}, {Di Venere}, {Dubois}, {Fegan}, {Focke}, {Franckowiak}, {Fukazawa}, {Funk}, {Fusco}, {Gargano}, {Gasparrini}, {Giglietto}, {Giordano}, {Giroletti}, {Glanzman}, {Green}, {Grondin}, {Guillemot}, {Guiriec}, {Harding}, {Horan}, {J{\'o}hannesson}, {Kamae}, {Kensei}, {Kuss}, {La Mura}, {Latronico}, {Lemoine-Goumard}, {Longo}, {Loparco}, {Lovellette},
  {Lubrano}, {Magill}, {Maldera}, {Manfreda}, {Mazziotta}, {McEnery}, {Meyer}, {Michelson}, {Mirabal}, {Monzani}, {Moretti}, {Morselli}, {Moskalenko}, {Negro}, {Nuss}, {Ojha}, {Omodei}, {Orienti}, {Orlando}, {Palatiello}, {Paliya}, {Paneque}, {Pesce-Rollins}, {Piron}, {Porter}, {Principe}, {Rain{\`o}}, {Rando}, {Razzano}, {Razzaque}, {Reimer}, {Reimer}, {Reposeur}, {Rochester}, {Saz Parkinson}, {Sgr{\`o}}, {Siskind}, {Spada}, {Spandre}, {Suson}, {Takahashi}, {Tanaka}, {Thayer}, {Thayer}, {Thompson}, {Tibaldo}, {Torres}, {Torresi}, {Troja}, {Venters}, {Vianello}, {Zaharijas}, {Fermi Large Area Telescope Collaboration}, {Allison}, {Bannister}, {Dobie}, {Kaplan}, {Lenc}, {Lynch}, {Murphy}, {Sadler}, {Australia Telescope Compact Array}, {Hotan}, {James}, {Oslowski}, {Raja}, {Shannon}, {Whiting}, {Australian SKA Pathfinder}, {Arcavi}, {Howell}, {McCully}, {Hosseinzadeh}, {Hiramatsu}, {Poznanski}, {Barnes}, {Zaltzman}, {Vasylyev}, {Maoz}, {Las Cumbres Observatory Group}, {Cooke}, {Bailes}, {Wolf}, {Deller},
  {Lidman}, {Wang}, {Gendre}, {Andreoni}, {Ackley}, {Pritchard}, {Bessell}, {Chang}, {M{\"o}ller}, {Onken}, {Scalzo}, {Ridden-Harper}, {Sharp}, {Tucker}, {Farrell}, {Elmer}, {Johnston}, {Venkatraman Krishnan}, {Keane}, {Green}, {Jameson}, {Hu}, {Ma}, {Sun}, {Wu}, {Wang}, {Shang}, {Hu}, {Ashley}, {Yuan}, {Li}, {Tao}, {Zhu}, {Zhang}, {Suntzeff}, {Zhou}, {Yang}, {Orange}, {Morris}, {Cucchiara}, {Giblin}, {Klotz}, {Staff}, {Thierry}, {Schmidt}, {OzGrav}, {(Deeper}, {Wider}, {program}, {AST3}, {CAASTRO Collaborations}, {Tanvir}, {Levan}, {Cano}, {de Ugarte-Postigo}, {Gonz{\'a}lez-Fern{\'a}ndez}, {Greiner}, {Hjorth}, {Irwin}, {Kr{\"u}hler}, {Mandel}, {Milvang-Jensen}, {O'Brien}, {Rol}, {Rosetti}, {Rosswog}, {Rowlinson}, {Steeghs}, {Th{\"o}ne}, {Ulaczyk}, {Watson}, {Bruun}, {Cutter}, {Figuera Jaimes}, {Fujii}, {Fruchter}, {Gompertz}, {Jakobsson}, {Hodosan}, {J{\`e}rgensen}, {Kangas}, {Kann}, {Rabus}, {Schr{\o}der}, {Stanway}, {Wijers}, {VINROUGE Collaboration}, {Lipunov}, {Gorbovskoy}, {Kornilov}, {Tyurina},
  {Balanutsa}, {Kuznetsov}, {Vlasenko}, {Podesta}, {Lopez}, {Podesta}, {Levato}, {Saffe}, {Mallamaci}, {Budnev}, {Gress}, {Kuvshinov}, {Gorbunov}, {Vladimirov}, {Zimnukhov}, {Gabovich}, {Yurkov}, {Sergienko}, {Rebolo}, {Serra-Ricart}, {Tlatov}, {Ishmuhametova}, {MASTER Collaboration}, {Abe}, {Aoki}, {Aoki}, {Asakura}, {Baar}, {Barway}, {Bond}, {Doi}, {Finet}, {Fujiyoshi}, {Furusawa}, {Honda}, {Itoh}, {Kanda}, {Kawabata}, {Kawabata}, {Kim}, {Koshida}, {Kuroda}, {Lee}, {Liu}, {Matsubayashi}, {Miyazaki}, {Morihana}, {Morokuma}, {Motohara}, {Murata}, {Nagai}, {Nagashima}, {Nagayama}, {Nakaoka}, {Nakata}, {Ohsawa}, {Ohshima}, {Ohta}, {Okita}, {Saito}, {Saito}, {Sako}, {Sekiguchi}, {Sumi}, {Tajitsu}, {Takahashi}, {Takayama}, {Tamura}, {Tanaka}, {Tanaka}, {Terai}, {Tominaga}, {Tristram}, {Uemura}, {Utsumi}, {Yamaguchi}, {Yasuda}, {Yoshida}, {Zenko}, {J-GEM}, {Adams}, {Anupama}, {Bally}, {Barway}, {Bellm}, {Blagorodnova}, {Cannella}, {Chandra}, {Chatterjee}, {Clarke}, {Cobb}, {Cook}, {Copperwheat}, {De}, {Emery},
  {Feindt}, {Foster}, {Fox}, {Frail}, {Fremling}, {Frohmaier}, {Garcia}, {Ghosh}, {Giacintucci}, {Goobar}, {Gottlieb}, {Grefenstette}, {Hallinan}, {Harrison}, {Heida}, {Helou}, {Ho}, {Horesh}, {Hotokezaka}, {Ip}, {Itoh}, {Jacobs}, {Jencson}, {Kasen}, {Kasliwal}, {Kassim}, {Kim}, {Kiran}, {Kuin}, {Kulkarni}, {Kupfer}, {Lau}, {Madsen}, {Mazzali}, {Miller}, {Miyasaka}, {Mooley}, {Myers}, {Nakar}, {Ngeow}, {Nugent}, {Ofek}, {Palliyaguru}, {Pavana}, {Perley}, {Peters}, {Pike}, {Piran}, {Qi}, {Quimby}, {Rana}, {Rosswog}, {Rusu}, {Sadler}, {Van Sistine}, {Sollerman}, {Xu}, {Yan}, {Yatsu}, {Yu}, {Zhang}, {Zhao}, {GROWTH}, {JAGWAR}, {Caltech-NRAO}, {TTU-NRAO}, {NuSTAR Collaborations}, {Chambers}, {Huber}, {Schultz}, {Bulger}, {Flewelling}, {Magnier}, {Lowe}, {Wainscoat}, {Waters}, {Willman}, {Pan-STARRS}, {Ebisawa}, {Hanyu}, {Harita}, {Hashimoto}, {Hidaka}, {Hori}, {Ishikawa}, {Isobe}, {Iwakiri}, {Kawai}, {Kawai}, {Kawamuro}, {Kawase}, {Kitaoka}, {Makishima}, {Matsuoka}, {Mihara}, {Morita}, {Morita}, {Nakahira},
  {Nakajima}, {Nakamura}, {Negoro}, {Oda}, {Sakamaki}, {Sasaki}, {Serino}, {Shidatsu}, {Shimomukai}, {Sugawara}, {Sugita}, {Sugizaki}, {Tachibana}, {Takao}, {Tanimoto}, {Tomida}, {Tsuboi}, {Tsunemi}, {Ueda}, {Ueno}, {Yamada}, {Yamaoka}, {Yamauchi}, {Yatabe}, {Yoneyama}, {Yoshii}, {MAXI Team}, {Coward}, {Crisp}, {Macpherson}, {Andreoni}, {Laugier}, {Noysena}, {Klotz}, {Gendre}, {Thierry}, {Turpin}, {Consortium}, {Im}, {Choi}, {Kim}, {Yoon}, {Lim}, {Lee}, {Lee}, {Kim}, {Ko}, {Joe}, {Kwon}, {Kim}, {Lim}, {Choi}, {KU Collaboration}, {Fynbo}, {Malesani}, {Xu}, {Optical Telescope}, {Smartt}, {Jerkstrand}, {Kankare}, {Sim}, {Fraser}, {Inserra}, {Maguire}, {Leloudas}, {Magee}, {Shingles}, {Smith}, {Young}, {Kotak}, {Gal-Yam}, {Lyman}, {Homan}, {Agliozzo}, {Anderson}, {Angus}, {Ashall}, {Barbarino}, {Bauer}, {Berton}, {Botticella}, {Bulla}, {Cannizzaro}, {Cartier}, {Cikota}, {Clark}, {De Cia}, {Della Valle}, {Dennefeld}, {Dessart}, {Dimitriadis}, {Elias-Rosa}, {Firth}, {Fl{\"o}rs}, {Frohmaier}, {Galbany},
  {Gonz{\'a}lez-Gait{\'a}n}, {Gromadzki}, {Guti{\'e}rrez}, {Hamanowicz}, {Harmanen}, {Heintz}, {Hernandez}, {Hodgkin}, {Hook}, {Izzo}, {James}, {Jonker}, {Kerzendorf}, {Kostrzewa-Rutkowska}, {Kromer}, {Kuncarayakti}, {Lawrence}, {Manulis}, {Mattila}, {McBrien}, {M{\"u}ller}, {Nordin}, {O'Neill}, {Onori}, {Palmerio}, {Pastorello}, {Patat}, {Pignata}, {Podsiadlowski}, {Razza}, {Reynolds}, {Roy}, {Ruiter}, {Rybicki}, {Salmon}, {Pumo}, {Prentice}, {Seitenzahl}, {Smith}, {Sollerman}, {Sullivan}, {Szegedi}, {Taddia}, {Taubenberger}, {Terreran}, {Van Soelen}, {Vos}, {Walton}, {Wright}, {Wyrzykowski}, {Yaron}, {pre=''(''>ePESSTO}, {Chen}, {Kr{\"u}hler}, {Schady}, {Wiseman}, {Greiner}, {Rau}, {Schweyer}, {Klose}, {Nicuesa Guelbenzu}, {GROND}, {Palliyaguru}, {Tech University}, {Shara}, {Williams}, {Vaisanen}, {Potter}, {Romero Colmenero}, {Crawford}, {Buckley}, {Mao}, {SALT Group}, {D{\'\i}az}, {Macri}, {Garc{\'\i}a Lambas}, {Mendes de Oliveira}, {Nilo Castell{\'o}n}, {Ribeiro}, {S{\'a}nchez}, {Schoenell}, {Abramo},
  {Akras}, {Alcaniz}, {Artola}, {Beroiz}, {Bonoli}, {Cabral}, {Camuccio}, {Chavushyan}, {Coelho}, {Colazo}, {Costa-Duarte}, {Cuevas Larenas}, {Dom{\'\i}nguez Romero}, {Dultzin}, {Fern{\'a}ndez}, {Garc{\'\i}a}, {Girardini}, {Gon{\c{c}}alves}, {Gon{\c{c}}alves}, {Gurovich}, {Jim{\'e}nez-Teja}, {Kanaan}, {Lares}, {Lopes de Oliveira}, {L{\'o}pez-Cruz}, {Melia}, {Molino}, {Padilla}, {Pe{\~n}uela}, {Placco}, {Qui{\~n}ones}, {Ram{\'\i}rez Rivera}, {Renzi}, {Riguccini}, {R{\'\i}os-L{\'o}pez}, {Rodriguez}, {Sampedro}, {Schneiter}, {Sodr{\'e}}, {Starck}, {Torres-Flores}, {Tornatore}, {Zadro{\.z}ny}, {Castillo}, {TOROS: Transient Robotic Observatory of South Collaboration}, {Castro-Tirado}, {Tello}, {Hu}, {Zhang}, {Cunniffe}, {Castell{\'o}n}, {Hiriart}, {Caballero-Garc{\'\i}a}, {Jel{\'\i}nek}, {Kub{\'a}nek}, {P{\'e}rez del Pulgar}, {Park}, {Jeong}, {Castro Cer{\'o}n}, {Pandey}, {Yock}, {Querel}, {Fan}, {Wang}, {BOOTES Collaboration}, {Beardsley}, {Brown}, {Crosse}, {Emrich}, {Franzen}, {Gaensler}, {Horsley},
  {Johnston-Hollitt}, {Kenney}, {Morales}, {Pallot}, {Sokolowski}, {Steele}, {Tingay}, {Trott}, {Walker}, {Wayth}, {Williams}, {Wu}, {Murchison Widefield Array}, {Yoshida}, {Sakamoto}, {Kawakubo}, {Yamaoka}, {Takahashi}, {Asaoka}, {Ozawa}, {Torii}, {Shimizu}, {Tamura}, {Ishizaki}, {Cherry}, {Ricciarini}, {Penacchioni}, {Marrocchesi}, {CALET Collaboration}, {Pozanenko}, {Volnova}, {Mazaeva}, {Minaev}, {Krugov}, {Kusakin}, {Reva}, {Moskvitin}, {Rumyantsev}, {Inasaridze}, {Klunko}, {Tungalag}, {Schmalz}, {Burhonov}, {IKI-GW Follow-up Collaboration}, {Abdalla}, {Abramowski}, {Aharonian}, {Ait Benkhali}, {Ang{\"u}ner}, {Arakawa}, {Arrieta}, {Aubert}, {Backes}, {Balzer}, {Barnard}, {Becherini}, {Becker Tjus}, {Berge}, {Bernhard}, {Bernl{\"o}hr}, {Blackwell}, {B{\"o}ttcher}, {Boisson}, {Bolmont}, {Bonnefoy}, {Bordas}, {Bregeon}, {Brun}, {Brun}, {Bryan}, {B{\"u}chele}, {Bulik}, {Capasso}, {Caroff}, {Carosi}, {Casanova}, {Cerruti}, {Chakraborty}, {Chaves}, {Chen}, {Chevalier}, {Colafrancesco}, {Condon}, {Conrad},
  {Davids}, {Decock}, {Deil}, {Devin}, {deWilt}, {Dirson}, {Djannati-Ata{\"\i}}, {Donath}, {O'C. Drury}, {Dutson}, {Dyks}, {Edwards}, {Egberts}, {Emery}, {Ernenwein}, {Eschbach}, {Farnier}, {Fegan}, {Fernandes}, {Fiasson}, {Fontaine}, {Funk}, {F{\"u}ssling}, {Gabici}, {Gallant}, {Garrigoux}, {Gat{\'e}}, {Giavitto}, {Giebels}, {Glawion}, {Glicenstein}, {Gottschall}, {Grondin}, {Hahn}, {Haupt}, {Hawkes}, {Heinzelmann}, {Henri}, {Hermann}, {Hinton}, {Hofmann}, {Hoischen}, {Holch}, {Holler}, {Horns}, {Ivascenko}, {Iwasaki}, {Jacholkowska}, {Jamrozy}, {Jankowsky}, {Jankowsky}, {Jingo}, {Jouvin}, {Jung-Richardt}, {Kastendieck}, {Katarzy{\'n}ski}, {Katsuragawa}, {Kerszberg}, {Khangulyan}, {Kh{\'e}lifi}, {King}, {Klepser}, {Klochkov}, {Klu{\'z}niak}, {Komin}, {Kosack}, {Krakau}, {Kraus}, {Kr{\"u}ger}, {Laffon}, {Lamanna}, {Lau}, {Lees}, {Lefaucheur}, {Lemi{\`e}re}, {Lemoine-Goumard}, {Lenain}, {Leser}, {Lohse}, {Lorentz}, {Liu}, {Lypova}, {Malyshev}, {Marandon}, {Marcowith}, {Mariaud}, {Marx}, {Maurin}, {Maxted},
  {Mayer}, {Meintjes}, {Meyer}, {Mitchell}, {Moderski}, {Mohamed}, {Mohrmann}, {Mor{\r{a}}}, {Moulin}, {Murach}, {Nakashima}, {de Naurois}, {Ndiyavala}, {Niederwanger}, {Niemiec}, {Oakes}, {O'Brien}, {Odaka}, {Ohm}, {Ostrowski}, {Oya}, {Padovani}, {Panter}, {Parsons}, {Pekeur}, {Pelletier}, {Perennes}, {Petrucci}, {Peyaud}, {Piel}, {Pita}, {Poireau}, {Poon}, {Prokhorov}, {Prokoph}, {P{\"u}hlhofer}, {Punch}, {Quirrenbach}, {Raab}, {Rauth}, {Reimer}, {Reimer}, {Renaud}, {de los Reyes}, {Rieger}, {Rinchiuso}, {Romoli}, {Rowell}, {Rudak}, {Rulten}, {Sahakian}, {Saito}, {Sanchez}, {Santangelo}, {Sasaki}, {Schlickeiser}, {Sch{\"u}ssler}, {Schulz}, {Schwanke}, {Schwemmer}, {Seglar-Arroyo}, {Settimo}, {Seyffert}, {Shafi}, {Shilon}, {Shiningayamwe}, {Simoni}, {Sol}, {Spanier}, {Spir-Jacob}, {Stawarz}, {Steenkamp}, {Stegmann}, {Steppa}, {Sushch}, {Takahashi}, {Tavernet}, {Tavernier}, {Taylor}, {Terrier}, {Tibaldo}, {Tiziani}, {Tluczykont}, {Trichard}, {Tsirou}, {Tsuji}, {Tuffs}, {Uchiyama}, {van der Walt}, {van Eldik},
  {van Rensburg}, {van Soelen}, {Vasileiadis}, {Veh}, {Venter}, {Viana}, {Vincent}, {Vink}, {Voisin}, {V{\"o}lk}, {Vuillaume}, {Wadiasingh}, {Wagner}, {Wagner}, {Wagner}, {White}, {Wierzcholska}, {Willmann}, {W{\"o}rnlein}, {Wouters}, {Yang}, {Zaborov}, {Zacharias}, {Zanin}, {Zdziarski}, {Zech}, {Zefi}, {Ziegler}, {Zorn}, {{\.Z}ywucka}, {H.~E.~S.~S. Collaboration}, {Fender}, {Broderick}, {Rowlinson}, {Wijers}, {Stewart}, {ter Veen}, {Shulevski}, {LOFAR Collaboration}, {Kavic}, {Simonetti}, {League}, {Tsai}, {Obenberger}, {Nathaniel}, {Taylor}, {Dowell}, {Liebling}, {Estes}, {Lippert}, {Sharma}, {Vincent}, {Farella}, {Wavelength Array}, {Abeysekara}, {Albert}, {Alfaro}, {Alvarez}, {Arceo}, {Arteaga-Vel{\'a}zquez}, {Avila Rojas}, {Ayala Solares}, {Barber}, {Becerra Gonzalez}, {Becerril}, {Belmont-Moreno}, {BenZvi}, {Berley}, {Bernal}, {Braun}, {Brisbois}, {Caballero-Mora}, {Capistr{\'a}n}, {Carrami{\~n}ana}, {Casanova}, {Castillo}, {Cotti}, {Cotzomi}, {Couti{\~n}o de Le{\'o}n}, {De Le{\'o}n}, {De la Fuente},
  {Diaz Hernandez}, {Dichiara}, {Dingus}, {DuVernois}, {D{\'\i}az-V{\'e}lez}, {Ellsworth}, {Engel}, {Enr{\'\i}quez-Rivera}, {Fiorino}, {Fleischhack}, {Fraija}, {Garc{\'\i}a-Gonz{\'a}lez}, {Garfias}, {Gerhardt}, {Gonz{\~o}lez Mu{\~n}oz}, {Gonz{\'a}lez}, {Goodman}, {Hampel-Arias}, {Harding}, {Hernandez}, {Hernandez-Almada}, {Hona}, {H{\"u}ntemeyer}, {Iriarte}, {Jardin-Blicq}, {Joshi}, {Kaufmann}, {Kieda}, {Lara}, {Lauer}, {Lennarz}, {Le{\'o}n Vargas}, {Linnemann}, {Longinotti}, {Raya}, {Luna-Garc{\'\i}a}, {L{\'o}pez-Coto}, {Malone}, {Marinelli}, {Martinez}, {Martinez-Castellanos}, {Mart{\'\i}nez-Castro}, {Mart{\'\i}nez-Huerta}, {Matthews}, {Miranda-Romagnoli}, {Moreno}, {Mostaf{\'a}}, {Nellen}, {Newbold}, {Nisa}, {Noriega-Papaqui}, {Pelayo}, {Pretz}, {P{\'e}rez-P{\'e}rez}, {Ren}, {Rho}, {Rivi{\`e}re}, {Rosa-Gonz{\'a}lez}, {Rosenberg}, {Ruiz-Velasco}, {Salazar}, {Salesa Greus}, {Sandoval}, {Schneider}, {Schoorlemmer}, {Sinnis}, {Smith}, {Springer}, {Surajbali}, {Tibolla}, {Tollefson}, {Torres}, {Ukwatta},
  {Weisgarber}, {Westerhoff}, {Wisher}, {Wood}, {Yapici}, {Yodh}, {Younk}, {Zhou}, {{\'A}lvarez}, {HAWC Collaboration}, {Aab}, {Abreu}, {Aglietta}, {Albuquerque}, {Albury}, {Allekotte}, {Almela}, {Alvarez Castillo}, {Alvarez-Mu{\~n}iz}, {Anastasi}, {Anchordoqui}, {Andrada}, {Andringa}, {Aramo}, {Arsene}, {Asorey}, {Assis}, {Avila}, {Badescu}, {Balaceanu}, {Barbato}, {Barreira Luz}, {Becker}, {Bellido}, {Berat}, {Bertaina}, {Bertou}, {Biermann}, {Biteau}, {Blaess}, {Blanco}, {Blazek}, {Bleve}, {Boh{\'a}{\v{c}}ov{\'a}}, {Bonifazi}, {Borodai}, {Botti}, {Brack}, {Brancus}, {Bretz}, {Bridgeman}, {Briechle}, {Buchholz}, {Bueno}, {Buitink}, {Buscemi}, {Caballero-Mora}, {Caccianiga}, {Cancio}, {Canfora}, {Caruso}, {Castellina}, {Catalani}, {Cataldi}, {Cazon}, {Chavez}, {Chinellato}, {Chudoba}, {Clay}, {Cobos Cerutti}, {Colalillo}, {Coleman}, {Collica}, {Coluccia}, {Concei{\c{c}}{\~a}o}, {Consolati}, {Contreras}, {Cooper}, {Coutu}, {Covault}, {Cronin}, {D'Amico}, {Daniel}, {Dasso}, {Daumiller}, {Dawson}, {Day}, {de
  Almeida}, {de Jong}, {De Mauro}, {de Mello Neto}, {De Mitri}, {de Oliveira}, {de Souza}, {Debatin}, {Deligny}, {D{\'\i}az Castro}, {Diogo}, {Dobrigkeit}, {D'Olivo}, {Dorosti}, {Dos Anjos}, {Dova}, {Dundovic}, {Ebr}, {Engel}, {Erdmann}, {Erfani}, {Escobar}, {Espadanal}, {Etchegoyen}, {Falcke}, {Farmer}, {Farrar}, {Fauth}, {Fazzini}, {Feldbusch}, {Fenu}, {Fick}, {Figueira}, {Filip{\v{c}}i{\v{c}}}, {Freire}, {Fujii}, {Fuster}, {Ga{\"\i}or}, {Garc{\'\i}a}, {Gat{\'e}}, {Gemmeke}, {Gherghel-Lascu}, {Ghia}, {Giaccari}, {Giammarchi}, {Giller}, {G{\l}as}, {Glaser}, {Golup}, {G{\'o}mez Berisso}, {G{\'o}mez Vitale}, {Gonz{\'a}lez}, {Gorgi}, {Gottowik}, {Grillo}, {Grubb}, {Guarino}, {Guedes}, {Halliday}, {Hampel}, {Hansen}, {Harari}, {Harrison}, {Harvey}, {Haungs}, {Hebbeker}, {Heck}, {Heimann}, {Herve}, {Hill}, {Hojvat}, {Holt}, {Homola}, {H{\"o}randel}, {Horvath}, {Hrabovsk{\'y}}, {Huege}, {Hulsman}, {Insolia}, {Isar}, {Jandt}, {Johnsen}, {Josebachuili}, {Jurysek}, {K{\"a}{\"a}p{\"a}}, {Kampert}, {Keilhauer},
  {Kemmerich}, {Kemp}, {Kieckhafer}, {Klages}, {Kleifges}, {Kleinfeller}, {Krause}, {Krohm}, {Kuempel}, {Kukec Mezek}, {Kunka}, {Kuotb Awad}, {Lago}, {LaHurd}, {Lang}, {Lauscher}, {Legumina}, {Leigui de Oliveira}, {Letessier-Selvon}, {Lhenry-Yvon}, {Link}, {Lo Presti}, {Lopes}, {L{\'o}pez}, {L{\'o}pez Casado}, {Lorek}, {Luce}, {Lucero}, {Malacari}, {Mallamaci}, {Mandat}, {Mantsch}, {Mariazzi}, {Maris}, {Marsella}, {Martello}, {Martinez}, {Mart{\'\i}nez Bravo}, {Mas{\'\i}as Meza}, {Mathes}, {Mathys}, {Matthews}, {Matthiae}, {Mayotte}, {Mazur}, {Medina}, {Medina-Tanco}, {Melo}, {Menshikov}, {Merenda}, {Michal}, {Micheletti}, {Middendorf}, {Miramonti}, {Mitrica}, {Mockler}, {Mollerach}, {Montanet}, {Morello}, {Morlino}, {M{\"u}ller}, {M{\"u}ller}, {Muller}, {M{\"u}ller}, {Mussa}, {Naranjo}, {Nguyen}, {Niculescu-Oglinzanu}, {Niechciol}, {Niemietz}, {Niggemann}, {Nitz}, {Nosek}, {Novotny}, {No{\v{z}}ka}, {N{\'u}{\~n}ez}, {Oikonomou}, {Olinto}, {Palatka}, {Pallotta}, {Papenbreer}, {Parente}, {Parra}, {Paul},
  {Pech}, {Pedreira}, {P{\c{e}}kala}, {Pe{\~n}a-Rodriguez}, {Pereira}, {Perlin}, {Perrone}, {Peters}, {Petrera}, {Phuntsok}, {Pierog}, {Pimenta}, {Pirronello}, {Platino}, {Plum}, {Poh}, {Porowski}, {Prado}, {Privitera}, {Prouza}, {Quel}, {Querchfeld}, {Quinn}, {Ramos-Pollan}, {Rautenberg}, {Ravignani}, {Ridky}, {Riehn}, {Risse}, {Ristori}, {Rizi}, {Rodrigues de Carvalho}, {Rodriguez Fernandez}, {Rodriguez Rojo}, {Roncoroni}, {Roth}, {Roulet}, {Rovero}, {Ruehl}, {Saffi}, {Saftoiu}, {Salamida}, {Salazar}, {Saleh}, {Salina}, {S{\'a}nchez}, {Sanchez-Lucas}, {Santos}, {Santos}, {Sarazin}, {Sarmento}, {Sarmiento-Cano}, {Sato}, {Schauer}, {Scherini}, {Schieler}, {Schimp}, {Schmidt}, {Scholten}, {Schov{\'a}nek}, {Schr{\"o}der}, {Schr{\"o}der}, {Schulz}, {Schumacher}, {Sciutto}, {Segreto}, {Shadkam}, {Shellard}, {Sigl}, {Silli}, {{\v{S}}m{\'\i}da}, {Snow}, {Sommers}, {Sonntag}, {Soriano}, {Squartini}, {Stanca}, {Stani{\v{c}}}, {Stasielak}, {Stassi}, {Stolpovskiy}, {Strafella}, {Streich}, {Suarez}, {Suarez-Dur{\'a}n},
  {Sudholz}, {Suomij{\"a}rvi}, {Supanitsky}, {{\v{S}}up{\'\i}k}, {Swain}, {Szadkowski}, {Taboada}, {Taborda}, {Timmermans}, {Todero Peixoto}, {Tomankova}, {Tom{\'e}}, {Torralba Elipe}, {Travnicek}, {Trini}, {Tueros}, {Ulrich}, {Unger}, {Urban}, {Vald{\'e}s Galicia}, {Vali{\~n}o}, {Valore}, {van Aar}, {van Bodegom}, {van den Berg}, {van Vliet}, {Varela}, {Vargas C{\'a}rdenas}, {V{\'a}zquez}, {Veberi{\v{c}}}, {Ventura}, {Vergara Quispe}, {Verzi}, {Vicha}, {Villase{\~n}or}, {Vorobiov}, {Wahlberg}, {Wainberg}, {Walz}, {Watson}, {Weber}, {Weindl}, {Wiede{\'n}ski}, {Wiencke}, {Wilczy{\'n}ski}, {Wirtz}, {Wittkowski}, {Wundheiler}, {Yang}, {Yushkov}, {Zas}, {Zavrtanik}, {Zavrtanik}, {Zepeda}, {Zimmermann}, {Ziolkowski}, {Zong}, {Zuccarello}, {Pierre Auger Collaboration}, {Kim}, {Schulze}, {Bauer}, {Corral-Santana}, {de Gregorio-Monsalvo}, {Gonz{\'a}lez-L{\'o}pez}, {Hartmann}, {Ishwara-Chandra}, {Mart{\'\i}n}, {Mehner}, {Misra}, {Micha{\l}owski}, {Resmi}, {ALMA Collaboration}, {Paragi}, {Agudo}, {An}, {Beswick},
  {Casadio}, {Frey}, {Jonker}, {Kettenis}, {Marcote}, {Moldon}, {Szomoru}, {van Langevelde}, {Yang}, {Euro VLBI Team}, {Cwiek}, {Cwiok}, {Czyrkowski}, {Dabrowski}, {Kasprowicz}, {Mankiewicz}, {Nawrocki}, {Opiela}, {Piotrowski}, {Wrochna}, {Zaremba}, {{\.Z}arnecki}, {Pi of Sky Collaboration}, {Haggard}, {Nynka}, {Ruan}, {Chandra Team at McGill University}, {Bland}, {Booler}, {Devillepoix}, {de Gois}, {Hancock}, {Howie}, {Paxman}, {Sansom}, {Towner}, {Desert Fireball Network}, {Tonry}, {Coughlin}, {Stubbs}, {Denneau}, {Heinze}, {Stalder}, {Weiland}, {ATLAS}, {Eatough}, {Kramer}, {Kraus}, {Time Resolution Universe Survey}, {Troja}, {Piro}, {Becerra Gonz{\'a}lez}, {Butler}, {Fox}, {Khandrika}, {Kutyrev}, {Lee}, {Ricci}, {Ryan}, {S{\'a}nchez-Ram{\'\i}rez}, {Veilleux}, {Watson}, {Wieringa}, {Burgess}, {van Eerten}, {Fontes}, {Fryer}, {Korobkin}, {Wollaeger}, {RIMAS}, {RATIR}, {Camilo}, {Foley}, {Goedhart}, {Makhathini}, {Oozeer}, {Smirnov}, {Fender}, {Woudt}, \& {South Africa/MeerKAT}}]{Abbott17b}
---. 2017{\natexlab{b}}, \apjl, 848, L12, \dodoi{10.3847/2041-8213/aa91c9}

\bibitem[{{Abbott} {et~al.}(2020){Abbott}, {Abbott}, {Abbott}, {Abraham}, {Acernese}, {Ackley}, {Adams}, {Adya}, {Affeldt}, {Agathos}, {Agatsuma}, {Aggarwal}, {Aguiar}, {Aiello}, {Ain}, {Ajith}, {Akutsu}, {Allen}, {Allocca}, {Aloy}, {Altin}, {Amato}, {Ananyeva}, {Anderson}, {Anderson}, {Ando}, {Angelova}, {Antier}, {Appert}, {Arai}, {Arai}, {Arai}, {Araki}, {Araya}, {Araya}, {Areeda}, {Ar{\`e}ne}, {Aritomi}, {Arnaud}, {Arun}, {Ascenzi}, {Ashton}, {Aso}, {Aston}, {Astone}, {Aubin}, {Aufmuth}, {Aultoneal}, {Austin}, {Avendano}, {Avila-Alvarez}, {Babak}, {Bacon}, {Badaracco}, {Bader}, {Bae}, {Bae}, {Baiotti}, {Bajpai}, {Baker}, {Baldaccini}, {Ballardin}, {Ballmer}, {Banagiri}, {Barayoga}, {Barclay}, {Barish}, {Barker}, {Barkett}, {Barnum}, {Barone}, {Barr}, {Barsotti}, {Barsuglia}, {Barta}, {Bartlett}, {Barton}, {Bartos}, {Bassiri}, {Basti}, {Bawaj}, {Bayley}, {Bazzan}, {B{\'e}csy}, {Bejger}, {Belahcene}, {Bell}, {Beniwal}, {Berger}, {Bergmann}, {Bernuzzi}, {Bero}, {Berry}, {Bersanetti}, {Bertolini},
  {Betzwieser}, {Bhandare}, {Bidler}, {Bilenko}, {Bilgili}, {Billingsley}, {Birch}, {Birney}, {Birnholtz}, {Biscans}, {Biscoveanu}, {Bisht}, {Bitossi}, {Bizouard}, {Blackburn}, {Blair}, {Blair}, {Blair}, {Bloemen}, {Bode}, {Boer}, {Boetzel}, {Bogaert}, {Bondu}, {Bonilla}, {Bonnand}, {Booker}, {Boom}, {Booth}, {Bork}, {Boschi}, {Bose}, {Bossie}, {Bossilkov}, {Bosveld}, {Bouffanais}, {Bozzi}, {Bradaschia}, {Brady}, {Bramley}, {Branchesi}, {Brau}, {Briant}, {Briggs}, {Brighenti}, {Brillet}, {Brinkmann}, {Brisson}, {Brockill}, {Brooks}, {Brown}, {Brown}, {Brunett}, {Buikema}, {Bulik}, {Bulten}, {Buonanno}, {Buskulic}, {Buy}, {Byer}, {Cabero}, {Cadonati}, {Cagnoli}, {Cahillane}, {Bustillo}, {Callister}, {Calloni}, {Camp}, {Campbell}, {Canepa}, {Cannon}, {Cannon}, {Cao}, {Cao}, {Capocasa}, {Carbognani}, {Caride}, {Carney}, {Carullo}, {Casanueva Diaz}, {Casentini}, {Caudill}, {Cavagli{\`a}}, {Cavalier}, {Cavalieri}, {Cella}, {Cerd{\'a}-Dur{\'a}n}, {Cerretani}, {Cesarini}, {Chaibi}, {Chakravarti}, {Chamberlin},
  {Chan}, {Chan}, {Chao}, {Charlton}, {Chase}, {Chassande-Mottin}, {Chatterjee}, {Chaturvedi}, {Chatziioannou}, {Cheeseboro}, {Chen}, {Chen}, {Chen}, {Chen}, {Chen}, {Chen}, {Cheng}, {Cheong}, {Chia}, {Chincarini}, {Chiummo}, {Cho}, {Cho}, {Cho}, {Christensen}, {Chu}, {Chu}, {Chu}, {Chua}, {Chung}, {Chung}, {Ciani}, {Ciobanu}, {Ciolfi}, {Cipriano}, {Cirone}, {Clara}, {Clark}, {Clearwater}, {Cleva}, {Cocchieri}, {Coccia}, {Cohadon}, {Cohen}, {Colgan}, {Colleoni}, {Collette}, {Collins}, {Cominsky}, {Constancio}, {Conti}, {Cooper}, {Corban}, {Corbitt}, {Cordero-Carri{\'o}n}, {Corley}, {Cornish}, {Corsi}, {Cortese}, {Costa}, {Cotesta}, {Coughlin}, {Coughlin}, {Coulon}, {Countryman}, {Couvares}, {Covas}, {Cowan}, {Coward}, {Cowart}, {Coyne}, {Coyne}, {Creighton}, {Creighton}, {Cripe}, {Croquette}, {Crowder}, {Cullen}, {Cumming}, {Cunningham}, {Cuoco}, {Dal Canton}, {D{\'a}lya}, {Danilishin}, {D'Antonio}, {Danzmann}, {Dasgupta}, {da Silva Costa}, {Datrier}, {Dattilo}, {Dave}, {Davier}, {Davis}, {Daw}, {Debra},
  {Deenadayalan}, {Degallaix}, {de Laurentis}, {Del{\'e}glise}, {Pozzo}, {Demarchi}, {Demos}, {Dent}, {de Pietri}, {Derby}, {De Rosa}, {de Rossi}, {Desalvo}, {de Varona}, {Dhurandhar}, {D{\'\i}az}, {Dietrich}, {di Fiore}, {di Giovanni}, {di Girolamo}, {di Lieto}, {Ding}, {di Pace}, {di Palma}, {di Renzo}, {Dmitriev}, {Doctor}, {Doi}, {Donovan}, {Dooley}, {Doravari}, {Dorrington}, {Downes}, {Drago}, {Driggers}, {Du}, {Ducoin}, {Dupej}, {Dwyer}, {Easter}, {Edo}, {Edwards}, {Effler}, {Eguchi}, {Ehrens}, {Eichholz}, {Eikenberry}, {Eisenmann}, {Eisenstein}, {Enomoto}, {Essick}, {Estelles}, {Estevez}, {Etienne}, {Etzel}, {Evans}, {Evans}, {Fafone}, {Fair}, {Fairhurst}, {Fan}, {Farinon}, {Farr}, {Farr}, {Fauchon-Jones}, {Favata}, {Fays}, {Fazio}, {Fee}, {Feicht}, {Fejer}, {Feng}, {Fernandez-Galiana}, {Ferrante}, {Ferreira}, {Ferreira}, {Ferrini}, {Fidecaro}, {Fiori}, {Fiorucci}, {Fishbach}, {Fisher}, {Fishner}, {Fitz-Axen}, {Flaminio}, {Fletcher}, {Flynn}, {Fong}, {Font}, {Forsyth}, {Fournier}, {Frasca}, {Frasconi},
  {Frei}, {Freise}, {Frey}, {Frey}, {Fritschel}, {Frolov}, {Fujii}, {Fukunaga}, {Fukushima}, {Fulda}, {Fyffe}, {Gabbard}, {Gadre}, {Gaebel}, {Gair}, {Gammaitoni}, {Ganija}, {Gaonkar}, {Garcia}, {Garc{\'\i}a-Quir{\'o}s}, {Garufi}, {Gateley}, {Gaudio}, {Gaur}, {Gayathri}, {Ge}, {Gemme}, {Genin}, {Gennai}, {George}, {George}, {Gergely}, {Germain}, {Ghonge}, {Ghosh}, {Ghosh}, {Ghosh}, {Giacomazzo}, {Giaime}, {Giardina}, {Giazotto}, {Gill}, {Giordano}, {Glover}, {Godwin}, {Goetz}, {Goetz}, {Goncharov}, {Gonz{\'a}lez}, {Gonzalez Castro}, {Gopakumar}, {Gorodetsky}, {Gossan}, {Gosselin}, {Gouaty}, {Grado}, {Graef}, {Granata}, {Grant}, {Gras}, {Grassia}, {Gray}, {Gray}, {Greco}, {Green}, {Green}, {Gretarsson}, {Groot}, {Grote}, {Grunewald}, {Gruning}, {Guidi}, {Gulati}, {Guo}, {Gupta}, {Gupta}, {Gustafson}, {Gustafson}, {Haegel}, {Hagiwara}, {Haino}, {Halim}, {Hall}, {Hall}, {Hamilton}, {Hammond}, {Haney}, {Hanke}, {Hanks}, {Hanna}, {Hannam}, {Hannuksela}, {Hanson}, {Hardwick}, {Haris}, {Harms}, {Harry}, {Harry},
  {Hasegawa}, {Haster}, {Haughian}, {Hayakawa}, {Hayama}, {Hayes}, {Healy}, {Heidmann}, {Heintze}, {Heitmann}, {Hello}, {Hemming}, {Hendry}, {Heng}, {Hennig}, {Heptonstall}, {Heurs}, {Hild}, {Himemoto}, {Hinderer}, {Hiranuma}, {Hirata}, {Hirose}, {Hoak}, {Hochheim}, {Hofman}, {Holgado}, {Holland}, {Holt}, {Holz}, {Hong}, {Hopkins}, {Horst}, {Hough}, {Howell}, {Hoy}, {Hreibi}, {Hsieh}, {Huang}, {Huang}, {Huang}, {Huerta}, {Huet}, {Hughey}, {Hulko}, {Husa}, {Huttner}, {Huynh-Dinh}, {Idzkowski}, {Iess}, {Ikenoue}, {Imam}, {Inayoshi}, {Ingram}, {Inoue}, {Inta}, {Intini}, {Ioka}, {Irwin}, {Isa}, {Isac}, {Isi}, {Itoh}, {Iyer}, {Izumi}, {Jacqmin}, {Jadhav}, {Jani}, {Janthalur}, {Jaranowski}, {Jenkins}, {Jiang}, {Johnson}, {Jones}, {Jones}, {Jones}, {Jonker}, {Ju}, {Jung}, {Jung}, {Junker}, {Kajita}, {Kalaghatgi}, {Kalogera}, {Kamai}, {Kamiizumi}, {Kanda}, {Kandhasamy}, {Kang}, {Kanner}, {Kapadia}, {Karki}, {Karvinen}, {Kashyap}, {Kasprzack}, {Katsanevas}, {Katsavounidis}, {Katzman}, {Kaufer}, {Kawabe}, {Kawaguchi},
  {Kawai}, {Kawasaki}, {Keerthana}, {K{\'e}f{\'e}lian}, {Keitel}, {Kennedy}, {Key}, {Khalili}, {Khan}, {Khan}, {Khan}, {Khan}, {Khazanov}, {Khursheed}, {Kijbunchoo}, {Kim}, {Kim}, {Kim}, {Kim}, {Kim}, {Kim}, {Kim}, {Kim}, {Kimball}, {Kimura}, {King}, {King}, {Kinley-Hanlon}, {Kirchhoff}, {Kissel}, {Kita}, {Kitazawa}, {Kleybolte}, {Klika}, {Klimenko}, {Knowles}, {Knyazev}, {Koch}, {Koehlenbeck}, {Koekoek}, {Kojima}, {Kokeyama}, {Koley}, {Komori}, {Kondrashov}, {Kong}, {Kontos}, {Koper}, {Korobko}, {Korth}, {Kotake}, {Kowalska}, {Kozak}, {Kozakai}, {Kozu}, {Kringel}, {Krishnendu}, {Kr{\'o}lak}, {Kuehn}, {Kumar}, {Kumar}, {Kumar}, {Kumar}, {Kumar}, {Kume}, {Kuo}, {Kuo}, {Kuo}, {Kuroyanagi}, {Kusayanagi}, {Kutynia}, {Kwak}, {Kwang}, {Lackey}, {Lai}, {Lam}, {Landry}, {Lane}, {Lang}, {Lange}, {Lantz}, {Lanza}, {Lartaux-Vollard}, {Lasky}, {Laxen}, {Lazzarini}, {Lazzaro}, {Leaci}, {Leavey}, {Lecoeuche}, {Lee}, {Lee}, {Lee}, {Lee}, {Lee}, {Lee}, {Lee}, {Lehmann}, {Lenon}, {Leonardi}, {Leroy}, {Letendre}, {Levin},
  {Li}, {Li}, {Li}, {Li}, {Lin}, {Lin}, {Lin}, {Lin}, {Linde}, {Linker}, {Littenberg}, {Liu}, {Liu}, {Liu}, {Lo}, {Lockerbie}, {London}, {Longo}, {Lorenzini}, {Loriette}, {Lormand}, {Losurdo}, {Lough}, {Lousto}, {Lovelace}, {Lower}, {L{\"u}ck}, {Lumaca}, {Lundgren}, {Luo}, {Lynch}, {Ma}, {Macas}, {Macfoy}, {Macinnis}, {MacLeod}, {Macquet}, {Maga{\~n}a-Sandoval}, {Zertuche}, {Magee}, {Majorana}, {Maksimovic}, {Malik}, {Man}, {Mandic}, {Mangano}, {Mansell}, {Manske}, {Mantovani}, {Marchesoni}, {Marchio}, {Marion}, {M{\'a}rka}, {M{\'a}rka}, {Markakis}, {Markosyan}, {Markowitz}, {Maros}, {Marquina}, {Marsat}, {Martelli}, {Martin}, {Martin}, {Martynov}, {Mason}, {Massera}, {Masserot}, {Massinger}, {Masso-Reid}, {Mastrogiovanni}, {Matas}, {Matichard}, {Matone}, {Mavalvala}, {Mazumder}, {McCann}, {McCarthy}, {McClelland}, {McCormick}, {McCuller}, {McGuire}, {McIver}, {McManus}, {McRae}, {McWilliams}, {Meacher}, {Meadors}, {Mehmet}, {Mehta}, {Meidam}, {Melatos}, {Mendell}, {Mercer}, {Mereni}, {Merilh}, {Merzougui},
  {Meshkov}, {Messenger}, {Messick}, {Metzdorff}, {Meyers}, {Miao}, {Michel}, {Michimura}, {Middleton}, {Mikhailov}, {Milano}, {Miller}, {Miller}, {Millhouse}, {Mills}, {Milovich-Goff}, {Minazzoli}, {Minenkov}, {Mio}, {Mishkin}, {Mishra}, {Mistry}, {Mitra}, {Mitrofanov}, {Mitselmakher}, {Mittleman}, {Miyakawa}, {Miyamoto}, {Miyazaki}, {Miyo}, {Miyoki}, {Mo}, {Moffa}, {Mogushi}, {Mohapatra}, {Montani}, {Moore}, {Moraru}, {Moreno}, {Morisaki}, {Moriwaki}, {Mours}, {Mow-Lowry}, {Mukherjee}, {Mukherjee}, {Mukherjee}, {Mukund}, {Mullavey}, {Munch}, {Mu{\~n}iz}, {Muratore}, {Murray}, {Nagano}, {Nagano}, {Nagar}, {Nakamura}, {Nakano}, {Nakano}, {Nakashima}, {Nardecchia}, {Narikawa}, {Naticchioni}, {Nayak}, {Negishi}, {Neilson}, {Nelemans}, {Nelson}, {Nery}, {Neunzert}, {Ng}, {Ng}, {Nguyen}, {Ni}, {Nichols}, {Nishizawa}, {Nissanke}, {Nocera}, {North}, {Nuttall}, {Obergaulinger}, {Oberling}, {O'Brien}, {Obuchi}, {O'Dea}, {Ogaki}, {Ogin}, {Oh}, {Oh}, {Ohashi}, {Ohishi}, {Ohkawa}, {Ohme}, {Ohta}, {Okada}, {Okutomi},
  {Oliver}, {Oohara}, {Ooi}, {Oppermann}, {Oram}, {O'Reilly}, {Ormiston}, {Ortega}, {O'Shaughnessy}, {Oshino}, {Ossokine}, {Ottaway}, {Overmier}, {Owen}, {Pace}, {Pagano}, {Page}, {Pai}, {Pai}, {Palamos}, {Palashov}, {Palomba}, {Pal-Singh}, {Pan}, {Pan}, {Pang}, {Pang}, {Pang}, {Pankow}, {Pannarale}, {Pant}, {Paoletti}, {Paoli}, {Papa}, {Parida}, {Park}, {Parker}, {Pascucci}, {Pasqualetti}, {Passaquieti}, {Passuello}, {Patil}, {Patricelli}, {Pearlstone}, {Pedersen}, {Pedraza}, {Pedurand}, {Pele}, {Arellano}, {Penn}, {Perez}, {Perreca}, {Pfeiffer}, {Phelps}, {Phukon}, {Piccinni}, {Pichot}, {Piergiovanni}, {Pillant}, {Pinard}, {Pinto}, {Pirello}, {Pitkin}, {Poggiani}, {Pong}, {Ponrathnam}, {Popolizio}, {Porter}, {Powell}, {Prajapati}, {Prasad}, {Prasai}, {Prasanna}, {Pratten}, {Prestegard}, {Privitera}, {Prodi}, {Prokhorov}, {Puncken}, {Punturo}, {Puppo}, {P{\"u}rrer}, {Qi}, {Quetschke}, {Quinonez}, {Quintero}, {Quitzow-James}, {Raab}, {Radkins}, {Radulescu}, {Raffai}, {Raja}, {Rajan}, {Rajbhandari},
  {Rakhmanov}, {Ramirez}, {Ramos-Buades}, {Rana}, {Rao}, {Rapagnani}, {Raymond}, {Razzano}, {Read}, {Regimbau}, {Rei}, {Reid}, {Reitze}, {Ren}, {Ricci}, {Richardson}, {Richardson}, {Ricker}, {Riles}, {Rizzo}, {Robertson}, {Robie}, {Robinet}, {Rocchi}, {Rolland}, {Rollins}, {Roma}, {Romanelli}, {Romano}, {Romel}, {Romie}, {Rose}, {Rosi{\'n}ska}, {Rosofsky}, {Ross}, {Rowan}, {R{\"u}diger}, {Ruggi}, {Rutins}, {Ryan}, {Sachdev}, {Sadecki}, {Sago}, {Saito}, {Saito}, {Sakai}, {Sakai}, {Sakamoto}, {Sakellariadou}, {Sakuno}, {Salconi}, {Saleem}, {Samajdar}, {Sammut}, {Sanchez}, {Sanchez}, {Sanchis-Gual}, {Sandberg}, {Sanders}, {Santiago}, {Sarin}, {Sassolas}, {Sathyaprakash}, {Sato}, {Sato}, {Sauter}, {Savage}, {Sawada}, {Schale}, {Scheel}, {Scheuer}, {Schmidt}, {Schnabel}, {Schofield}, {Sch{\"o}nbeck}, {Schreiber}, {Schulte}, {Schutz}, {Schwalbe}, {Scott}, {Scott}, {Seidel}, {Sekiguchi}, {Sekiguchi}, {Sellers}, {Sengupta}, {Sennett}, {Sentenac}, {Sequino}, {Sergeev}, {Setyawati}, {Shaddock}, {Shaffer}, {Shahriar},
  {Shaner}, {Shao}, {Sharma}, {Shawhan}, {Shen}, {Shibagaki}, {Shimizu}, {Shimoda}, {Shimode}, {Shink}, {Shinkai}, {Shishido}, {Shoda}, {Shoemaker}, {Shoemaker}, {Shyamsundar}, {Siellez}, {Sieniawska}, {Sigg}, {Silva}, {Singer}, {Singh}, {Singhal}, {Sintes}, {Sitmukhambetov}, {Skliris}, {Slagmolen}, {Slaven-Blair}, {Smith}, {Smith}, {Somala}, {Somiya}, {Son}, {Sorazu}, {Sorrentino}, {Sotani}, {Souradeep}, {Sowell}, {Spencer}, {Srivastava}, {Srivastava}, {Staats}, {Stachie}, {Standke}, {Steer}, {Steinke}, {Steinlechner}, {Steinlechner}, {Steinmeyer}, {Stevenson}, {Stocks}, {Stone}, {Stops}, {Strain}, {Stratta}, {Strigin}, {Strunk}, {Sturani}, {Stuver}, {Sudhir}, {Sugimoto}, {Summerscales}, {Sun}, {Sunil}, {Suresh}, {Sutton}, {Suzuki}, {Suzuki}, {Swinkels}, {Szczepa{\'n}czyk}, {Tacca}, {Tagoshi}, {Tait}, {Takahashi}, {Takahashi}, {Takamori}, {Takano}, {Takeda}, {Takeda}, {Talbot}, {Talukder}, {Tanaka}, {Tanaka}, {Tanaka}, {Tanaka}, {Tanaka}, {Tanioka}, {Tanner}, {T{\'a}pai}, {Tapia San Martin}, {Taracchini},
  {Tasson}, {Taylor}, {Telada}, {Thies}, {Thomas}, {Thomas}, {Thondapu}, {Thorne}, {Thrane}, {Tiwari}, {Tiwari}, {Tiwari}, {Toland}, {Tomaru}, {Tomigami}, {Tomura}, {Tonelli}, {Tornasi}, {Torres-Forn{\'e}}, {Torrie}, {T{\"o}yr{\"a}}, {Travasso}, {Traylor}, {Tringali}, {Trovato}, {Trozzo}, {Trudeau}, {Tsang}, {Tsang}, {Tse}, {Tso}, {Tsubono}, {Tsuchida}, {Tsukada}, {Tsuna}, {Tsuzuki}, {Tuyenbayev}, {Uchikata}, {Uchiyama}, {Ueda}, {Uehara}, {Ueno}, {Ueshima}, {Ugolini}, {Unnikrishnan}, {Uraguchi}, {Urban}, {Ushiba}, {Usman}, {Vahlbruch}, {Vajente}, {Valdes}, {van Bakel}, {van Beuzekom}, {van den Brand}, {van den Broeck}, {Vander-Hyde}, {van der Schaaf}, {van Heijningen}, {van Putten}, {van Veggel}, {Vardaro}, {Varma}, {Vass}, {Vas{\'u}th}, {Vecchio}, {Vedovato}, {Veitch}, {Veitch}, {Venkateswara}, {Venugopalan}, {Verkindt}, {Vetrano}, {Vicer{\'e}}, {Viets}, {Vine}, {Vinet}, {Vitale}, {Vivanco}, {Vo}, {Vocca}, {Vorvick}, {Vyatchanin}, {Wade}, {Wade}, {Wade}, {Walet}, {Walker}, {Wallace}, {Walsh}, {Wang}, {Wang},
  {Wang}, {Wang}, {Wang}, {Wang}, {Ward}, {Warden}, {Warner}, {Was}, {Watchi}, {Weaver}, {Wei}, {Weinert}, {Weinstein}, {Weiss}, {Wellmann}, {Wen}, {Wessel}, {We{\ss}els}, {Westhouse}, {Wette}, {Whelan}, {Whiting}, {Whittle}, {Wilken}, {Williams}, {Williamson}, {Willis}, {Willke}, {Wimmer}, {Winkler}, {Wipf}, {Wittel}, {Woan}, {Woehler}, {Wofford}, {Worden}, {Wright}, {Wu}, {Wu}, {Wu}, {Wu}, {Wysocki}, {Xiao}, {Xu}, {Yamada}, {Yamamoto}, {Yamamoto}, {Yamamoto}, {Yamamoto}, {Yancey}, {Yang}, {Yap}, {Yazback}, {Yeeles}, {Yokogawa}, {Yokoyama}, {Yokozawa}, {Yoshioka}, {Yu}, {Yu}, {Yuen}, {Yuzurihara}, {Yvert}, {Zadro{\.z}ny}, {Zanolin}, {Zeidler}, {Zelenova}, {Zendri}, {Zevin}, {Zhang}, {Zhang}, {Zhang}, {Zhao}, {Zhao}, {Zhou}, {Zhou}, {Zhu}, {Zhu}, {Zimmerman}, {Zucker}, {Zweizig}, {Kagra Collaboration}, \& {VIRGO Collaboration}}]{Abbott20}
---. 2020, Living Reviews in Relativity, 23, 3, \dodoi{10.1007/s41114-020-00026-9}

\bibitem[{{Abbott} {et~al.}(2023{\natexlab{a}}){Abbott}, {Abbott}, {Acernese}, {Ackley}, {Adams}, {Adhikari}, {Adhikari}, {Adya}, {Affeldt}, {Agarwal}, {Agathos}, {Agatsuma}, {Aggarwal}, {Aguiar}, {Aiello}, {Ain}, {Ajith}, {Akcay}, {Akutsu}, {Albanesi}, {Allocca}, {Altin}, {Amato}, {Anand}, {Anand}, {Ananyeva}, {Anderson}, {Anderson}, {Ando}, {Andrade}, {Andres}, {Andri{\'c}}, {Angelova}, {Ansoldi}, {Antelis}, {Antier}, {Appert}, {Arai}, {Arai}, {Arai}, {Araki}, {Araya}, {Araya}, {Areeda}, {Ar{\`e}ne}, {Aritomi}, {Arnaud}, {Arogeti}, {Aronson}, {Arun}, {Asada}, {Asali}, {Ashton}, {Aso}, {Assiduo}, {Aston}, {Astone}, {Aubin}, {Austin}, {Babak}, {Badaracco}, {Bader}, {Badger}, {Bae}, {Bae}, {Baer}, {Bagnasco}, {Bai}, {Baiotti}, {Baird}, {Bajpai}, {Ball}, {Ballardin}, {Ballmer}, {Balsamo}, {Baltus}, {Banagiri}, {Bankar}, {Barayoga}, {Barbieri}, {Barish}, {Barker}, {Barneo}, {Barone}, {Barr}, {Barsotti}, {Barsuglia}, {Barta}, {Bartlett}, {Barton}, {Bartos}, {Bassiri}, {Basti}, {Bawaj}, {Bayley}, {Baylor},
  {Bazzan}, {B{\'e}csy}, {Bedakihale}, {Bejger}, {Belahcene}, {Benedetto}, {Beniwal}, {Bennett}, {Bentley}, {Benyaala}, {Bergamin}, {Berger}, {Bernuzzi}, {Berry}, {Bersanetti}, {Bertolini}, {Betzwieser}, {Beveridge}, {Bhandare}, {Bhardwaj}, {Bhattacharjee}, {Bhaumik}, {Bilenko}, {Billingsley}, {Bini}, {Birney}, {Birnholtz}, {Biscans}, {Bischi}, {Biscoveanu}, {Bisht}, {Biswas}, {Bitossi}, {Bizouard}, {Blackburn}, {Blair}, {Blair}, {Blair}, {Bobba}, {Bode}, {Boer}, {Bogaert}, {Boldrini}, {Bonavena}, {Bondu}, {Bonilla}, {Bonnand}, {Booker}, {Boom}, {Bork}, {Boschi}, {Bose}, {Bose}, {Bossilkov}, {Boudart}, {Bouffanais}, {Bozzi}, {Bradaschia}, {Brady}, {Bramley}, {Branch}, {Branchesi}, {Brandt}, {Brau}, {Breschi}, {Briant}, {Briggs}, {Brillet}, {Brinkmann}, {Brockill}, {Brooks}, {Brooks}, {Brown}, {Brunett}, {Bruno}, {Bruntz}, {Bryant}, {Bulik}, {Bulten}, {Buonanno}, {Buscicchio}, {Buskulic}, {Buy}, {Byer}, {Davies}, {Cadonati}, {Cagnoli}, {Cahillane}, {Bustillo}, {Callaghan}, {Callister}, {Calloni}, {Cameron},
  {Camp}, {Canepa}, {Canevarolo}, {Cannavacciuolo}, {Cannon}, {Cao}, {Cao}, {Capocasa}, {Capote}, {Carapella}, {Carbognani}, {Carlin}, {Carney}, {Carpinelli}, {Carrillo}, {Carullo}, {Carver}, {Diaz}, {Casentini}, {Castaldi}, {Caudill}, {Cavagli{\`a}}, {Cavalier}, {Cavalieri}, {Ceasar}, {Cella}, {Cerd{\'a}-Dur{\'a}n}, {Cesarini}, {Chaibi}, {Chakravarti}, {Subrahmanya}, {Champion}, {Chan}, {Chan}, {Chan}, {Chan}, {Chan}, {Chandra}, {Chanial}, {Chao}, {Chapman-Bird}, {Charlton}, {Chase}, {Chassande-Mottin}, {Chatterjee}, {Chatterjee}, {Chatterjee}, {Chaturvedi}, {Chaty}, {Chatziioannou}, {Chen}, {Chen}, {Chen}, {Chen}, {Chen}, {Chen}, {Chen}, {Chen}, {Cheng}, {Cheong}, {Cheung}, {Chia}, {Chiadini}, {Chiang}, {Chiarini}, {Chierici}, {Chincarini}, {Chiofalo}, {Chiummo}, {Cho}, {Cho}, {Choudhary}, {Choudhary}, {Christensen}, {Chu}, {Chu}, {Chu}, {Chua}, {Chung}, {Ciani}, {Ciecielag}, {Cie{\'s}lar}, {Cifaldi}, {Ciobanu}, {Ciolfi}, {Cipriano}, {Cirone}, {Clara}, {Clark}, {Clark}, {Clarke}, {Clearwater}, {Clesse},
  {Cleva}, {Coccia}, {Codazzo}, {Cohadon}, {Cohen}, {Cohen}, {Colleoni}, {Collette}, {Colombo}, {Colpi}, {Compton}, {Constancio}, {Conti}, {Cooper}, {Corban}, {Corbitt}, {Cordero-Carri{\'o}n}, {Corezzi}, {Corley}, {Cornish}, {Corre}, {Corsi}, {Cortese}, {Costa}, {Cotesta}, {Coughlin}, {Coulon}, {Countryman}, {Cousins}, {Couvares}, {Coward}, {Cowart}, {Coyne}, {Coyne}, {Creighton}, {Creighton}, {Criswell}, {Croquette}, {Crowder}, {Cudell}, {Cullen}, {Cumming}, {Cummings}, {Cunningham}, {Cuoco}, {Cury{\l}o}, {Dabadie}, {Canton}, {Dall'Osso}, {D{\'a}lya}, {Dana}, {Daneshgaranbajastani}, {D'Angelo}, {Danila}, {Danilishin}, {D'Antonio}, {Danzmann}, {Darsow-Fromm}, {Dasgupta}, {Datrier}, {Dattilo}, {Dave}, {Davier}, {Davis}, {Davis}, {Daw}, {de Alarc{\'o}n}, {Dean}, {Debra}, {Deenadayalan}, {Degallaix}, {de Laurentis}, {Del{\'e}glise}, {Del Favero}, {de Lillo}, {de Lillo}, {Del Pozzo}, {Demarchi}, {de Matteis}, {D'Emilio}, {Demos}, {Dent}, {Depasse}, {de Pietri}, {De Rosa}, {de Rossi}, {Desalvo}, {de Simone},
  {Dhurandhar}, {D{\'\i}az}, {Diaz-Ortiz}, {Didio}, {Dietrich}, {di Fiore}, {di Fronzo}, {di Giorgio}, {di Giovanni}, {di Giovanni}, {di Girolamo}, {di Lieto}, {Ding}, {di Pace}, {di Palma}, {di Renzo}, {Divakarla}, {Dmitriev}, {Doctor}, {D'Onofrio}, {Donovan}, {Dooley}, {Doravari}, {Dorrington}, {Drago}, {Driggers}, {Drori}, {Ducoin}, {Dupej}, {Durante}, {D'Urso}, {Duverne}, {Dwyer}, {Eassa}, {Easter}, {Ebersold}, {Eckhardt}, {Eddolls}, {Edelman}, {Edo}, {Edy}, {Effler}, {Eguchi}, {Eichholz}, {Eikenberry}, {Eisenmann}, {Eisenstein}, {Ejlli}, {Engelby}, {Enomoto}, {Errico}, {Essick}, {Estell{\'e}s}, {Estevez}, {Etienne}, {Etzel}, {Evans}, {Evans}, {Ewing}, {Fafone}, {Fair}, {Fairhurst}, {Farah}, {Farinon}, {Farr}, {Farr}, {Farrow}, {Fauchon-Jones}, {Favaro}, {Favata}, {Fays}, {Fazio}, {Feicht}, {Fejer}, {Fenyvesi}, {Ferguson}, {Fernandez-Galiana}, {Ferrante}, {Ferreira}, {Fidecaro}, {Figura}, {Fiori}, {Fishbach}, {Fisher}, {Fittipaldi}, {Fiumara}, {Flaminio}, {Floden}, {Fong}, {Font}, {Fornal}, {Forsyth},
  {Franke}, {Frasca}, {Frasconi}, {Frederick}, {Freed}, {Frei}, {Freise}, {Frey}, {Fritschel}, {Frolov}, {Fronz{\'e}}, {Fujii}, {Fujikawa}, {Fukunaga}, {Fukushima}, {Fulda}, {Fyffe}, {Gabbard}, {Gabella}, {Gadre}, {Gair}, {Gais}, {Galaudage}, {Gamba}, {Ganapathy}, {Ganguly}, {Gao}, {Gaonkar}, {Garaventa}, {Garc{\'\i}a}, {Garc{\'\i}a-N{\'u}{\~n}ez}, {Garc{\'\i}a-Quir{\'o}s}, {Garufi}, {Gateley}, {Gaudio}, {Gayathri}, {Ge}, {Gemme}, {Gennai}, {George}, {George}, {Gerberding}, {Gergely}, {Gewecke}, {Ghonge}, {Ghosh}, {Ghosh}, {Ghosh}, {Ghosh}, {Giacomazzo}, {Giacoppo}, {Giaime}, {Giardina}, {Gibson}, {Gier}, {Giesler}, {Giri}, {Gissi}, {Glanzer}, {Gleckl}, {Godwin}, {Goetz}, {Goetz}, {Gohlke}, {Golomb}, {Goncharov}, {Gonz{\'a}lez}, {Gopakumar}, {Gosselin}, {Gouaty}, {Gould}, {Grace}, {Grado}, {Granata}, {Granata}, {Grant}, {Gras}, {Grassia}, {Gray}, {Gray}, {Greco}, {Green}, {Green}, {Gretarsson}, {Gretarsson}, {Griffith}, {Griffiths}, {Griggs}, {Grignani}, {Grimaldi}, {Grimm}, {Grote}, {Grunewald}, {Gruning},
  {Guerra}, {Guidi}, {Guimaraes}, {Guix{\'e}}, {Gulati}, {Guo}, {Guo}, {Gupta}, {Gupta}, {Gupta}, {Gustafson}, {Gustafson}, {Guzman}, {Ha}, {Haegel}, {Hagiwara}, {Haino}, {Halim}, {Hall}, {Hamilton}, {Hammond}, {Han}, {Haney}, {Hanks}, {Hanna}, {Hannam}, {Hannuksela}, {Hansen}, {Hansen}, {Hanson}, {Harder}, {Hardwick}, {Haris}, {Harms}, {Harry}, {Harry}, {Hartwig}, {Hasegawa}, {Haskell}, {Hasskew}, {Haster}, {Hattori}, {Haughian}, {Hayakawa}, {Hayama}, {Hayes}, {Healy}, {Heidmann}, {Heidt}, {Heintze}, {Heinze}, {Heinzel}, {Heitmann}, {Hellman}, {Hello}, {Helmling-Cornell}, {Hemming}, {Hendry}, {Heng}, {Hennes}, {Hennig}, {Hennig}, {Hernandez}, {Hernandez Vivanco}, {Heurs}, {Hild}, {Hill}, {Himemoto}, {Hines}, {Hiranuma}, {Hirata}, {Hirose}, {Hochheim}, {Hofman}, {Hohmann}, {Holcomb}, {Holland}, {Holley-Bockelmann}, {Hollows}, {Holmes}, {Holt}, {Holz}, {Hong}, {Hopkins}, {Hough}, {Hourihane}, {Howell}, {Hoy}, {Hoyland}, {Hreibi}, {Hsieh}, {Hsu}, {Huang}, {Huang}, {Huang}, {Huang}, {Huang}, {Huang},
  {H{\"u}bner}, {Huddart}, {Hughey}, {Hui}, {Hui}, {Husa}, {Huttner}, {Huxford}, {Huynh-Dinh}, {Ide}, {Idzkowski}, {Iess}, {Ikenoue}, {Imam}, {Inayoshi}, {Ingram}, {Inoue}, {Ioka}, {Isi}, {Isleif}, {Ito}, {Itoh}, {Iyer}, {Izumi}, {Jaberianhamedan}, {Jacqmin}, {Jadhav}, {Jadhav}, {James}, {Jan}, {Jani}, {Janquart}, {Janssens}, {Janthalur}, {Jaranowski}, {Jariwala}, {Jaume}, {Jenkins}, {Jenner}, {Jeon}, {Jeunon}, {Jia}, {Jin}, {Johns}, {Johnson-McDaniel}, {Jones}, {Jones}, {Jones}, {Jones}, {Jones}, {Jonker}, {Ju}, {Jung}, {Jung}, {Junker}, {Juste}, {Kaihotsu}, {Kajita}, {Kakizaki}, {Kalaghatgi}, {Kalogera}, {Kamai}, {Kamiizumi}, {Kanda}, {Kandhasamy}, {Kang}, {Kanner}, {Kao}, {Kapadia}, {Kapasi}, {Karat}, {Karathanasis}, {Karki}, {Kashyap}, {Kasprzack}, {Kastaun}, {Katsanevas}, {Katsavounidis}, {Katzman}, {Kaur}, {Kawabe}, {Kawaguchi}, {Kawai}, {Kawasaki}, {K{\'e}f{\'e}lian}, {Keitel}, {Key}, {Khadka}, {Khalili}, {Khan}, {Khazanov}, {Khetan}, {Khursheed}, {Kijbunchoo}, {Kim}, {Kim}, {Kim}, {Kim}, {Kim}, {Kim},
  {Kimball}, {Kimura}, {Kinley-Hanlon}, {Kirchhoff}, {Kissel}, {Kita}, {Kitazawa}, {Kleybolte}, {Klimenko}, {Knee}, {Knowles}, {Knyazev}, {Koch}, {Koekoek}, {Kojima}, {Kokeyama}, {Koley}, {Kolitsidou}, {Kolstein}, {Komori}, {Kondrashov}, {Kong}, {Kontos}, {Koper}, {Korobko}, {Kotake}, {Kovalam}, {Kozak}, {Kozakai}, {Kozu}, {Kringel}, {Krishnendu}, {Kr{\'o}lak}, {Kuehn}, {Kuei}, {Kuijer}, {Kulkarni}, {Kumar}, {Kumar}, {Kumar}, {Kumar}, {Kume}, {Kuns}, {Kuo}, {Kuo}, {Kuromiya}, {Kuroyanagi}, {Kusayanagi}, {Kuwahara}, {Kwak}, {Lagabbe}, {Laghi}, {Lalande}, {Lam}, {Lamberts}, {Landry}, {Lane}, {Lang}, {Lange}, {Lantz}, {La Rosa}, {Lartaux-Vollard}, {Lasky}, {Laxen}, {Lazzarini}, {Lazzaro}, {Leaci}, {Leavey}, {Lecoeuche}, {Lee}, {Lee}, {Lee}, {Lee}, {Lee}, {Lee}, {Lehmann}, {Lema{\^\i}tre}, {Leonardi}, {Leroy}, {Letendre}, {Levesque}, {Levin}, {Leviton}, {Leyde}, {Li}, {Li}, {Li}, {Li}, {Li}, {Li}, {Lin}, {Lin}, {Lin}, {Lin}, {Lin}, {Linde}, {Linker}, {Linley}, {Littenberg}, {Liu}, {Liu}, {Liu}, {Liu}, {Llamas},
  {Llorens-Monteagudo}, {Lo}, {Lockwood}, {Loh}, {London}, {Longo}, {Lopez}, {Portilla}, {Lorenzini}, {Loriette}, {Lormand}, {Losurdo}, {Lott}, {Lough}, {Lousto}, {Lovelace}, {Lucaccioni}, {L{\"u}ck}, {Lumaca}, {Lundgren}, {Luo}, {Lynam}, {Macas}, {Macinnis}, {MacLeod}, {MacMillan}, {Macquet}, {Hernandez}, {Magazz{\`u}}, {Magee}, {Maggiore}, {Magnozzi}, {Mahesh}, {Majorana}, {Makarem}, {Maksimovic}, {Maliakal}, {Malik}, {Man}, {Mandic}, {Mangano}, {Mango}, {Mansell}, {Manske}, {Mantovani}, {Mapelli}, {Marchesoni}, {Marchio}, {Marion}, {Mark}, {M{\'a}rka}, {M{\'a}rka}, {Markakis}, {Markosyan}, {Markowitz}, {Maros}, {Marquina}, {Marsat}, {Martelli}, {Martin}, {Martin}, {Martinez}, {Martinez}, {Martinez}, {Martinovic}, {Martynov}, {Marx}, {Masalehdan}, {Mason}, {Massera}, {Masserot}, {Massinger}, {Masso-Reid}, {Mastrogiovanni}, {Matas}, {Mateu-Lucena}, {Matichard}, {Matiushechkina}, {Mavalvala}, {McCann}, {McCarthy}, {McClelland}, {McClincy}, {McCormick}, {McCuller}, {McGhee}, {McGuire}, {McIsaac}, {McIver},
  {McRae}, {McWilliams}, {Meacher}, {Mehmet}, {Mehta}, {Meijer}, {Melatos}, {Melchor}, {Mendell}, {Menendez-Vazquez}, {Menoni}, {Mercer}, {Mereni}, {Merfeld}, {Merilh}, {Merritt}, {Merzougui}, {Meshkov}, {Messenger}, {Messick}, {Meyers}, {Meylahn}, {Mhaske}, {Miani}, {Miao}, {Michaloliakos}, {Michel}, {Michimura}, {Middleton}, {Milano}, {Miller}, {Miller}, {Miller}, {Millhouse}, {Mills}, {Milotti}, {Minazzoli}, {Minenkov}, {Mio}, {Mir}, {Miravet-Ten{\'e}s}, {Mishra}, {Mishra}, {Mistry}, {Mitra}, {Mitrofanov}, {Mitselmakher}, {Mittleman}, {Miyakawa}, {Miyamoto}, {Miyazaki}, {Miyo}, {Miyoki}, {Mo}, {Modafferi}, {Moguel}, {Mogushi}, {Mohapatra}, {Mohite}, {Molina}, {Molina-Ruiz}, {Mondin}, {Montani}, {Moore}, {Moraru}, {Morawski}, {More}, {Moreno}, {Moreno}, {Mori}, {Morisaki}, {Moriwaki}, {Morr{\'a}s}, {Mours}, {Mow-Lowry}, {Mozzon}, {Muciaccia}, {Mukherjee}, {Mukherjee}, {Mukherjee}, {Mukherjee}, {Mukherjee}, {Mukund}, {Mullavey}, {Munch}, {Mu{\~n}iz}, {Murray}, {Musenich}, {Muusse}, {Nadji}, {Nagano},
  {Nagano}, {Nagar}, {Nakamura}, {Nakano}, {Nakano}, {Nakashima}, {Nakayama}, {Napolano}, {Nardecchia}, {Narikawa}, {Naticchioni}, {Nayak}, {Nayak}, {Negishi}, {Neil}, {Neilson}, {Nelemans}, {Nelson}, {Nery}, {Neubauer}, {Neunzert}, {Ng}, {Ng}, {Nguyen}, {Nguyen}, {Nguyen}, {Quynh}, {Ni}, {Nichols}, {Nishizawa}, {Nissanke}, {Nitoglia}, {Nocera}, {Norman}, {North}, {Nozaki}, {Siles}, {Nuttall}, {Oberling}, {O'Brien}, {Obuchi}, {O'Dell}, {Oelker}, {Ogaki}, {Oganesyan}, {Oh}, {Oh}, {Oh}, {Ohashi}, {Ohishi}, {Ohkawa}, {Ohme}, {Ohta}, {Okada}, {Okutani}, {Okutomi}, {Olivetto}, {Oohara}, {Ooi}, {Oram}, {O'Reilly}, {Ormiston}, {Ormsby}, {Ortega}, {O'Shaughnessy}, {O'Shea}, {Oshino}, {Ossokine}, {Osthelder}, {Otabe}, {Ottaway}, {Overmier}, {Pace}, {Pagano}, {Page}, {Pagliaroli}, {Pai}, {Pai}, {Palamos}, {Palashov}, {Palomba}, {Pan}, {Pan}, {Panda}, {Pang}, {Pang}, {Pankow}, {Pannarale}, {Pant}, {Panther}, {Paoletti}, {Paoli}, {Paolone}, {Parisi}, {Park}, {Park}, {Parker}, {Pascucci}, {Pasqualetti}, {Passaquieti},
  {Passuello}, {Patel}, {Pathak}, {Patricelli}, {Patron}, {Paul}, {Payne}, {Pedraza}, {Pegoraro}, {Pele}, {Arellano}, {Penn}, {Perego}, {Pereira}, {Pereira}, {Perez}, {P{\'e}rigois}, {Perkins}, {Perreca}, {Perri{\`e}s}, {Petermann}, {Petterson}, {Pfeiffer}, {Pham}, {Phukon}, {Piccinni}, {Pichot}, {Piendibene}, {Piergiovanni}, {Pierini}, {Pierro}, {Pillant}, {Pillas}, {Pilo}, {Pinard}, {Pinto}, {Pinto}, {Piotrzkowski}, {Piotrzkowski}, {Pirello}, {Pitkin}, {Placidi}, {Planas}, {Plastino}, {Pluchar}, {Poggiani}, {Polini}, {Pong}, {Ponrathnam}, {Popolizio}, {Porter}, {Poulton}, {Powell}, {Pracchia}, {Pradier}, {Prajapati}, {Prasai}, {Prasanna}, {Pratten}, {Principe}, {Prodi}, {Prokhorov}, {Prosposito}, {Prudenzi}, {Puecher}, {Punturo}, {Puosi}, {Puppo}, {P{\"u}rrer}, {Qi}, {Quetschke}, {Quitzow-James}, {Qutob}, {Raab}, {Raaijmakers}, {Radkins}, {Radulesco}, {Raffai}, {Rail}, {Raja}, {Rajan}, {Ramirez}, {Ramirez}, {Ramos-Buades}, {Rana}, {Rapagnani}, {Rapol}, {Ray}, {Raymond}, {Raza}, {Razzano}, {Read}, {Rees},
  {Regimbau}, {Rei}, {Reid}, {Reid}, {Reitze}, {Relton}, {Renzini}, {Rettegno}, {Reza}, {Rezac}, {Ricci}, {Richards}, {Richardson}, {Richardson}, {Riemenschneider}, {Riles}, {Rinaldi}, {Rink}, {Rizzo}, {Robertson}, {Robie}, {Robinet}, {Rocchi}, {Rodriguez}, {Rolland}, {Rollins}, {Romanelli}, {Romano}, {Romel}, {Romero-Rodr{\'\i}guez}, {Romero-Shaw}, {Romie}, {Ronchini}, {Rosa}, {Rose}, {Rosi{\'n}ska}, {Ross}, {Rowan}, {Rowlinson}, {Roy}, {Roy}, {Roy}, {Rozza}, {Ruggi}, {Ruiz-Rocha}, {Ryan}, {Sachdev}, {Sadecki}, {Sadiq}, {Sago}, {Saito}, {Saito}, {Sakai}, {Sakai}, {Sakellariadou}, {Sakuno}, {Salafia}, {Salconi}, {Saleem}, {Salemi}, {Samajdar}, {Sanchez}, {Sanchez}, {Sanchez}, {Sanchis-Gual}, {Sanders}, {Sanuy}, {Saravanan}, {Sarin}, {Sassolas}, {Satari}, {Sathyaprakash}, {Sato}, {Sato}, {Sauter}, {Savage}, {Sawada}, {Sawant}, {Sawant}, {Sayah}, {Schaetzl}, {Scheel}, {Scheuer}, {Schiworski}, {Schmidt}, {Schmidt}, {Schnabel}, {Schneewind}, {Schofield}, {Sch{\"o}nbeck}, {Schulte}, {Schutz}, {Schwartz}, {Scott},
  {Scott}, {Seglar-Arroyo}, {Sekiguchi}, {Sekiguchi}, {Sellers}, {Sengupta}, {Sentenac}, {Seo}, {Sequino}, {Sergeev}, {Setyawati}, {Shaffer}, {Shahriar}, {Shams}, {Shao}, {Sharma}, {Sharma}, {Shawhan}, {Shcheblanov}, {Shibagaki}, {Shikauchi}, {Shimizu}, {Shimoda}, {Shimode}, {Shinkai}, {Shishido}, {Shoda}, {Shoemaker}, {Shoemaker}, {Shyamsundar}, {Sieniawska}, {Sigg}, {Singer}, {Singh}, {Singh}, {Singha}, {Sintes}, {Sipala}, {Skliris}, {Slagmolen}, {Slaven-Blair}, {Smetana}, {Smith}, {Smith}, {Soldateschi}, {Somala}, {Somiya}, {Son}, {Soni}, {Soni}, {Sordini}, {Sorrentino}, {Sorrentino}, {Sotani}, {Soulard}, {Souradeep}, {Sowell}, {Spagnuolo}, {Spencer}, {Spera}, {Srinivasan}, {Srivastava}, {Srivastava}, {Staats}, {Stachie}, {Steer}, {Steinhoff}, {Steinlechner}, {Steinlechner}, {Stevenson}, {Stops}, {Stover}, {Strain}, {Strang}, {Stratta}, {Strunk}, {Sturani}, {Stuver}, {Sudhagar}, {Sudhir}, {Sugimoto}, {Suh}, {Sullivan}, {Sullivan}, {Summerscales}, {Sun}, {Sun}, {Sunil}, {Sur}, {Suresh}, {Sutton}, {Suzuki},
  {Suzuki}, {Swinkels}, {Szczepa{\'n}czyk}, {Szewczyk}, {Tacca}, {Tagoshi}, {Tait}, {Takahashi}, {Takahashi}, {Takamori}, {Takano}, {Takeda}, {Takeda}, {Talbot}, {Talbot}, {Tanaka}, {Tanaka}, {Tanaka}, {Tanaka}, {Tanaka}, {Tanasijczuk}, {Tanioka}, {Tanner}, {Tao}, {Tao}, {Mart{\'\i}n}, {Taranto}, {Tasson}, {Telada}, {Tenorio}, {Terhune}, {Terkowski}, {Thirugnanasambandam}, {Thomas}, {Thomas}, {Thomas}, {Thompson}, {Thondapu}, {Thorne}, {Thrane}, {Tiwari}, {Tiwari}, {Tiwari}, {Toivonen}, {Toland}, {Tolley}, {Tomaru}, {Tomigami}, {Tomura}, {Tonelli}, {Torres-Forn{\'e}}, {Torrie}, {E Melo}, {T{\"o}yr{\"a}}, {Trapananti}, {Travasso}, {Traylor}, {Trevor}, {Tringali}, {Tripathee}, {Troiano}, {Trovato}, {Trozzo}, {Trudeau}, {Tsai}, {Tsai}, {Tsang}, {Tsang}, {Tsao}, {Tse}, {Tso}, {Tsubono}, {Tsuchida}, {Tsukada}, {Tsuna}, {Tsutsui}, {Tsuzuki}, {Turbang}, {Turconi}, {Tuyenbayev}, {Ubhi}, {Uchikata}, {Uchiyama}, {Udall}, {Ueda}, {Uehara}, {Ueno}, {Ueshima}, {Unnikrishnan}, {Uraguchi}, {Urban}, {Ushiba}, {Utina},
  {Vahlbruch}, {Vajente}, {Vajpeyi}, {Valdes}, {Valentini}, {Valsan}, {van Bakel}, {van Beuzekom}, {van den Brand}, {van den Broeck}, {Vander-Hyde}, {van der Schaaf}, {van Heijningen}, {Vanosky}, {van Putten}, {van Remortel}, {Vardaro}, {Vargas}, {Varma}, {Vas{\'u}th}, {Vecchio}, {Vedovato}, {Veitch}, {Veitch}, {Venneberg}, {Venugopalan}, {Verkindt}, {Verma}, {Verma}, {Veske}, {Vetrano}, {Vicer{\'e}}, {Vidyant}, {Viets}, {Vijaykumar}, {Villa-Ortega}, {Vinet}, {Virtuoso}, {Vitale}, {Vo}, {Vocca}, {von Reis}, {von Wrangel}, {Vorvick}, {Vyatchanin}, {Wade}, {Wade}, {Wagner}, {Walet}, {Walker}, {Wallace}, {Wallace}, {Walsh}, {Wang}, {Wang}, {Wang}, {Ward}, {Warner}, {Was}, {Washimi}, {Washington}, {Watchi}, {Weaver}, {Webster}, {Weinert}, {Weinstein}, {Weiss}, {Weller}, {Weller}, {Wellmann}, {Wen}, {We{\ss}els}, {Wette}, {Whelan}, {White}, {Whiting}, {Whittle}, {Wilken}, {Williams}, {Williams}, {Williams}, {Williamson}, {Willis}, {Willke}, {Wilson}, {Winkler}, {Wipf}, {Wlodarczyk}, {Woan}, {Woehler}, {Wofford},
  {Wong}, {Wu}, {Wu}, {Wu}, {Wu}, {Wysocki}, {Xiao}, {Xu}, {Yamada}, {Yamamoto}, {Yamamoto}, {Yamamoto}, {Yamamoto}, {Yamashita}, {Yamazaki}, {Yang}, {Yang}, {Yang}, {Yang}, {Yang}, {Yap}, {Yeeles}, {Yelikar}, {Ying}, {Yokogawa}, {Yokoyama}, {Yokozawa}, {Yoo}, {Yoshioka}, {Yu}, {Yu}, {Yuzurihara}, {Zadro{\.z}ny}, {Zanolin}, {Zeidler}, {Zelenova}, {Zendri}, {Zevin}, {Zhan}, {Zhang}, {Zhang}, {Zhang}, {Zhang}, {Zhang}, {Zhao}, {Zhao}, {Zhao}, {Zhao}, {Zheng}, {Zhou}, {Zhou}, {Zhu}, {Zhu}, {Zimmerman}, {Zlochower}, {Zucker}, {Zweizig}, {Ligo Scientific Collaboration}, {VIRGO Collaboration}, \& {Kagra Collaboration}}]{GWTC3}
{Abbott}, R., {Abbott}, T.~D., {Acernese}, F., {et~al.} 2023{\natexlab{a}}, Physical Review X, 13, 041039, \dodoi{10.1103/PhysRevX.13.041039}

\bibitem[{{Abbott} {et~al.}(2023{\natexlab{b}}){Abbott}, {Abbott}, {Acernese}, {Ackley}, {Adams}, {Adhikari}, {Adhikari}, {Adya}, {Affeldt}, {Agarwal}, {Agathos}, {Agatsuma}, {Aggarwal}, {Aguiar}, {Aiello}, {Ain}, {Ajith}, {Akutsu}, {de Alarc{\'o}n}, {Akcay}, {Albanesi}, {Allocca}, {Altin}, {Amato}, {Anand}, {Anand}, {Ananyeva}, {Anderson}, {Anderson}, {Ando}, {Andrade}, {Andres}, {Andri{\'c}}, {Angelova}, {Ansoldi}, {Antelis}, {Antier}, {Antonini}, {Appert}, {Arai}, {Arai}, {Arai}, {Araki}, {Araya}, {Araya}, {Areeda}, {Ar{\`e}ne}, {Aritomi}, {Arnaud}, {Arogeti}, {Aronson}, {Arun}, {Asada}, {Asali}, {Ashton}, {Aso}, {Assiduo}, {Aston}, {Astone}, {Aubin}, {Austin}, {Babak}, {Badaracco}, {Bader}, {Badger}, {Bae}, {Bae}, {Baer}, {Bagnasco}, {Bai}, {Baiotti}, {Baird}, {Bajpai}, {Ball}, {Ballardin}, {Ballmer}, {Balsamo}, {Baltus}, {Banagiri}, {Bankar}, {Barayoga}, {Barbieri}, {Barish}, {Barker}, {Barneo}, {Barone}, {Barr}, {Barsotti}, {Barsuglia}, {Barta}, {Bartlett}, {Barton}, {Bartos}, {Bassiri}, {Basti},
  {Bawaj}, {Bayley}, {Baylor}, {Bazzan}, {B{\'e}csy}, {Bedakihale}, {Bejger}, {Belahcene}, {Benedetto}, {Beniwal}, {Bennett}, {Bentley}, {Benyaala}, {Bergamin}, {Berger}, {Bernuzzi}, {Berry}, {Bersanetti}, {Bertolini}, {Betzwieser}, {Beveridge}, {Bhandare}, {Bhardwaj}, {Bhattacharjee}, {Bhaumik}, {Bilenko}, {Billingsley}, {Bini}, {Birney}, {Birnholtz}, {Biscans}, {Bischi}, {Biscoveanu}, {Bisht}, {Biswas}, {Bitossi}, {Bizouard}, {Blackburn}, {Blair}, {Blair}, {Blair}, {Bobba}, {Bode}, {Boer}, {Bogaert}, {Boldrini}, {Bonavena}, {Bondu}, {Bonilla}, {Bonnand}, {Booker}, {Boom}, {Bork}, {Boschi}, {Bose}, {Bose}, {Bossilkov}, {Boudart}, {Bouffanais}, {Bozzi}, {Bradaschia}, {Brady}, {Bramley}, {Branch}, {Branchesi}, {Brandt}, {Brau}, {Breschi}, {Briant}, {Briggs}, {Brillet}, {Brinkmann}, {Brockill}, {Brooks}, {Brooks}, {Brown}, {Brunett}, {Bruno}, {Bruntz}, {Bryant}, {Bulik}, {Bulten}, {Buonanno}, {Buscicchio}, {Buskulic}, {Buy}, {Byer}, {Cadonati}, {Cagnoli}, {Cahillane}, {Bustillo}, {Callaghan}, {Callister},
  {Calloni}, {Cameron}, {Camp}, {Canepa}, {Canevarolo}, {Cannavacciuolo}, {Cannon}, {Cao}, {Cao}, {Capocasa}, {Capote}, {Carapella}, {Carbognani}, {Carlin}, {Carney}, {Carpinelli}, {Carrillo}, {Carullo}, {Carver}, {Diaz}, {Casentini}, {Castaldi}, {Caudill}, {Cavagli{\`a}}, {Cavalier}, {Cavalieri}, {Ceasar}, {Cella}, {Cerd{\'a}-Dur{\'a}n}, {Cesarini}, {Chaibi}, {Chakravarti}, {Subrahmanya}, {Champion}, {Chan}, {Chan}, {Chan}, {Chan}, {Chan}, {Chandra}, {Chanial}, {Chao}, {Chapman-Bird}, {Charlton}, {Chase}, {Chassande-Mottin}, {Chatterjee}, {Chatterjee}, {Chatterjee}, {Chaturvedi}, {Chaty}, {Chatziioannou}, {Chen}, {Chen}, {Chen}, {Chen}, {Chen}, {Chen}, {Chen}, {Chen}, {Cheng}, {Cheong}, {Cheung}, {Chia}, {Chiadini}, {Chiang}, {Chiarini}, {Chierici}, {Chincarini}, {Chiofalo}, {Chiummo}, {Cho}, {Cho}, {Choudhary}, {Choudhary}, {Christensen}, {Chu}, {Chu}, {Chu}, {Chua}, {Chung}, {Ciani}, {Ciecielag}, {Cie{\'s}lar}, {Cifaldi}, {Ciobanu}, {Ciolfi}, {Cipriano}, {Cirone}, {Clara}, {Clark}, {Clark}, {Clarke},
  {Clearwater}, {Clesse}, {Cleva}, {Coccia}, {Codazzo}, {Cohadon}, {Cohen}, {Cohen}, {Colleoni}, {Collette}, {Colombo}, {Colpi}, {Compton}, {Constancio}, {Conti}, {Cooper}, {Corban}, {Corbitt}, {Cordero-Carri{\'o}n}, {Corezzi}, {Corley}, {Cornish}, {Corre}, {Corsi}, {Cortese}, {Costa}, {Cotesta}, {Coughlin}, {Coulon}, {Countryman}, {Cousins}, {Couvares}, {Coward}, {Cowart}, {Coyne}, {Coyne}, {Creighton}, {Creighton}, {Criswell}, {Croquette}, {Crowder}, {Cudell}, {Cullen}, {Cumming}, {Cummings}, {Cunningham}, {Cuoco}, {Cury{\l}o}, {Dabadie}, {Canton}, {Dall'Osso}, {D{\'a}lya}, {Dana}, {Daneshgaranbajastani}, {D'Angelo}, {Danila}, {Danilishin}, {D'Antonio}, {Danzmann}, {Darsow-Fromm}, {Dasgupta}, {Datrier}, {Datta}, {Dattilo}, {Dave}, {Davier}, {Davies}, {Davis}, {Davis}, {Daw}, {Dean}, {Debra}, {Deenadayalan}, {Degallaix}, {de Laurentis}, {Del{\'e}glise}, {Del Favero}, {de Lillo}, {de Lillo}, {Del Pozzo}, {Demarchi}, {de Matteis}, {D'Emilio}, {Demos}, {Dent}, {Depasse}, {de Pietri}, {De Rosa}, {de Rossi},
  {Desalvo}, {de Simone}, {Dhurandhar}, {D{\'\i}az}, {Diaz-Ortiz}, {Didio}, {Dietrich}, {di Fiore}, {di Fronzo}, {di Giorgio}, {di Giovanni}, {di Giovanni}, {di Girolamo}, {di Lieto}, {Ding}, {di Pace}, {di Palma}, {di Renzo}, {Divakarla}, {Dmitriev}, {Doctor}, {D'Onofrio}, {Donovan}, {Dooley}, {Doravari}, {Dorrington}, {Drago}, {Driggers}, {Drori}, {Ducoin}, {Dupej}, {Durante}, {D'Urso}, {Duverne}, {Dwyer}, {Eassa}, {Easter}, {Ebersold}, {Eckhardt}, {Eddolls}, {Edelman}, {Edo}, {Edy}, {Effler}, {Eguchi}, {Eichholz}, {Eikenberry}, {Eisenmann}, {Eisenstein}, {Ejlli}, {Engelby}, {Enomoto}, {Errico}, {Essick}, {Estell{\'e}s}, {Estevez}, {Etienne}, {Etzel}, {Evans}, {Evans}, {Ewing}, {Fafone}, {Fair}, {Fairhurst}, {Farah}, {Farinon}, {Farr}, {Farr}, {Farrow}, {Fauchon-Jones}, {Favaro}, {Favata}, {Fays}, {Fazio}, {Feicht}, {Fejer}, {Fenyvesi}, {Ferguson}, {Fernandez-Galiana}, {Ferrante}, {Ferreira}, {Fidecaro}, {Figura}, {Fiori}, {Fishbach}, {Fisher}, {Fittipaldi}, {Fiumara}, {Flaminio}, {Floden}, {Fong}, {Font},
  {Fornal}, {Forsyth}, {Franke}, {Frasca}, {Frasconi}, {Frederick}, {Freed}, {Frei}, {Freise}, {Frey}, {Fritschel}, {Frolov}, {Fronz{\'e}}, {Fujii}, {Fujikawa}, {Fukunaga}, {Fukushima}, {Fulda}, {Fyffe}, {Gabbard}, {Gadre}, {Gair}, {Gais}, {Galaudage}, {Gamba}, {Ganapathy}, {Ganguly}, {Gao}, {Gaonkar}, {Garaventa}, {Garc{\'\i}a}, {Garc{\'\i}a-N{\'u}{\~n}ez}, {Garc{\'\i}a-Quir{\'o}s}, {Garufi}, {Gateley}, {Gaudio}, {Gayathri}, {Ge}, {Gemme}, {Gennai}, {George}, {George}, {Gerberding}, {Gergely}, {Gewecke}, {Ghonge}, {Ghosh}, {Ghosh}, {Ghosh}, {Ghosh}, {Giacomazzo}, {Giacoppo}, {Giaime}, {Giardina}, {Gibson}, {Gier}, {Giesler}, {Giri}, {Gissi}, {Glanzer}, {Gleckl}, {Godwin}, {Golomb}, {Goetz}, {Goetz}, {Gohlke}, {Goncharov}, {Gonz{\'a}lez}, {Gopakumar}, {Gosselin}, {Gouaty}, {Gould}, {Grace}, {Grado}, {Granata}, {Granata}, {Grant}, {Gras}, {Grassia}, {Gray}, {Gray}, {Greco}, {Green}, {Green}, {Gretarsson}, {Gretarsson}, {Griffith}, {Griffiths}, {Griggs}, {Grignani}, {Grimaldi}, {Grimm}, {Grote}, {Grunewald},
  {Gruning}, {Guerra}, {Guidi}, {Guimaraes}, {Guix{\'e}}, {Gulati}, {Guo}, {Guo}, {Gupta}, {Gupta}, {Gupta}, {Gustafson}, {Gustafson}, {Guzman}, {Ha}, {Haegel}, {Hagiwara}, {Haino}, {Halim}, {Hall}, {Hamilton}, {Hammond}, {Han}, {Haney}, {Hanks}, {Hanna}, {Hannam}, {Hannuksela}, {Hansen}, {Hansen}, {Hanson}, {Harder}, {Hardwick}, {Haris}, {Harms}, {Harry}, {Harry}, {Hartwig}, {Hasegawa}, {Haskell}, {Hasskew}, {Haster}, {Hattori}, {Haughian}, {Hayakawa}, {Hayama}, {Hayes}, {Healy}, {Heidmann}, {Heidt}, {Heintze}, {Heinze}, {Heinzel}, {Heitmann}, {Hellman}, {Hello}, {Helmling-Cornell}, {Hemming}, {Hendry}, {Heng}, {Hennes}, {Hennig}, {Hennig}, {Hernandez}, {Vivanco}, {Heurs}, {Hild}, {Hill}, {Himemoto}, {Hines}, {Hiranuma}, {Hirata}, {Hirose}, {Hochheim}, {Hofman}, {Hohmann}, {Holcomb}, {Holland}, {Hollows}, {Holmes}, {Holt}, {Holz}, {Hong}, {Hopkins}, {Hough}, {Hourihane}, {Howell}, {Hoy}, {Hoyland}, {Hreibi}, {Hsieh}, {Hsu}, {Huang}, {Huang}, {Huang}, {Huang}, {Huang}, {Huang}, {H{\"u}bner}, {Huddart},
  {Hughey}, {Hui}, {Hui}, {Husa}, {Huttner}, {Huxford}, {Huynh-Dinh}, {Ide}, {Idzkowski}, {Iess}, {Ikenoue}, {Imam}, {Inayoshi}, {Ingram}, {Inoue}, {Ioka}, {Isi}, {Isleif}, {Ito}, {Itoh}, {Iyer}, {Izumi}, {Jaberianhamedan}, {Jacqmin}, {Jadhav}, {Jadhav}, {James}, {Jan}, {Jani}, {Janquart}, {Janssens}, {Janthalur}, {Jaranowski}, {Jariwala}, {Jaume}, {Jenkins}, {Jenner}, {Jeon}, {Jeunon}, {Jia}, {Jin}, {Johns}, {Jones}, {Jones}, {Jones}, {Jones}, {Jones}, {Jonker}, {Ju}, {Jung}, {Jung}, {Junker}, {Juste}, {Kaihotsu}, {Kajita}, {Kakizaki}, {Kalaghatgi}, {Kalogera}, {Kamai}, {Kamiizumi}, {Kanda}, {Kandhasamy}, {Kang}, {Kanner}, {Kao}, {Kapadia}, {Kapasi}, {Karat}, {Karathanasis}, {Karki}, {Kashyap}, {Kasprzack}, {Kastaun}, {Katsanevas}, {Katsavounidis}, {Katzman}, {Kaur}, {Kawabe}, {Kawaguchi}, {Kawai}, {Kawasaki}, {K{\'e}f{\'e}lian}, {Keitel}, {Key}, {Khadka}, {Khalili}, {Khan}, {Khazanov}, {Khetan}, {Khursheed}, {Kijbunchoo}, {Kim}, {Kim}, {Kim}, {Kim}, {Kim}, {Kim}, {Kimball}, {Kimura}, {Kinley-Hanlon},
  {Kirchhoff}, {Kissel}, {Kita}, {Kitazawa}, {Kleybolte}, {Klimenko}, {Knee}, {Knowles}, {Knyazev}, {Koch}, {Koekoek}, {Kojima}, {Kokeyama}, {Koley}, {Kolitsidou}, {Kolstein}, {Komori}, {Kondrashov}, {Kong}, {Kontos}, {Koper}, {Korobko}, {Kotake}, {Kovalam}, {Kozak}, {Kozakai}, {Kozu}, {Kringel}, {Krishnendu}, {Kr{\'o}lak}, {Kuehn}, {Kuei}, {Kuijer}, {Kulkarni}, {Kumar}, {Kumar}, {Kumar}, {Kumar}, {Kume}, {Kuns}, {Kuo}, {Kuo}, {Kuromiya}, {Kuroyanagi}, {Kusayanagi}, {Kuwahara}, {Kwak}, {Lagabbe}, {Laghi}, {Lalande}, {Lam}, {Lamberts}, {Landry}, {Landry}, {Lane}, {Lang}, {Lange}, {Lantz}, {La Rosa}, {Lartaux-Vollard}, {Lasky}, {Laxen}, {Lazzarini}, {Lazzaro}, {Leaci}, {Leavey}, {Lecoeuche}, {Lee}, {Lee}, {Lee}, {Lee}, {Lee}, {Lee}, {Lehmann}, {Lema{\^\i}tre}, {Leonardi}, {Leroy}, {Letendre}, {Levesque}, {Levin}, {Leviton}, {Leyde}, {Li}, {Li}, {Li}, {Li}, {Li}, {Li}, {Lin}, {Lin}, {Lin}, {Lin}, {Lin}, {Linde}, {Linker}, {Linley}, {Littenberg}, {Liu}, {Liu}, {Liu}, {Liu}, {Llamas}, {Llorens-Monteagudo}, {Lo},
  {Lockwood}, {Loh}, {London}, {Longo}, {Lopez}, {Portilla}, {Lorenzini}, {Loriette}, {Lormand}, {Losurdo}, {Lott}, {Lough}, {Lousto}, {Lovelace}, {Lucaccioni}, {L{\"u}ck}, {Lumaca}, {Lundgren}, {Luo}, {Lynam}, {Macas}, {Macinnis}, {MacLeod}, {MacMillan}, {Macquet}, {Hernandez}, {Magazz{\`u}}, {Magee}, {Maggiore}, {Magnozzi}, {Mahesh}, {Majorana}, {Makarem}, {Maksimovic}, {Maliakal}, {Malik}, {Man}, {Mandic}, {Mangano}, {Mango}, {Mansell}, {Manske}, {Mantovani}, {Mapelli}, {Marchesoni}, {Marchio}, {Marion}, {Mark}, {M{\'a}rka}, {M{\'a}rka}, {Markakis}, {Markosyan}, {Markowitz}, {Maros}, {Marquina}, {Marsat}, {Martelli}, {Martin}, {Martin}, {Martinez}, {Martinez}, {Martinez}, {Martinovic}, {Martynov}, {Marx}, {Masalehdan}, {Mason}, {Massera}, {Masserot}, {Massinger}, {Masso-Reid}, {Mastrogiovanni}, {Matas}, {Mateu-Lucena}, {Matichard}, {Matiushechkina}, {Mavalvala}, {McCann}, {McCarthy}, {McClelland}, {McClincy}, {McCormick}, {McCuller}, {McGhee}, {McGuire}, {McIsaac}, {McIver}, {McRae}, {McWilliams},
  {Meacher}, {Mehmet}, {Mehta}, {Meijer}, {Melatos}, {Melchor}, {Mendell}, {Menendez-Vazquez}, {Menoni}, {Mercer}, {Mereni}, {Merfeld}, {Merilh}, {Merritt}, {Merzougui}, {Meshkov}, {Messenger}, {Messick}, {Meyers}, {Meylahn}, {Mhaske}, {Miani}, {Miao}, {Michaloliakos}, {Michel}, {Michimura}, {Middleton}, {Milano}, {Miller}, {Miller}, {Miller}, {Miller}, {Millhouse}, {Mills}, {Milotti}, {Minazzoli}, {Minenkov}, {Mio}, {Mir}, {Miravet-Ten{\'e}s}, {Mishra}, {Mishra}, {Mistry}, {Mitra}, {Mitrofanov}, {Mitselmakher}, {Mittleman}, {Miyakawa}, {Miyamoto}, {Miyazaki}, {Miyo}, {Miyoki}, {Mo}, {Modafferi}, {Moguel}, {Mogushi}, {Mohapatra}, {Mohite}, {Molina}, {Molina-Ruiz}, {Mondin}, {Montani}, {Moore}, {Moraru}, {Morawski}, {More}, {Moreno}, {Moreno}, {Mori}, {Morisaki}, {Moriwaki}, {Morr{\'a}s}, {Mours}, {Mow-Lowry}, {Mozzon}, {Muciaccia}, {Mukherjee}, {Mukherjee}, {Mukherjee}, {Mukherjee}, {Mukherjee}, {Mukund}, {Mullavey}, {Munch}, {Mu{\~n}iz}, {Murray}, {Musenich}, {Muusse}, {Nadji}, {Nagano}, {Nagano}, {Nagar},
  {Nakamura}, {Nakano}, {Nakano}, {Nakashima}, {Nakayama}, {Napolano}, {Nardecchia}, {Narikawa}, {Naticchioni}, {Nayak}, {Nayak}, {Negishi}, {Neil}, {Neilson}, {Nelemans}, {Nelson}, {Nery}, {Neubauer}, {Neunzert}, {Ng}, {Ng}, {Nguyen}, {Nguyen}, {Nguyen}, {Quynh}, {Ni}, {Nichols}, {Nishizawa}, {Nissanke}, {Nitoglia}, {Nocera}, {Norman}, {North}, {Nozaki}, {Siles}, {Nuttall}, {Oberling}, {O'Brien}, {Obuchi}, {O'Dell}, {Oelker}, {Ogaki}, {Oganesyan}, {Oh}, {Oh}, {Oh}, {Ohashi}, {Ohishi}, {Ohkawa}, {Ohme}, {Ohta}, {Okada}, {Okutani}, {Okutomi}, {Olivetto}, {Oohara}, {Ooi}, {Oram}, {O'Reilly}, {Ormiston}, {Ormsby}, {Ortega}, {O'Shaughnessy}, {O'Shea}, {Oshino}, {Ossokine}, {Osthelder}, {Otabe}, {Ottaway}, {Overmier}, {Pace}, {Pagano}, {Page}, {Pagliaroli}, {Pai}, {Pai}, {Palamos}, {Palashov}, {Palomba}, {Pan}, {Pan}, {Panda}, {Pang}, {Pang}, {Pankow}, {Pannarale}, {Pant}, {Panther}, {Paoletti}, {Paoli}, {Paolone}, {Parisi}, {Park}, {Park}, {Parker}, {Pascucci}, {Pasqualetti}, {Passaquieti}, {Passuello}, {Patel},
  {Pathak}, {Patricelli}, {Patron}, {Paul}, {Payne}, {Pedraza}, {Pegoraro}, {Pele}, {Arellano}, {Penn}, {Perego}, {Pereira}, {Pereira}, {Perez}, {P{\'e}rigois}, {Perkins}, {Perreca}, {Perri{\`e}s}, {Petermann}, {Petterson}, {Pfeiffer}, {Pham}, {Phukon}, {Piccinni}, {Pichot}, {Piendibene}, {Piergiovanni}, {Pierini}, {Pierro}, {Pillant}, {Pillas}, {Pilo}, {Pinard}, {Pinto}, {Pinto}, {Piotrzkowski}, {Piotrzkowski}, {Pirello}, {Pitkin}, {Placidi}, {Planas}, {Plastino}, {Pluchar}, {Poggiani}, {Polini}, {Pong}, {Ponrathnam}, {Popolizio}, {Porter}, {Poulton}, {Powell}, {Pracchia}, {Pradier}, {Prajapati}, {Prasai}, {Prasanna}, {Pratten}, {Principe}, {Prodi}, {Prokhorov}, {Prosposito}, {Prudenzi}, {Puecher}, {Punturo}, {Puosi}, {Puppo}, {P{\"u}rrer}, {Qi}, {Quetschke}, {Quitzow-James}, {Raab}, {Raaijmakers}, {Radkins}, {Radulesco}, {Raffai}, {Rail}, {Raja}, {Rajan}, {Ramirez}, {Ramirez}, {Ramos-Buades}, {Rana}, {Rapagnani}, {Rapol}, {Ray}, {Raymond}, {Raza}, {Razzano}, {Read}, {Rees}, {Regimbau}, {Rei}, {Reid},
  {Reid}, {Reitze}, {Relton}, {Renzini}, {Rettegno}, {Reza}, {Rezac}, {Ricci}, {Richards}, {Richardson}, {Richardson}, {Riemenschneider}, {Riles}, {Rinaldi}, {Rink}, {Rizzo}, {Robertson}, {Robie}, {Robinet}, {Rocchi}, {Rodriguez}, {Rolland}, {Rollins}, {Romanelli}, {Romano}, {Romel}, {Romero-Rodr{\'\i}guez}, {Romero-Shaw}, {Romie}, {Ronchini}, {Rosa}, {Rose}, {Rosi{\'n}ska}, {Ross}, {Rowan}, {Rowlinson}, {Roy}, {Roy}, {Roy}, {Rozza}, {Ruggi}, {Ryan}, {Sachdev}, {Sadecki}, {Sadiq}, {Sago}, {Saito}, {Saito}, {Sakai}, {Sakai}, {Sakellariadou}, {Sakuno}, {Salafia}, {Salconi}, {Saleem}, {Salemi}, {Samajdar}, {Sanchez}, {Sanchez}, {Sanchez}, {Sanchis-Gual}, {Sanders}, {Sanuy}, {Saravanan}, {Sarin}, {Sassolas}, {Satari}, {Sathyaprakash}, {Sato}, {Sato}, {Sauter}, {Savage}, {Sawada}, {Sawant}, {Sawant}, {Sayah}, {Schaetzl}, {Scheel}, {Scheuer}, {Schiworski}, {Schmidt}, {Schmidt}, {Schnabel}, {Schneewind}, {Schofield}, {Sch{\"o}nbeck}, {Schulte}, {Schutz}, {Schwartz}, {Scott}, {Scott}, {Seglar-Arroyo}, {Sekiguchi},
  {Sekiguchi}, {Sellers}, {Sengupta}, {Sentenac}, {Seo}, {Sequino}, {Sergeev}, {Setyawati}, {Shaffer}, {Shahriar}, {Shams}, {Shao}, {Sharma}, {Sharma}, {Shawhan}, {Shcheblanov}, {Shibagaki}, {Shikauchi}, {Shimizu}, {Shimoda}, {Shimode}, {Shinkai}, {Shishido}, {Shoda}, {Shoemaker}, {Shoemaker}, {Shyamsundar}, {Sieniawska}, {Sigg}, {Singer}, {Singh}, {Singh}, {Singha}, {Sintes}, {Sipala}, {Skliris}, {Slagmolen}, {Slaven-Blair}, {Smetana}, {Smith}, {Smith}, {Soldateschi}, {Somala}, {Somiya}, {Son}, {Soni}, {Soni}, {Sordini}, {Sorrentino}, {Sorrentino}, {Sotani}, {Soulard}, {Souradeep}, {Sowell}, {Spagnuolo}, {Spencer}, {Spera}, {Srinivasan}, {Srivastava}, {Srivastava}, {Staats}, {Stachie}, {Steer}, {Steinhoff}, {Steinlechner}, {Steinlechner}, {Stevenson}, {Stops}, {Stover}, {Strain}, {Strang}, {Stratta}, {Strunk}, {Sturani}, {Stuver}, {Sudhagar}, {Sudhir}, {Sugimoto}, {Suh}, {Sullivan}, {Summerscales}, {Sun}, {Sun}, {Sunil}, {Sur}, {Suresh}, {Sutton}, {Suzuki}, {Suzuki}, {Swinkels}, {Szczepa{\'n}czyk},
  {Szewczyk}, {Tacca}, {Tagoshi}, {Tait}, {Takahashi}, {Takahashi}, {Takamori}, {Takano}, {Takeda}, {Takeda}, {Talbot}, {Talbot}, {Tanaka}, {Tanaka}, {Tanaka}, {Tanaka}, {Tanaka}, {Tanasijczuk}, {Tanioka}, {Tanner}, {Tao}, {Tao}, {Mart{\'\i}n}, {Taranto}, {Tasson}, {Telada}, {Tenorio}, {Terhune}, {Terkowski}, {Thirugnanasambandam}, {Thomas}, {Thomas}, {Thomas}, {Thompson}, {Thondapu}, {Thorne}, {Thrane}, {Tiwari}, {Tiwari}, {Tiwari}, {Toivonen}, {Toland}, {Tolley}, {Tomaru}, {Tomigami}, {Tomura}, {Tonelli}, {Torres-Forn{\'e}}, {Torrie}, {E Melo}, {T{\"o}yr{\"a}}, {Trapananti}, {Travasso}, {Traylor}, {Trevor}, {Tringali}, {Tripathee}, {Troiano}, {Trovato}, {Trozzo}, {Trudeau}, {Tsai}, {Tsai}, {Tsang}, {Tsang}, {Tsao}, {Tse}, {Tso}, {Tsubono}, {Tsuchida}, {Tsukada}, {Tsuna}, {Tsutsui}, {Tsuzuki}, {Turbang}, {Turconi}, {Tuyenbayev}, {Ubhi}, {Uchikata}, {Uchiyama}, {Udall}, {Ueda}, {Uehara}, {Ueno}, {Ueshima}, {Unnikrishnan}, {Uraguchi}, {Urban}, {Ushiba}, {Utina}, {Vahlbruch}, {Vajente}, {Vajpeyi}, {Valdes},
  {Valentini}, {Valsan}, {van Bakel}, {van Beuzekom}, {van den Brand}, {van den Broeck}, {Vander-Hyde}, {van der Schaaf}, {van Heijningen}, {Vanosky}, {van Putten}, {van Remortel}, {Vardaro}, {Vargas}, {Varma}, {Vas{\'u}th}, {Vecchio}, {Vedovato}, {Veitch}, {Veitch}, {Venneberg}, {Venugopalan}, {Verkindt}, {Verma}, {Verma}, {Veske}, {Vetrano}, {Vicer{\'e}}, {Vidyant}, {Viets}, {Vijaykumar}, {Villa-Ortega}, {Vinet}, {Virtuoso}, {Vitale}, {Vo}, {Vocca}, {von Reis}, {von Wrangel}, {Vorvick}, {Vyatchanin}, {Wade}, {Wade}, {Wagner}, {Walet}, {Walker}, {Wallace}, {Wallace}, {Walsh}, {Wang}, {Wang}, {Wang}, {Ward}, {Warner}, {Was}, {Washimi}, {Washington}, {Watchi}, {Weaver}, {Webster}, {Weinert}, {Weinstein}, {Weiss}, {Weller}, {Wellmann}, {Wen}, {We{\ss}els}, {Wette}, {Whelan}, {White}, {Whiting}, {Whittle}, {Wilken}, {Williams}, {Williams}, {Williamson}, {Willis}, {Willke}, {Wilson}, {Winkler}, {Wipf}, {Wlodarczyk}, {Woan}, {Woehler}, {Wofford}, {Wong}, {Wu}, {Wu}, {Wu}, {Wu}, {Wysocki}, {Xiao}, {Xu}, {Yamada},
  {Yamamoto}, {Yamamoto}, {Yamamoto}, {Yamamoto}, {Yamashita}, {Yamazaki}, {Yang}, {Yang}, {Yang}, {Yang}, {Yang}, {Yap}, {Yeeles}, {Yelikar}, {Ying}, {Yokogawa}, {Yokoyama}, {Yokozawa}, {Yoo}, {Yoshioka}, {Yu}, {Yu}, {Yuzurihara}, {Zadro{\.z}ny}, {Zanolin}, {Zeidler}, {Zelenova}, {Zendri}, {Zevin}, {Zhan}, {Zhang}, {Zhang}, {Zhang}, {Zhang}, {Zhang}, {Zhao}, {Zhao}, {Zhao}, {Zhao}, {Zheng}, {Zhou}, {Zhou}, {Zhu}, {Zhu}, {Zimmerman}, {Zlochower}, {Zucker}, {Zweizig}, {LIGO Scientific Collaboration}, {VIRGO Collaboration}, \& {KAGRA Collaboration}}]{GWTC23}
---. 2023{\natexlab{b}}, Physical Review X, 13, 011048, \dodoi{10.1103/PhysRevX.13.011048}

\bibitem[{{Aerts}(2021)}]{Aerts21}
{Aerts}, C. 2021, Reviews of Modern Physics, 93, 015001, \dodoi{10.1103/RevModPhys.93.015001}

\bibitem[{{Andreoni} {et~al.}(2020){Andreoni}, {Kool}, {Sagu{\'e}s Carracedo}, {Kasliwal}, {Bulla}, {Ahumada}, {Coughlin}, {Anand}, {Sollerman}, {Goobar}, {Kaplan}, {Loveridge}, {Karambelkar}, {Cooke}, {Bagdasaryan}, {Bellm}, {Cenko}, {Cook}, {De}, {Dekany}, {Delacroix}, {Drake}, {Duev}, {Fremling}, {Golkhou}, {Graham}, {Hale}, {Kulkarni}, {Kupfer}, {Laher}, {Mahabal}, {Masci}, {Rusholme}, {Smith}, {Tzanidakis}, {Van Sistine}, \& {Yao}}]{Andreoni20}
{Andreoni}, I., {Kool}, E.~C., {Sagu{\'e}s Carracedo}, A., {et~al.} 2020, \apj, 904, 155, \dodoi{10.3847/1538-4357/abbf4c}

\bibitem[{{Arcavi} {et~al.}(2017){Arcavi}, {Hosseinzadeh}, {Howell}, {McCully}, {Poznanski}, {Kasen}, {Barnes}, {Zaltzman}, {Vasylyev}, {Maoz}, \& {Valenti}}]{Arcavi17}
{Arcavi}, I., {Hosseinzadeh}, G., {Howell}, D.~A., {et~al.} 2017, \nat, 551, 64, \dodoi{10.1038/nature24291}

\bibitem[{{Asquini} {et~al.}(2024){Asquini}, {Landoni}, {Campana}, {Reguitti}, {Benetti}, \& {Farias}}]{Asquini24}
{Asquini}, L., {Landoni}, M., {Campana}, S., {et~al.} 2024, Transient Name Server AstroNote, 258, 1

\bibitem[{{Bellm} {et~al.}(2019){Bellm}, {Kulkarni}, {Graham}, {Dekany}, {Smith}, {Riddle}, {Masci}, {Helou}, {Prince}, {Adams}, {Barbarino}, {Barlow}, {Bauer}, {Beck}, {Belicki}, {Biswas}, {Blagorodnova}, {Bodewits}, {Bolin}, {Brinnel}, {Brooke}, {Bue}, {Bulla}, {Burruss}, {Cenko}, {Chang}, {Connolly}, {Coughlin}, {Cromer}, {Cunningham}, {De}, {Delacroix}, {Desai}, {Duev}, {Eadie}, {Farnham}, {Feeney}, {Feindt}, {Flynn}, {Franckowiak}, {Frederick}, {Fremling}, {Gal-Yam}, {Gezari}, {Giomi}, {Goldstein}, {Golkhou}, {Goobar}, {Groom}, {Hacopians}, {Hale}, {Henning}, {Ho}, {Hover}, {Howell}, {Hung}, {Huppenkothen}, {Imel}, {Ip}, {Ivezi{\'c}}, {Jackson}, {Jones}, {Juric}, {Kasliwal}, {Kaspi}, {Kaye}, {Kelley}, {Kowalski}, {Kramer}, {Kupfer}, {Landry}, {Laher}, {Lee}, {Lin}, {Lin}, {Lunnan}, {Giomi}, {Mahabal}, {Mao}, {Miller}, {Monkewitz}, {Murphy}, {Ngeow}, {Nordin}, {Nugent}, {Ofek}, {Patterson}, {Penprase}, {Porter}, {Rauch}, {Rebbapragada}, {Reiley}, {Rigault}, {Rodriguez}, {van Roestel}, {Rusholme}, {van
  Santen}, {Schulze}, {Shupe}, {Singer}, {Soumagnac}, {Stein}, {Surace}, {Sollerman}, {Szkody}, {Taddia}, {Terek}, {Van Sistine}, {van Velzen}, {Vestrand}, {Walters}, {Ward}, {Ye}, {Yu}, {Yan}, \& {Zolkower}}]{ztf19}
{Bellm}, E.~C., {Kulkarni}, S.~R., {Graham}, M.~J., {et~al.} 2019, \pasp, 131, 018002, \dodoi{10.1088/1538-3873/aaecbe}

\bibitem[{{Blagorodnova} {et~al.}(2017){Blagorodnova}, {Kotak}, {Polshaw}, {Kasliwal}, {Cao}, {Cody}, {Doran}, {Elias-Rosa}, {Fraser}, {Fremling}, {Gonzalez-Fernandez}, {Harmanen}, {Jencson}, {Kankare}, {Kudritzki}, {Kulkarni}, {Magnier}, {Manulis}, {Masci}, {Mattila}, {Nugent}, {Ochner}, {Pastorello}, {Reynolds}, {Smith}, {Sollerman}, {Taddia}, {Terreran}, {Tomasella}, {Turatto}, {Vreeswijk}, {Wozniak}, \& {Zaggia}}]{Blagorodnova2017}
{Blagorodnova}, N., {Kotak}, R., {Polshaw}, J., {et~al.} 2017, \apj, 834, 107, \dodoi{10.3847/1538-4357/834/2/107}

\bibitem[{{Blagorodnova} {et~al.}(2021){Blagorodnova}, {Klencki}, {Pejcha}, {Vreeswijk}, {Bond}, {Burdge}, {De}, {Fremling}, {Gehrz}, {Jencson}, {Kasliwal}, {Kupfer}, {Lau}, {Masci}, \& {Rich}}]{Blagorodnova2021}
{Blagorodnova}, N., {Klencki}, J., {Pejcha}, O., {et~al.} 2021, \aap, 653, A134, \dodoi{10.1051/0004-6361/202140525}

\bibitem[{{Bloemen} {et~al.}(2016){Bloemen}, {Groot}, {Woudt}, {Klein Wolt}, {McBride}, {Nelemans}, {K{\"o}rding}, {Pretorius}, {Roelfsema}, {Bettonvil}, {Balster}, {Bakker}, {Dolron}, {van Elteren}, {Elswijk}, {Engels}, {Fender}, {Fokker}, {de Haan}, {Hagoort}, {de Hoog}, {ter Horst}, {van der Kevie}, {Koz{\l}owski}, {Kragt}, {Lech}, {Le Poole}, {Lesman}, {Morren}, {Navarro}, {Paalberends}, {Paterson}, {Paw{\l}aszek}, {Pessemier}, {Raskin}, {Rutten}, {Scheers}, {Schuil}, \& {Sybilski}}]{Bloemen16}
{Bloemen}, S., {Groot}, P., {Woudt}, P., {et~al.} 2016, in Society of Photo-Optical Instrumentation Engineers (SPIE) Conference Series, Vol. 9906, Ground-based and Airborne Telescopes VI, ed. H.~J. {Hall}, R.~{Gilmozzi}, \& H.~K. {Marshall}, 990664, \dodoi{10.1117/12.2232522}

\bibitem[{{Burdge} {et~al.}(2019){Burdge}, {Coughlin}, {Fuller}, {Kupfer}, {Bellm}, {Bildsten}, {Graham}, {Kaplan}, {Roestel}, {Dekany}, {Duev}, {Feeney}, {Giomi}, {Helou}, {Kaye}, {Laher}, {Mahabal}, {Masci}, {Riddle}, {Shupe}, {Soumagnac}, {Smith}, {Szkody}, {Walters}, {Kulkarni}, \& {Prince}}]{Burdge19}
{Burdge}, K.~B., {Coughlin}, M.~W., {Fuller}, J., {et~al.} 2019, \nat, 571, 528, \dodoi{10.1038/s41586-019-1403-0}

\bibitem[{{Burdge} {et~al.}(2020){Burdge}, {Coughlin}, {Fuller}, {Kaplan}, {Kulkarni}, {Marsh}, {Bellm}, {Dekany}, {Duev}, {Graham}, {Mahabal}, {Masci}, {Laher}, {Riddle}, {Soumagnac}, \& {Prince}}]{Burdge20}
---. 2020, \apjl, 905, L7, \dodoi{10.3847/2041-8213/abca91}

\bibitem[{{Burdge} {et~al.}(2023){Burdge}, {El-Badry}, {Rappaport}, {Sunny Wong}, {Bauer}, {Bildsten}, {Caiazzo}, {Chakrabarty}, {Chickles}, {Graham}, {Kara}, {Kulkarni}, {Marsh}, {Nynka}, {Prince}, {Simcoe}, {van Roestel}, {Vanderbosch}, {Bellm}, {Dekany}, {Drake}, {Helou}, {Masci}, {Milburn}, {Riddle}, {Rusholme}, \& {Smith}}]{Burdge23}
{Burdge}, K.~B., {El-Badry}, K., {Rappaport}, S., {et~al.} 2023, \apjl, 953, L1, \dodoi{10.3847/2041-8213/ace7cf}

\bibitem[{{Chambers} {et~al.}(2016){Chambers}, {Magnier}, {Metcalfe}, {Flewelling}, {Huber}, {Waters}, {Denneau}, {Draper}, {Farrow}, {Finkbeiner}, {Holmberg}, {Koppenhoefer}, {Price}, {Rest}, {Saglia}, {Schlafly}, {Smartt}, {Sweeney}, {Wainscoat}, {Burgett}, {Chastel}, {Grav}, {Heasley}, {Hodapp}, {Jedicke}, {Kaiser}, {Kudritzki}, {Luppino}, {Lupton}, {Monet}, {Morgan}, {Onaka}, {Shiao}, {Stubbs}, {Tonry}, {White}, {Ba{\~n}ados}, {Bell}, {Bender}, {Bernard}, {Boegner}, {Boffi}, {Botticella}, {Calamida}, {Casertano}, {Chen}, {Chen}, {Cole}, {Deacon}, {Frenk}, {Fitzsimmons}, {Gezari}, {Gibbs}, {Goessl}, {Goggia}, {Gourgue}, {Goldman}, {Grant}, {Grebel}, {Hambly}, {Hasinger}, {Heavens}, {Heckman}, {Henderson}, {Henning}, {Holman}, {Hopp}, {Ip}, {Isani}, {Jackson}, {Keyes}, {Koekemoer}, {Kotak}, {Le}, {Liska}, {Long}, {Lucey}, {Liu}, {Martin}, {Masci}, {McLean}, {Mindel}, {Misra}, {Morganson}, {Murphy}, {Obaika}, {Narayan}, {Nieto-Santisteban}, {Norberg}, {Peacock}, {Pier}, {Postman}, {Primak}, {Rae}, {Rai},
  {Riess}, {Riffeser}, {Rix}, {R{\"o}ser}, {Russel}, {Rutz}, {Schilbach}, {Schultz}, {Scolnic}, {Strolger}, {Szalay}, {Seitz}, {Small}, {Smith}, {Soderblom}, {Taylor}, {Thomson}, {Taylor}, {Thakar}, {Thiel}, {Thilker}, {Unger}, {Urata}, {Valenti}, {Wagner}, {Walder}, {Walter}, {Watters}, {Werner}, {Wood-Vasey}, \& {Wyse}}]{PS1_DR2}
{Chambers}, K.~C., {Magnier}, E.~A., {Metcalfe}, N., {et~al.} 2016, arXiv e-prints, arXiv:1612.05560, \dodoi{10.48550/arXiv.1612.05560}

\bibitem[{{Chase} {et~al.}(2022){Chase}, {O'Connor}, {Fryer}, {Troja}, {Korobkin}, {Wollaeger}, {Ristic}, {Fontes}, {Hungerford}, \& {Herring}}]{Chase22}
{Chase}, E.~A., {O'Connor}, B., {Fryer}, C.~L., {et~al.} 2022, \apj, 927, 163, \dodoi{10.3847/1538-4357/ac3d25}

\bibitem[{{C{\'o}rsico} {et~al.}(2019){C{\'o}rsico}, {Althaus}, {Miller Bertolami}, \& {Kepler}}]{Corsico19}
{C{\'o}rsico}, A.~H., {Althaus}, L.~G., {Miller Bertolami}, M.~M., \& {Kepler}, S.~O. 2019, \aapr, 27, 7, \dodoi{10.1007/s00159-019-0118-4}

\bibitem[{{de Wet} {et~al.}(2021){de Wet}, {Groot}, {Bloemen}, {Le Poole}, {Klein-Wolt}, {K{\"o}rding}, {McBride}, {Paterson}, {Pieterse}, {Vreeswijk}, \& {Woudt}}]{DeWet21}
{de Wet}, S., {Groot}, P.~J., {Bloemen}, S., {et~al.} 2021, \aap, 649, A72, \dodoi{10.1051/0004-6361/202040231}

\bibitem[{{de Wet} {et~al.}(2023){de Wet}, {Izzo}, {Groot}, {Bisero}, {D'Elia}, {De Pasquale}, {Hartmann}, {Heintz}, {Jakobsson}, {Laskar}, {Levan}, {Martin-Carrillo}, {Melandri}, {Nicuesa Guelbenzu}, {Pugliese}, {Rossi}, {Saccardi}, {Savaglio}, {Schady}, {Tanvir}, {van Eerten}, \& {Vergani}}]{DeWet23}
{de Wet}, S., {Izzo}, L., {Groot}, P.~J., {et~al.} 2023, \aap, 677, A32, \dodoi{10.1051/0004-6361/202347017}

\bibitem[{{Dekany} {et~al.}(2020){Dekany}, {Smith}, {Riddle}, {Feeney}, {Porter}, {Hale}, {Zolkower}, {Belicki}, {Kaye}, {Henning}, {Walters}, {Cromer}, {Delacroix}, {Rodriguez}, {Reiley}, {Mao}, {Hover}, {Murphy}, {Burruss}, {Baker}, {Kowalski}, {Reif}, {Mueller}, {Bellm}, {Graham}, \& {Kulkarni}}]{Dekany20}
{Dekany}, R., {Smith}, R.~M., {Riddle}, R., {et~al.} 2020, \pasp, 132, 038001, \dodoi{10.1088/1538-3873/ab4ca2}

\bibitem[{{Dichiara} {et~al.}(2022){Dichiara}, {Gropp}, {Kennea}, {Kuin}, {Lien}, {Marshall}, {Tohuvavohu}, {Williams}, \& {Neil Gehrels Swift Observatory Team}}]{Dichiara22}
{Dichiara}, S., {Gropp}, J.~D., {Kennea}, J.~A., {et~al.} 2022, GRB Coordinates Network, 32632, 1

\bibitem[{{Doi} {et~al.}(2010){Doi}, {Tanaka}, {Fukugita}, {Gunn}, {Yasuda}, {Ivezi{\'c}}, {Brinkmann}, {de Haars}, {Kleinman}, {Krzesinski}, \& {French Leger}}]{Doi10}
{Doi}, M., {Tanaka}, M., {Fukugita}, M., {et~al.} 2010, \aj, 139, 1628, \dodoi{10.1088/0004-6256/139/4/1628}

\bibitem[{{Drew} {et~al.}(2014){Drew}, {Gonzalez-Solares}, {Greimel}, {Irwin}, {K{\"u}pc{\"u} Yoldas}, {Lewis}, {Barentsen}, {Eisl{\"o}ffel}, {Farnhill}, {Martin}, {Walsh}, {Walton}, {Mohr-Smith}, {Raddi}, {Sale}, {Wright}, {Groot}, {Barlow}, {Corradi}, {Drake}, {Fabregat}, {Frew}, {G{\"a}nsicke}, {Knigge}, {Mampaso}, {Morris}, {Naylor}, {Parker}, {Phillipps}, {Ruhland}, {Steeghs}, {Unruh}, {Vink}, {Wesson}, \& {Zijlstra}}]{Drew14}
{Drew}, J.~E., {Gonzalez-Solares}, E., {Greimel}, R., {et~al.} 2014, \mnras, 440, 2036, \dodoi{10.1093/mnras/stu394}

\bibitem[{{Drew} {et~al.}(2016){Drew}, {Gonzales-Solares}, {Greimel}, {Irwin}, {Kupcu Yoldas}, {Lewis}, {Barentsen}, {Eisloffel}, {Farnhill}, {Martin}, {Walsh}, {Walton}, {Mohr-Smith}, {Raddi}, {Sale}, {Wright}, {Groot}, {Barlow}, {Corradi}, {Drake}, {Fabregat}, {Frew}, {Gansicke}, {Knigge}, {Mampaso}, {Morris}, {Naylor}, {Parker}, {Phillipps}, {Ruhland}, {Steeghs}, {Unruh}, {Vink}, {Wesson}, \& {Zijlstra}}]{VPHASDR2}
{Drew}, J.~E., {Gonzales-Solares}, E., {Greimel}, R., {et~al.} 2016, VizieR Online Data Catalog, II/341

\bibitem[{{Drlica-Wagner} {et~al.}(2022){Drlica-Wagner}, {Ferguson}, {Adam{\'o}w}, {Aguena}, {Allam}, {Andrade-Oliveira}, {Bacon}, {Bechtol}, {Bell}, {Bertin}, {Bilaji}, {Bocquet}, {Bom}, {Brooks}, {Burke}, {Carballo-Bello}, {Carlin}, {Carnero Rosell}, {Carrasco Kind}, {Carretero}, {Castander}, {Cerny}, {Chang}, {Choi}, {Conselice}, {Costanzi}, {Crnojevi{\'c}}, {da Costa}, {de Vicente}, {Desai}, {Esteves}, {Everett}, {Ferrero}, {Fitzpatrick}, {Flaugher}, {Friedel}, {Frieman}, {Garc{\'\i}a-Bellido}, {Gatti}, {Gaztanaga}, {Gerdes}, {Gruen}, {Gruendl}, {Gschwend}, {Hartley}, {Hernandez-Lang}, {Hinton}, {Hollowood}, {Honscheid}, {Hughes}, {Jacques}, {James}, {Johnson}, {Kuehn}, {Kuropatkin}, {Lahav}, {Li}, {Lidman}, {Lin}, {March}, {Marshall}, {Mart{\'\i}nez-Delgado}, {Mart{\'\i}nez-V{\'a}zquez}, {Massana}, {Mau}, {McNanna}, {Melchior}, {Menanteau}, {Miller}, {Miquel}, {Mohr}, {Morgan}, {Mutlu-Pakdil}, {Mu{\~n}oz}, {Neilsen}, {Nidever}, {Nikutta}, {Nilo Castellon}, {No{\"e}l}, {Ogando}, {Olsen}, {Pace},
  {Palmese}, {Paz-Chinch{\'o}n}, {Pereira}, {Pieres}, {Plazas Malag{\'o}n}, {Prat}, {Riley}, {Rodriguez-Monroy}, {Romer}, {Roodman}, {Sako}, {Sakowska}, {Sanchez}, {S{\'a}nchez}, {Sand}, {Santana-Silva}, {Santiago}, {Schubnell}, {Serrano}, {Sevilla-Noarbe}, {Simon}, {Smith}, {Soares-Santos}, {Stringfellow}, {Suchyta}, {Suson}, {Tan}, {Tarle}, {Tavangar}, {Thomas}, {To}, {Tollerud}, {Troxel}, {Tucker}, {Varga}, {Vivas}, {Walker}, {Weller}, {Wilkinson}, {Wu}, {Yanny}, {Zaborowski}, {Zenteno}, {Delve Collaboration}, {Des Collaboration}, \& {Astro Data Lab}}]{DELVEDR2}
{Drlica-Wagner}, A., {Ferguson}, P.~S., {Adam{\'o}w}, M., {et~al.} 2022, \apjs, 261, 38, \dodoi{10.3847/1538-4365/ac78eb}

\bibitem[{{Duarte} {et~al.}(2024){Duarte}, {Gonz{\'a}lez-Gait{\'a}n}, {Guti{\'e}rrez}, {Dennefeld}, {Kumar}, {Pessi}, {Zimmerman}, {Anderson}, {Chen}, {Gromadzki}, {Inserra}, {Kankare}, {Bravo}, {Nicholl}, {Yaron}, {Bruch}, {Young}, {Tonry}, {Denneau}, {Weiland}, {Lawrence}, {Siverd}, {Erasmus}, {Koorts}, {Jordan}, {Suc}, {Smartt}, {Smith}, {Srivastav}, {Fulton}, {McCollum}, {Moore}, {Weston}, {Shingles}, {Rhodes}, {Sommer}, {Rest}, \& {Stubbs}}]{Duarte24}
{Duarte}, J., {Gonz{\'a}lez-Gait{\'a}n}, S., {Guti{\'e}rrez}, C., {et~al.} 2024, Transient Name Server AstroNote, 194, 1

\bibitem[{{Fairhurst}(2011)}]{Fairhurst11}
{Fairhurst}, S. 2011, Classical and Quantum Gravity, 28, 105021, \dodoi{10.1088/0264-9381/28/10/105021}

\bibitem[{{Gaia Collaboration} {et~al.}(2023){Gaia Collaboration}, {Vallenari}, {Brown}, {Prusti}, {de Bruijne}, {Arenou}, {Babusiaux}, {Biermann}, {Creevey}, {Ducourant}, {Evans}, {Eyer}, {Guerra}, {Hutton}, {Jordi}, {Klioner}, {Lammers}, {Lindegren}, {Luri}, {Mignard}, {Panem}, {Pourbaix}, {Randich}, {Sartoretti}, {Soubiran}, {Tanga}, {Walton}, {Bailer-Jones}, {Bastian}, {Drimmel}, {Jansen}, {Katz}, {Lattanzi}, {van Leeuwen}, {Bakker}, {Cacciari}, {Casta{\~n}eda}, {De Angeli}, {Fabricius}, {Fouesneau}, {Fr{\'e}mat}, {Galluccio}, {Guerrier}, {Heiter}, {Masana}, {Messineo}, {Mowlavi}, {Nicolas}, {Nienartowicz}, {Pailler}, {Panuzzo}, {Riclet}, {Roux}, {Seabroke}, {Sordo}, {Th{\'e}venin}, {Gracia-Abril}, {Portell}, {Teyssier}, {Altmann}, {Andrae}, {Audard}, {Bellas-Velidis}, {Benson}, {Berthier}, {Blomme}, {Burgess}, {Busonero}, {Busso}, {C{\'a}novas}, {Carry}, {Cellino}, {Cheek}, {Clementini}, {Damerdji}, {Davidson}, {de Teodoro}, {Nu{\~n}ez Campos}, {Delchambre}, {Dell'Oro}, {Esquej},
  {Fern{\'a}ndez-Hern{\'a}ndez}, {Fraile}, {Garabato}, {Garc{\'\i}a-Lario}, {Gosset}, {Haigron}, {Halbwachs}, {Hambly}, {Harrison}, {Hern{\'a}ndez}, {Hestroffer}, {Hodgkin}, {Holl}, {Jan{\ss}en}, {Jevardat de Fombelle}, {Jordan}, {Krone-Martins}, {Lanzafame}, {L{\"o}ffler}, {Marchal}, {Marrese}, {Moitinho}, {Muinonen}, {Osborne}, {Pancino}, {Pauwels}, {Recio-Blanco}, {Reyl{\'e}}, {Riello}, {Rimoldini}, {Roegiers}, {Rybizki}, {Sarro}, {Siopis}, {Smith}, {Sozzetti}, {Utrilla}, {van Leeuwen}, {Abbas}, {{\'A}brah{\'a}m}, {Abreu Aramburu}, {Aerts}, {Aguado}, {Ajaj}, {Aldea-Montero}, {Altavilla}, {{\'A}lvarez}, {Alves}, {Anders}, {Anderson}, {Anglada Varela}, {Antoja}, {Baines}, {Baker}, {Balaguer-N{\'u}{\~n}ez}, {Balbinot}, {Balog}, {Barache}, {Barbato}, {Barros}, {Barstow}, {Bartolom{\'e}}, {Bassilana}, {Bauchet}, {Becciani}, {Bellazzini}, {Berihuete}, {Bernet}, {Bertone}, {Bianchi}, {Binnenfeld}, {Blanco-Cuaresma}, {Blazere}, {Boch}, {Bombrun}, {Bossini}, {Bouquillon}, {Bragaglia}, {Bramante}, {Breedt},
  {Bressan}, {Brouillet}, {Brugaletta}, {Bucciarelli}, {Burlacu}, {Butkevich}, {Buzzi}, {Caffau}, {Cancelliere}, {Cantat-Gaudin}, {Carballo}, {Carlucci}, {Carnerero}, {Carrasco}, {Casamiquela}, {Castellani}, {Castro-Ginard}, {Chaoul}, {Charlot}, {Chemin}, {Chiaramida}, {Chiavassa}, {Chornay}, {Comoretto}, {Contursi}, {Cooper}, {Cornez}, {Cowell}, {Crifo}, {Cropper}, {Crosta}, {Crowley}, {Dafonte}, {Dapergolas}, {David}, {David}, {de Laverny}, {De Luise}, {De March}, {De Ridder}, {de Souza}, {de Torres}, {del Peloso}, {del Pozo}, {Delbo}, {Delgado}, {Delisle}, {Demouchy}, {Dharmawardena}, {Di Matteo}, {Diakite}, {Diener}, {Distefano}, {Dolding}, {Edvardsson}, {Enke}, {Fabre}, {Fabrizio}, {Faigler}, {Fedorets}, {Fernique}, {Fienga}, {Figueras}, {Fournier}, {Fouron}, {Fragkoudi}, {Gai}, {Garcia-Gutierrez}, {Garcia-Reinaldos}, {Garc{\'\i}a-Torres}, {Garofalo}, {Gavel}, {Gavras}, {Gerlach}, {Geyer}, {Giacobbe}, {Gilmore}, {Girona}, {Giuffrida}, {Gomel}, {Gomez}, {Gonz{\'a}lez-N{\'u}{\~n}ez},
  {Gonz{\'a}lez-Santamar{\'\i}a}, {Gonz{\'a}lez-Vidal}, {Granvik}, {Guillout}, {Guiraud}, {Guti{\'e}rrez-S{\'a}nchez}, {Guy}, {Hatzidimitriou}, {Hauser}, {Haywood}, {Helmer}, {Helmi}, {Sarmiento}, {Hidalgo}, {Hilger}, {H{\l}adczuk}, {Hobbs}, {Holland}, {Huckle}, {Jardine}, {Jasniewicz}, {Jean-Antoine Piccolo}, {Jim{\'e}nez-Arranz}, {Jorissen}, {Juaristi Campillo}, {Julbe}, {Karbevska}, {Kervella}, {Khanna}, {Kontizas}, {Kordopatis}, {Korn}, {K{\'o}sp{\'a}l}, {Kostrzewa-Rutkowska}, {Kruszy{\'n}ska}, {Kun}, {Laizeau}, {Lambert}, {Lanza}, {Lasne}, {Le Campion}, {Lebreton}, {Lebzelter}, {Leccia}, {Leclerc}, {Lecoeur-Taibi}, {Liao}, {Licata}, {Lindstr{\o}m}, {Lister}, {Livanou}, {Lobel}, {Lorca}, {Loup}, {Madrero Pardo}, {Magdaleno Romeo}, {Managau}, {Mann}, {Manteiga}, {Marchant}, {Marconi}, {Marcos}, {Marcos Santos}, {Mar{\'\i}n Pina}, {Marinoni}, {Marocco}, {Marshall}, {Martin Polo}, {Mart{\'\i}n-Fleitas}, {Marton}, {Mary}, {Masip}, {Massari}, {Mastrobuono-Battisti}, {Mazeh}, {McMillan}, {Messina}, {Michalik},
  {Millar}, {Mints}, {Molina}, {Molinaro}, {Moln{\'a}r}, {Monari}, {Mongui{\'o}}, {Montegriffo}, {Montero}, {Mor}, {Mora}, {Morbidelli}, {Morel}, {Morris}, {Muraveva}, {Murphy}, {Musella}, {Nagy}, {Noval}, {Oca{\~n}a}, {Ogden}, {Ordenovic}, {Osinde}, {Pagani}, {Pagano}, {Palaversa}, {Palicio}, {Pallas-Quintela}, {Panahi}, {Payne-Wardenaar}, {Pe{\~n}alosa Esteller}, {Penttil{\"a}}, {Pichon}, {Piersimoni}, {Pineau}, {Plachy}, {Plum}, {Poggio}, {Pr{\v{s}}a}, {Pulone}, {Racero}, {Ragaini}, {Rainer}, {Raiteri}, {Rambaux}, {Ramos}, {Ramos-Lerate}, {Re Fiorentin}, {Regibo}, {Richards}, {Rios Diaz}, {Ripepi}, {Riva}, {Rix}, {Rixon}, {Robichon}, {Robin}, {Robin}, {Roelens}, {Rogues}, {Rohrbasser}, {Romero-G{\'o}mez}, {Rowell}, {Royer}, {Ruz Mieres}, {Rybicki}, {Sadowski}, {S{\'a}ez N{\'u}{\~n}ez}, {Sagrist{\`a} Sell{\'e}s}, {Sahlmann}, {Salguero}, {Samaras}, {Sanchez Gimenez}, {Sanna}, {Santove{\~n}a}, {Sarasso}, {Schultheis}, {Sciacca}, {Segol}, {Segovia}, {S{\'e}gransan}, {Semeux}, {Shahaf}, {Siddiqui}, {Siebert},
  {Siltala}, {Silvelo}, {Slezak}, {Slezak}, {Smart}, {Snaith}, {Solano}, {Solitro}, {Souami}, {Souchay}, {Spagna}, {Spina}, {Spoto}, {Steele}, {Steidelm{\"u}ller}, {Stephenson}, {S{\"u}veges}, {Surdej}, {Szabados}, {Szegedi-Elek}, {Taris}, {Taylor}, {Teixeira}, {Tolomei}, {Tonello}, {Torra}, {Torra}, {Torralba Elipe}, {Trabucchi}, {Tsounis}, {Turon}, {Ulla}, {Unger}, {Vaillant}, {van Dillen}, {van Reeven}, {Vanel}, {Vecchiato}, {Viala}, {Vicente}, {Voutsinas}, {Weiler}, {Wevers}, {Wyrzykowski}, {Yoldas}, {Yvard}, {Zhao}, {Zorec}, {Zucker}, \& {Zwitter}}]{GaiaDR3_23}
{Gaia Collaboration}, {Vallenari}, A., {Brown}, A.~G.~A., {et~al.} 2023, \aap, 674, A1, \dodoi{10.1051/0004-6361/202243940}

\bibitem[{{Ghosh} {et~al.}(2016){Ghosh}, {Bloemen}, {Nelemans}, {Groot}, \& {Price}}]{Ghosh16}
{Ghosh}, S., {Bloemen}, S., {Nelemans}, G., {Groot}, P.~J., \& {Price}, L.~R. 2016, \aap, 592, A82, \dodoi{10.1051/0004-6361/201527712}

\bibitem[{{Giammichele} {et~al.}(2018){Giammichele}, {Charpinet}, {Fontaine}, {Brassard}, {Green}, {Van Grootel}, {Bergeron}, {Zong}, \& {Dupret}}]{Giammichele18}
{Giammichele}, N., {Charpinet}, S., {Fontaine}, G., {et~al.} 2018, \nat, 554, 73, \dodoi{10.1038/nature25136}

\bibitem[{{Gompertz} {et~al.}(2020){Gompertz}, {Cutter}, {Steeghs}, {Galloway}, {Lyman}, {Ulaczyk}, {Dyer}, {Ackley}, {Dhillon}, {O'Brien}, {Ramsay}, {Poshyachinda}, {Kotak}, {Nuttall}, {Breton}, {Pall{\'e}}, {Pollacco}, {Thrane}, {Aukkaravittayapun}, {Awiphan}, {Brown}, {Burhanudin}, {Chote}, {Chrimes}, {Daw}, {Duffy}, {Eyles-Ferris}, {Heikkil{\"a}}, {Irawati}, {Kennedy}, {Killestein}, {Levan}, {Littlefair}, {Makrygianni}, {Marsh}, {Mata S{\'a}nchez}, {Mattila}, {Maund}, {McCormac}, {Mkrtichian}, {Mong}, {Mullaney}, {M{\"u}ller}, {Obradovic}, {Rol}, {Sawangwit}, {Stanway}, {Starling}, {Str{\o}m}, {Tooke}, {West}, \& {Wiersema}}]{Gompertz20}
{Gompertz}, B.~P., {Cutter}, R., {Steeghs}, D., {et~al.} 2020, \mnras, 497, 726, \dodoi{10.1093/mnras/staa1845}

\bibitem[{{Groot} {et~al.}(2024){Groot}, {Tranin}, {Van Roestel}, {Zimmerman}, {Faris}, {Wichern}, {Ramsay}, {Stoppa}, {Blagorodnova}, {Bloemen}, {Vreeswijk}, \& {Pieterse}}]{Groot24a}
{Groot}, P., {Tranin}, H., {Van Roestel}, J., {et~al.} 2024, Transient Name Server AstroNote, 255, 1

\bibitem[{{Groot} {et~al.}(2022){Groot}, {Vreeswijk}, {Ter Horst}, {Bloemen}, {Jonker}, {de Wet}, {Malesani}, {Pieterse}, \& {BlackGEM Consortium}}]{Groot22}
{Groot}, P.~J., {Vreeswijk}, P.~M., {Ter Horst}, R., {et~al.} 2022, GRB Coordinates Network, 32678, 1

\bibitem[{{Grossman} {et~al.}(2014){Grossman}, {Korobkin}, {Rosswog}, \& {Piran}}]{Grossman14}
{Grossman}, D., {Korobkin}, O., {Rosswog}, S., \& {Piran}, T. 2014, \mnras, 439, 757, \dodoi{10.1093/mnras/stt2503}

\bibitem[{{Harmer} \& {Wynne}(1976)}]{HarmerWynne76}
{Harmer}, C.~F.~W., \& {Wynne}, C.~G. 1976, \mnras, 177, 25P, \dodoi{10.1093/mnras/177.1.25P}

\bibitem[{{Hermes} {et~al.}(2017){Hermes}, {G{\"a}nsicke}, {Kawaler}, {Greiss}, {Tremblay}, {Gentile Fusillo}, {Raddi}, {Fanale}, {Bell}, {Dennihy}, {Fuchs}, {Dunlap}, {Clemens}, {Montgomery}, {Winget}, {Chote}, {Marsh}, \& {Redfield}}]{Hermes17}
{Hermes}, J.~J., {G{\"a}nsicke}, B.~T., {Kawaler}, S.~D., {et~al.} 2017, \apjs, 232, 23, \dodoi{10.3847/1538-4365/aa8bb5}

\bibitem[{{Horne}(1986)}]{Horne86}
{Horne}, K. 1986, \pasp, 98, 609, \dodoi{10.1086/131801}

\bibitem[{{Hosenie} {et~al.}(2021){Hosenie}, {Bloemen}, {Groot}, {Lyon}, {Scheers}, {Stappers}, {Stoppa}, {Vreeswijk}, {De Wet}, {Wolt}, {K{\"o}rding}, {McBride}, {Le Poole}, {Paterson}, {Pieterse}, \& {Woudt}}]{MeerCrab21}
{Hosenie}, Z., {Bloemen}, S., {Groot}, P., {et~al.} 2021, Experimental Astronomy, 51, 319, \dodoi{10.1007/s10686-021-09757-1}

\bibitem[{{Kasen} {et~al.}(2013){Kasen}, {Badnell}, \& {Barnes}}]{Kasen13}
{Kasen}, D., {Badnell}, N.~R., \& {Barnes}, J. 2013, \apj, 774, 25, \dodoi{10.1088/0004-637X/774/1/25}

\bibitem[{{Kasliwal} {et~al.}(2020){Kasliwal}, {Anand}, {Ahumada}, {Stein}, {Carracedo}, {Andreoni}, {Coughlin}, {Singer}, {Kool}, {De}, {Kumar}, {AlMualla}, {Yao}, {Bulla}, {Dobie}, {Reusch}, {Perley}, {Cenko}, {Bhalerao}, {Kaplan}, {Sollerman}, {Goobar}, {Copperwheat}, {Bellm}, {Anupama}, {Corsi}, {Nissanke}, {Agudo}, {Bagdasaryan}, {Barway}, {Belicki}, {Bloom}, {Bolin}, {Buckley}, {Burdge}, {Burruss}, {Caballero-Garc{\'\i}a}, {Cannella}, {Castro-Tirado}, {Cook}, {Cooke}, {Cunningham}, {Dahiwale}, {Deshmukh}, {Dichiara}, {Duev}, {Dutta}, {Feeney}, {Franckowiak}, {Frederick}, {Fremling}, {Gal-Yam}, {Gatkine}, {Ghosh}, {Goldstein}, {Golkhou}, {Graham}, {Graham}, {Hankins}, {Helou}, {Hu}, {Ip}, {Jaodand}, {Karambelkar}, {Kong}, {Kowalski}, {Khandagale}, {Kulkarni}, {Kumar}, {Laher}, {Li}, {Mahabal}, {Masci}, {Miller}, {Mogotsi}, {Mohite}, {Mooley}, {Mroz}, {Newman}, {Ngeow}, {Oates}, {Patil}, {Pandey}, {Pavana}, {Pian}, {Riddle}, {S{\'a}nchez-Ram{\'\i}rez}, {Sharma}, {Singh}, {Smith}, {Soumagnac}, {Taggart},
  {Tan}, {Tzanidakis}, {Troja}, {Valeev}, {Walters}, {Waratkar}, {Webb}, {Yu}, {Zhang}, {Zhou}, \& {Zolkower}}]{Kasliwal20}
{Kasliwal}, M.~M., {Anand}, S., {Ahumada}, T., {et~al.} 2020, \apj, 905, 145, \dodoi{10.3847/1538-4357/abc335}

\bibitem[{{Kawaguchi} {et~al.}(2020){Kawaguchi}, {Shibata}, \& {Tanaka}}]{Kawaguchi20}
{Kawaguchi}, K., {Shibata}, M., \& {Tanaka}, M. 2020, \apj, 889, 171, \dodoi{10.3847/1538-4357/ab61f6}

\bibitem[{{Koz{\l}owski} {et~al.}(2017){Koz{\l}owski}, {Sybilski}, {Konacki}, {Paw{\l}aszek}, {Ratajczak}, {He{\l}miniak}, \& {Litwicki}}]{Kozlowski17}
{Koz{\l}owski}, S.~K., {Sybilski}, P.~W., {Konacki}, M., {et~al.} 2017, \pasp, 129, 105001, \dodoi{10.1088/1538-3873/aa83aa}

\bibitem[{{Kupfer} {et~al.}(2021){Kupfer}, {Prince}, {van Roestel}, {Bellm}, {Bildsten}, {Coughlin}, {Drake}, {Graham}, {Klein}, {Kulkarni}, {Masci}, {Walters}, {Andreoni}, {Biswas}, {Bradshaw}, {Duev}, {Dekany}, {Guidry}, {Hermes}, {Laher}, \& {Riddle}}]{Kupfer21}
{Kupfer}, T., {Prince}, T.~A., {van Roestel}, J., {et~al.} 2021, \mnras, 505, 1254, \dodoi{10.1093/mnras/stab1344}

\bibitem[{{Kupfer} {et~al.}(2023){Kupfer}, {Korol}, {Littenberg}, {Shah}, {Savalle}, {Groot}, {Marsh}, {Le Jeune}, {Nelemans}, {Petiteau}, {Ramsay}, {Steeghs}, \& {Babak}}]{Kupfer23}
{Kupfer}, T., {Korol}, V., {Littenberg}, T.~B., {et~al.} 2023, arXiv e-prints, arXiv:2302.12719, \dodoi{10.48550/arXiv.2302.12719}

\bibitem[{{Lindegren} {et~al.}(2021){Lindegren}, {Klioner}, {Hern{\'a}ndez}, {Bombrun}, {Ramos-Lerate}, {Steidelm{\"u}ller}, {Bastian}, {Biermann}, {de Torres}, {Gerlach}, {Geyer}, {Hilger}, {Hobbs}, {Lammers}, {McMillan}, {Stephenson}, {Casta{\~n}eda}, {Davidson}, {Fabricius}, {Gracia-Abril}, {Portell}, {Rowell}, {Teyssier}, {Torra}, {Bartolom{\'e}}, {Clotet}, {Garralda}, {Gonz{\'a}lez-Vidal}, {Torra}, {Abbas}, {Altmann}, {Anglada Varela}, {Balaguer-N{\'u}{\~n}ez}, {Balog}, {Barache}, {Becciani}, {Bernet}, {Bertone}, {Bianchi}, {Bouquillon}, {Brown}, {Bucciarelli}, {Busonero}, {Butkevich}, {Buzzi}, {Cancelliere}, {Carlucci}, {Charlot}, {Cioni}, {Crosta}, {Crowley}, {del Peloso}, {del Pozo}, {Drimmel}, {Esquej}, {Fienga}, {Fraile}, {Gai}, {Garcia-Reinaldos}, {Guerra}, {Hambly}, {Hauser}, {Jan{\ss}en}, {Jordan}, {Kostrzewa-Rutkowska}, {Lattanzi}, {Liao}, {Licata}, {Lister}, {L{\"o}ffler}, {Marchant}, {Masip}, {Mignard}, {Mints}, {Molina}, {Mora}, {Morbidelli}, {Murphy}, {Pagani}, {Panuzzo}, {Pe{\~n}alosa
  Esteller}, {Poggio}, {Re Fiorentin}, {Riva}, {Sagrist{\`a} Sell{\'e}s}, {Sanchez Gimenez}, {Sarasso}, {Sciacca}, {Siddiqui}, {Smart}, {Souami}, {Spagna}, {Steele}, {Taris}, {Utrilla}, {van Reeven}, \& {Vecchiato}}]{GaiaEDR3_Astrometry21}
{Lindegren}, L., {Klioner}, S.~A., {Hern{\'a}ndez}, J., {et~al.} 2021, \aap, 649, A2, \dodoi{10.1051/0004-6361/202039709}

\bibitem[{{Macfarlane} {et~al.}(2015){Macfarlane}, {Toma}, {Ramsay}, {Groot}, {Woudt}, {Drew}, {Barentsen}, \& {Eisl{\"o}ffel}}]{Macfarlane15}
{Macfarlane}, S.~A., {Toma}, R., {Ramsay}, G., {et~al.} 2015, \mnras, 454, 507, \dodoi{10.1093/mnras/stv1989}

\bibitem[{{Mandel} \& {Broekgaarden}(2022)}]{Mandel22}
{Mandel}, I., \& {Broekgaarden}, F.~S. 2022, Living Reviews in Relativity, 25, 1, \dodoi{10.1007/s41114-021-00034-3}

\bibitem[{{McCarthy} \& {Petit}(2004)}]{McCarthyPetit2004}
{McCarthy}, D.~D., \& {Petit}, G. 2004, IERS Technical Note, 32, 1

\bibitem[{{Nissanke} {et~al.}(2013){Nissanke}, {Kasliwal}, \& {Georgieva}}]{Nissanke13}
{Nissanke}, S., {Kasliwal}, M., \& {Georgieva}, A. 2013, \apj, 767, 124, \dodoi{10.1088/0004-637X/767/2/124}

\bibitem[{{Onken} {et~al.}(2024){Onken}, {Wolf}, {Bessell}, {Chang}, {Luvaul}, {Tonry}, {White}, \& {Da Costa}}]{SM_DR4}
{Onken}, C.~A., {Wolf}, C., {Bessell}, M.~S., {et~al.} 2024, arXiv e-prints, arXiv:2402.02015, \dodoi{10.48550/arXiv.2402.02015}

\bibitem[{{Petrov} {et~al.}(2022){Petrov}, {Singer}, {Coughlin}, {Kumar}, {Almualla}, {Anand}, {Bulla}, {Dietrich}, {Foucart}, \& {Guessoum}}]{Petrov22}
{Petrov}, P., {Singer}, L.~P., {Coughlin}, M.~W., {et~al.} 2022, \apj, 924, 54, \dodoi{10.3847/1538-4357/ac366d}

\bibitem[{{Qin} {et~al.}(2023){Qin}, {Zhang}, {Bloom}, {Sollerman}, {Zimmerman}, {Irani}, {Schulze}, {Gal-Yam}, {Kasliwal}, {Coughlin}, {Perley}, {Fremling}, \& {Kulkarni}}]{Qin2023arXiv}
{Qin}, Y.-J., {Zhang}, K., {Bloom}, J., {et~al.} 2023, arXiv e-prints, arXiv:2309.10022, \dodoi{10.48550/arXiv.2309.10022}

\bibitem[{{Ramsay} \& {Hakala}(2005)}]{RATS1}
{Ramsay}, G., \& {Hakala}, P. 2005, \mnras, 360, 314, \dodoi{10.1111/j.1365-2966.2005.09035.x}

\bibitem[{{Ramsay} {et~al.}(2006){Ramsay}, {Napiwotzki}, {Hakala}, \& {Lehto}}]{RATS2}
{Ramsay}, G., {Napiwotzki}, R., {Hakala}, P., \& {Lehto}, H. 2006, \mnras, 371, 957, \dodoi{10.1111/j.1365-2966.2006.10728.x}

\bibitem[{{Ramsay} {et~al.}(2018){Ramsay}, {Green}, {Marsh}, {Kupfer}, {Breedt}, {Korol}, {Groot}, {Knigge}, {Nelemans}, {Steeghs}, {Woudt}, \& {Aungwerojwit}}]{Ramsay18}
{Ramsay}, G., {Green}, M.~J., {Marsh}, T.~R., {et~al.} 2018, \aap, 620, A141, \dodoi{10.1051/0004-6361/201834261}

\bibitem[{{Ranaivomanana} {et~al.}(2023){Ranaivomanana}, {Johnston}, {Groot}, {Aerts}, {Lees}, {IJspeert}, {Bloemen}, {Klein-Wolt}, {Woudt}, {K{\"o}rding}, {Le Poole}, \& {Pieterse}}]{Princy23}
{Ranaivomanana}, P., {Johnston}, C., {Groot}, P.~J., {et~al.} 2023, \aap, 672, A69, \dodoi{10.1051/0004-6361/202245560}

\bibitem[{{Raskin} {et~al.}(2016){Raskin}, {Morren}, {Pessemier}, {Bloemen}, {Klein-Wolt}, {Roelfsema}, {Groot}, \& {Aerts}}]{raskin16}
{Raskin}, G., {Morren}, J., {Pessemier}, W., {et~al.} 2016, in Society of Photo-Optical Instrumentation Engineers (SPIE) Conference Series, Vol. 9908, Ground-based and Airborne Instrumentation for Astronomy VI, ed. C.~J. {Evans}, L.~{Simard}, \& H.~{Takami}, 99084L, \dodoi{10.1117/12.2232485}

\bibitem[{{Raskin} {et~al.}(2013){Raskin}, {Bloemen}, {Morren}, {Perez Padilla}, {Prins}, {Pessemier}, {Vandersteen}, {Merges}, {{\O}stensen}, {Van Winckel}, \& {Aerts}}]{raskin13}
{Raskin}, G., {Bloemen}, S., {Morren}, J., {et~al.} 2013, \aap, 559, A26, \dodoi{10.1051/0004-6361/201322471}

\bibitem[{{Rosswog} {et~al.}(2014){Rosswog}, {Korobkin}, {Arcones}, {Thielemann}, \& {Piran}}]{Rosswog14a}
{Rosswog}, S., {Korobkin}, O., {Arcones}, A., {Thielemann}, F.~K., \& {Piran}, T. 2014, \mnras, 439, 744, \dodoi{10.1093/mnras/stt2502}

\bibitem[{{Scaringi} {et~al.}(2015){Scaringi}, {Maccarone}, {Kording}, {Knigge}, {Vaughan}, {Marsh}, {Aranzana}, {Dhillon}, \& {Barros}}]{Scaringi15}
{Scaringi}, S., {Maccarone}, T.~J., {Kording}, E., {et~al.} 2015, Science Advances, 1, e1500686, \dodoi{10.1126/sciadv.1500686}

\bibitem[{{Scaringi} {et~al.}(2022){Scaringi}, {de Martino}, {Buckley}, {Groot}, {Knigge}, {Fratta}, {I{\l}kiewicz}, {Littlefield}, \& {Papitto}}]{Scaringi22}
{Scaringi}, S., {de Martino}, D., {Buckley}, D.~A.~H., {et~al.} 2022, Nature Astronomy, 6, 98, \dodoi{10.1038/s41550-021-01494-x}

\bibitem[{{Setzer} {et~al.}(2023){Setzer}, {Peiris}, {Korobkin}, \& {Rosswog}}]{Setzer23}
{Setzer}, C.~N., {Peiris}, H.~V., {Korobkin}, O., \& {Rosswog}, S. 2023, \mnras, 520, 2829, \dodoi{10.1093/mnras/stad257}

\bibitem[{{Smartt}(2015)}]{Smartt2015PASA}
{Smartt}, S.~J. 2015, \pasa, 32, e016, \dodoi{10.1017/pasa.2015.17}

\bibitem[{{Smith} {et~al.}(2011){Smith}, {Li}, {Silverman}, {Ganeshalingam}, \& {Filippenko}}]{Smith2011MNRAS}
{Smith}, N., {Li}, W., {Silverman}, J.~M., {Ganeshalingam}, M., \& {Filippenko}, A.~V. 2011, \mnras, 415, 773, \dodoi{10.1111/j.1365-2966.2011.18763.x}

\bibitem[{{Srivastav} {et~al.}(2024){Srivastav}, {Smartt}, {Fulton}, {Smith}, {Young}, {Gillanders}, {Stoppa}, {Chen}, \& {Schmidt}}]{Srivastav24}
{Srivastav}, S., {Smartt}, S.~J., {Fulton}, M., {et~al.} 2024, Transient Name Server AstroNote, 265, 1

\bibitem[{{Steeghs} {et~al.}(2022){Steeghs}, {Galloway}, {Ackley}, {Dyer}, {Lyman}, {Ulaczyk}, {Cutter}, {Mong}, {Dhillon}, {O'Brien}, {Ramsay}, {Poshyachinda}, {Kotak}, {Nuttall}, {Pall{\'e}}, {Breton}, {Pollacco}, {Thrane}, {Aukkaravittayapun}, {Awiphan}, {Burhanudin}, {Chote}, {Chrimes}, {Daw}, {Duffy}, {Eyles-Ferris}, {Gompertz}, {Heikkil{\"a}}, {Irawati}, {Kennedy}, {Killestein}, {Kuncarayakti}, {Levan}, {Littlefair}, {Makrygianni}, {Marsh}, {Mata-Sanchez}, {Mattila}, {Maund}, {McCormac}, {Mkrtichian}, {Mullaney}, {Noysena}, {Patel}, {Rol}, {Sawangwit}, {Stanway}, {Starling}, {Str{\o}m}, {Tooke}, {West}, {White}, \& {Wiersema}}]{GOTO22}
{Steeghs}, D., {Galloway}, D.~K., {Ackley}, K., {et~al.} 2022, \mnras, 511, 2405, \dodoi{10.1093/mnras/stac013}

\bibitem[{{ter Horst} {et~al.}(2016){ter Horst}, {Kragt}, {Lesman}, \& {Navarro}}]{terhorst16}
{ter Horst}, R., {Kragt}, J., {Lesman}, D., \& {Navarro}, R. 2016, in Society of Photo-Optical Instrumentation Engineers (SPIE) Conference Series, Vol. 9912, Advances in Optical and Mechanical Technologies for Telescopes and Instrumentation II, ed. R.~{Navarro} \& J.~H. {Burge}, 99121J, \dodoi{10.1117/12.2232348}

\bibitem[{{Toma} {et~al.}(2016){Toma}, {Ramsay}, {Macfarlane}, {Groot}, {Woudt}, {Dhillon}, {Jeffery}, {Marsh}, {Nelemans}, \& {Steeghs}}]{ruxy16}
{Toma}, R., {Ramsay}, G., {Macfarlane}, S., {et~al.} 2016, \mnras, 463, 1099, \dodoi{10.1093/mnras/stw2079}

\bibitem[{{Tonry} {et~al.}(2018){Tonry}, {Denneau}, {Heinze}, {Stalder}, {Smith}, {Smartt}, {Stubbs}, {Weiland}, \& {Rest}}]{ATLAS18}
{Tonry}, J.~L., {Denneau}, L., {Heinze}, A.~N., {et~al.} 2018, \pasp, 130, 064505, \dodoi{10.1088/1538-3873/aabadf}

\bibitem[{{Tranin} {et~al.}(2024){Tranin}, {Roestel}, {Stoppa}, {Groot}, {Pieterse}, {Faris}, {Wichern}, {Ramsay}, {Blagorodnova}, \& {Vreeswijk}}]{Tranin2024}
{Tranin}, H., {Roestel}, J.~V., {Stoppa}, F., {et~al.} 2024, Transient Name Server Discovery Report, 2024-2393, 1

\bibitem[{{Uzundag} {et~al.}(2021){Uzundag}, {Vu{\v{c}}kovi{\'c}}, {N{\'e}meth}, {Miller Bertolami}, {Silvotti}, {Baran}, {Telting}, {Reed}, {Shoaf}, {{\O}stensen}, \& {Sahoo}}]{Uzundag21}
{Uzundag}, M., {Vu{\v{c}}kovi{\'c}}, M., {N{\'e}meth}, P., {et~al.} 2021, \aap, 651, A121, \dodoi{10.1051/0004-6361/202140961}

\bibitem[{{van Roestel} {et~al.}(2019{\natexlab{a}}){van Roestel}, {Groot}, {Kupfer}, {Verbeek}, {van Velzen}, {Bours}, {Nugent}, {Prince}, {Levitan}, {Nissanke}, {Kulkarni}, \& {Laher}}]{jvr19}
{van Roestel}, J., {Groot}, P.~J., {Kupfer}, T., {et~al.} 2019{\natexlab{a}}, \mnras, 484, 4507, \dodoi{10.1093/mnras/stz241}

\bibitem[{{van Roestel} {et~al.}(2019{\natexlab{b}}){van Roestel}, {Groot}, {Kupfer}, {Verbeek}, {van Velzen}, {Bours}, {Nugent}, {Prince}, {Levitan}, {Nissanke}, {Kulkarni}, \& {Laher}}]{VanRoestel19}
---. 2019{\natexlab{b}}, \mnras, 484, 4507, \dodoi{10.1093/mnras/stz241}

\bibitem[{{van Roestel} {et~al.}(2022){van Roestel}, {Kupfer}, {Green}, {Wong}, {Bildsten}, {Burdge}, {Prince}, {Marsh}, {Szkody}, {Fremling}, {Graham}, {Dhillon}, {Littlefair}, {Bellm}, {Coughlin}, {Duev}, {Goldstein}, {Laher}, {Rusholme}, {Riddle}, {Dekany}, \& {Kulkarni}}]{JvR22}
{van Roestel}, J., {Kupfer}, T., {Green}, M.~J., {et~al.} 2022, \mnras, 512, 5440, \dodoi{10.1093/mnras/stab2421}

\bibitem[{{Zackay} {et~al.}(2016){Zackay}, {Ofek}, \& {Gal-Yam}}]{ZOGY16}
{Zackay}, B., {Ofek}, E.~O., \& {Gal-Yam}, A. 2016, \apj, 830, 27, \dodoi{10.3847/0004-637X/830/1/27}

\end{thebibliography}

\appendix 

\section{Filter and total throughput measurements \label{app:filterthroughput}}
\begin{table}
  \caption[]{Relative percentile transmission of the BlackGEM filter curves, in 5nm bins. A full resolution version is provided electronically. \label{tab:filters}}
  \tiny{
  \begin{tabular}{lrrrrrr}
      \hline
$\lambda$~(nm)&   $u_{BG}$&$g_{BG}$&$q_{BG}$&$r_{BG}$&$i_{BG}$&$z_{BG}$\\ \hline
292.25    &   0.000    &   0.000    &   0.000    &   0.000    &   0.000    &   0.000   \\
 297.25    &   0.000    &   0.000    &   0.000    &   0.000    &   0.000    &   0.000   \\
 302.25    &   0.000    &   0.000    &   0.000    &   0.000    &   0.000    &   0.000   \\
 307.25    &   0.000    &   0.000    &   0.000    &   0.000    &   0.000    &   0.000   \\
 312.25    &   0.000    &   0.001    &   0.000    &   0.000    &   0.000    &   0.000   \\
 317.25    &   0.000    &   0.011    &   0.001    &   0.002    &   0.000    &   0.000   \\
 322.25    &   0.000    &   0.000    &   0.000    &   0.000    &   0.000    &   0.000   \\
 327.25    &   0.000    &   0.000    &   0.000    &   0.001    &   0.000    &   0.001   \\
 332.25    &   0.001    &   0.000    &   0.000    &   0.000    &   0.000    &   0.006   \\
 337.25    &   0.002    &   0.000    &   0.000    &   0.000    &   0.000    &   0.001   \\
 342.25    &   0.096    &   0.000    &   0.000    &   0.000    &   0.000    &   0.000   \\
 347.25    &  22.166    &   0.000    &   0.000    &   0.000    &   0.000    &   0.000   \\
 352.25    &  92.958    &   0.000    &   0.000    &   0.000    &   0.000    &   0.000   \\
 357.25    &  96.683    &   0.000    &   0.000    &   0.000    &   0.000    &   0.000   \\
 362.25    &  97.085    &   0.000    &   0.000    &   0.000    &   0.000    &   0.000   \\
 367.25    &  97.545    &   0.000    &   0.000    &   0.000    &   0.000    &   0.001   \\
 372.25    &  97.296    &   0.000    &   0.000    &   0.000    &   0.001    &   0.000   \\
 377.25    &  97.328    &   0.000    &   0.000    &   0.001    &   0.000    &   0.000   \\
 382.25    &  97.208    &   0.000    &   0.000    &   0.000    &   0.000    &   0.000   \\
 387.25    &  96.446    &   0.000    &   0.000    &   0.000    &   0.000    &   0.000   \\
 392.25    &  96.367    &   0.000    &   0.000    &   0.000    &   0.000    &   0.000   \\
 397.25    &  96.642    &   0.000    &   0.000    &   0.000    &   0.000    &   0.000   \\
 402.25    &  96.305    &   0.006    &   0.000    &   0.000    &   0.000    &   0.000   \\
 407.25    &  57.628    &   0.466    &   0.000    &   0.000    &   0.000    &   0.000   \\
 412.25    &   0.030    &  70.832    &   0.000    &   0.000    &   0.000    &   0.001   \\
 417.25    &   0.000    &  96.467    &   0.000    &   0.000    &   0.000    &   0.002   \\
 422.25    &   0.000    &  96.786    &   0.000    &   0.000    &   0.000    &   0.000   \\
 427.25    &   0.000    &  97.458    &   0.000    &   0.000    &   0.000    &   0.000   \\
 432.25    &   0.000    &  97.964    &   0.002    &   0.000    &   0.000    &   0.000   \\
 437.25    &   0.000    &  98.265    &   5.539    &   0.000    &   0.000    &   0.000   \\
 442.25    &   0.001    &  98.545    &  82.212    &   0.000    &   0.000    &   0.000   \\
 447.25    &   0.000    &  98.629    &  97.310    &   0.000    &   0.000    &   0.000   \\
 452.25    &   0.000    &  98.357    &  97.653    &   0.000    &   0.000    &   0.000   \\
 457.25    &   0.000    &  98.402    &  97.549    &   0.000    &   0.000    &   0.000   \\
 462.25    &   0.000    &  98.275    &  97.799    &   0.000    &   0.000    &   0.000   \\
 467.25    &   0.000    &  97.986    &  97.747    &   0.000    &   0.000    &   0.000   \\
 472.25    &   0.000    &  98.032    &  97.708    &   0.000    &   0.000    &   0.001   \\
 477.25    &   0.000    &  98.148    &  97.624    &   0.000    &   0.000    &   0.001   \\
 482.25    &   0.000    &  97.995    &  97.772    &   0.000    &   0.000    &   0.000   \\
 487.25    &   0.000    &  98.326    &  97.856    &   0.000    &   0.000    &   0.000   \\
 492.25    &   0.000    &  98.237    &  97.740    &   0.000    &   0.000    &   0.000   \\
 497.25    &   0.000    &  98.239    &  97.831    &   0.000    &   0.000    &   0.000   \\
 502.25    &   0.000    &  98.556    &  97.965    &   0.000    &   0.000    &   0.000   \\
 507.25    &   0.000    &  98.201    &  97.385    &   0.000    &   0.000    &   0.000   \\
 512.25    &   0.001    &  98.172    &  97.685    &   0.000    &   0.000    &   0.000   \\
 517.25    &   0.000    &  98.363    &  97.062    &   0.000    &   0.000    &   0.000   \\
 522.25    &   0.000    &  98.098    &  97.442    &   0.000    &   0.000    &   0.000   \\
 527.25    &   0.000    &  97.788    &  97.577    &   0.000    &   0.000    &   0.000   \\
 532.25    &   0.000    &  98.153    &  97.882    &   0.000    &   0.000    &   0.000   \\
 537.25    &   0.000    &  98.021    &  97.794    &   0.000    &   0.000    &   0.000   \\
 542.25    &   0.000    &  97.273    &  97.573    &   0.000    &   0.000    &   0.000   \\
 547.25    &   0.000    &  79.142    &  97.722    &   0.000    &   0.000    &   0.000   \\
 552.25    &   0.000    &   1.203    &  97.751    &   0.000    &   0.000    &   0.000   \\
 557.25    &   0.000    &   0.002    &  97.261    &   0.239    &   0.000    &   0.000   \\
 562.25    &   0.000    &   0.001    &  97.608    &  63.827    &   0.000    &   0.000   \\
 567.25    &   0.000    &   0.001    &  97.916    &  98.437    &   0.000    &   0.000   \\
 572.25    &   0.000    &   0.000    &  97.612    &  98.487    &   0.000    &   0.000   \\
 577.25    &   0.000    &   0.000    &  97.233    &  98.939    &   0.000    &   0.000   \\
 582.25    &   0.001    &   0.000    &  97.514    &  98.933    &   0.000    &   0.000   \\
 587.25    &   0.000    &   0.000    &  97.649    &  98.950    &   0.000    &   0.000   \\
 592.25    &   0.000    &   0.000    &  97.833    &  98.979    &   0.000    &   0.000   \\
 597.25    &   0.001    &   0.000    &  97.715    &  98.960    &   0.000    &   0.000   \\
 602.25    &   0.001    &   0.000    &  97.891    &  99.075    &   0.000    &   0.000   \\
 607.25    &   0.002    &   0.000    &  97.905    &  99.080    &   0.000    &   0.000   \\
 612.25    &   0.001    &   0.000    &  97.949    &  99.067    &   0.000    &   0.000   \\
 617.25    &   0.001    &   0.000    &  98.255    &  99.112    &   0.000    &   0.000   \\
 622.25    &   0.001    &   0.000    &  98.371    &  99.110    &   0.000    &   0.000   \\
 627.25    &   0.001    &   0.000    &  98.272    &  99.097    &   0.000    &   0.000   \\
 632.25    &   0.000    &   0.000    &  98.081    &  99.012    &   0.000    &   0.000   \\
 637.25    &   0.001    &   0.000    &  98.117    &  99.029    &   0.000    &   0.000   \\
 642.25    &   0.000    &   0.000    &  98.101    &  98.988    &   0.000    &   0.000   \\
 647.25    &   0.000    &   0.000    &  98.182    &  98.896    &   0.000    &   0.000   \\
 652.25    &   0.000    &   0.000    &  98.467    &  98.854    &   0.000    &   0.000   
  \end{tabular}
 }
 \end{table}    
\addtocounter{table}{-1}
\begin{table}
  \caption[]{, continued}
    \tiny{
    \begin{tabular}{lrrrrrr}
      \hline
      $\lambda$ (nm)&   $u_{BG}$&$g_{BG}$&$q_{BG}$&$r_{BG}$&$i_{BG}$&$z_{BG}$\\ \hline
      657.25    &   0.000    &   0.000    &  98.539    &  98.878    &   0.001    &   0.000   \\
 662.25    &   0.000    &   0.000    &  98.334    &  98.986    &   0.000    &   0.000   \\
 667.25    &   0.001    &   0.000    &  98.127    &  98.974    &   0.000    &   0.000   \\
 672.25    &   0.001    &   0.000    &  98.199    &  98.340    &   0.000    &   0.000   \\
 677.25    &   0.001    &   0.000    &  98.150    &  98.793    &   0.000    &   0.000   \\
 682.25    &   0.000    &   0.000    &  98.123    &  97.319    &   0.002    &   0.000   \\
 687.25    &   0.001    &   0.000    &  98.161    &  57.822    &   7.231    &   0.000   \\
 692.25    &   0.001    &   0.000    &  98.432    &   0.174    &  91.458    &   0.000   \\
 697.25    &   0.001    &   0.000    &  98.174    &   0.001    &  98.784    &   0.000   \\
 702.25    &   0.001    &   0.000    &  97.971    &   0.001    &  99.037    &   0.000   \\
 707.25    &   0.001    &   0.000    &  97.917    &   0.000    &  99.088    &   0.000   \\
 712.25    &   0.002    &   0.000    &  96.086    &   0.000    &  99.115    &   0.000   \\
 717.25    &   0.001    &   0.000    &  85.231    &   0.000    &  99.034    &   0.000   \\
 722.25    &   0.000    &   0.000    &   8.205    &   0.000    &  99.275    &   0.000   \\
 727.25    &   0.001    &   0.000    &   0.024    &   0.000    &  99.307    &   0.000   \\
 732.25    &   0.002    &   0.000    &   0.005    &   0.000    &  99.285    &   0.000   \\
 737.25    &   0.001    &   0.000    &   0.000    &   0.000    &  99.238    &   0.000   \\
 742.25    &   0.001    &   0.000    &   0.000    &   0.000    &  99.343    &   0.001   \\
 747.25    &   0.001    &   0.000    &   0.000    &   0.001    &  99.344    &   0.001   \\
 752.25    &   0.001    &   0.000    &   0.001    &   0.000    &  99.358    &   0.000   \\
 757.25    &   0.001    &   0.000    &   0.005    &   0.000    &  99.293    &   0.000   \\
 762.25    &   0.001    &   0.000    &   0.001    &   0.000    &  99.382    &   0.000   \\
 767.25    &   0.001    &   0.000    &   0.000    &   0.000    &  99.427    &   0.000   \\
 772.25    &   0.001    &   0.000    &   0.000    &   0.002    &  99.395    &   0.000   \\
 777.25    &   0.001    &   0.000    &   0.001    &   0.000    &  99.322    &   0.000   \\
 782.25    &   0.001    &   0.000    &   0.001    &   0.000    &  99.306    &   0.000   \\
 787.25    &   0.001    &   0.000    &   0.000    &   0.001    &  99.344    &   0.000   \\
 792.25    &   0.000    &   0.000    &   0.000    &   0.000    &  99.258    &   0.000   \\
 797.25    &   0.000    &   0.000    &   0.000    &   0.000    &  99.325    &   0.000   \\
 802.25    &   0.000    &   0.000    &   0.000    &   0.000    &  98.956    &   0.000   \\
 807.25    &   0.000    &   0.000    &   0.000    &   0.001    &  99.023    &   0.000   \\
 812.25    &   0.000    &   0.000    &   0.000    &   0.000    &  99.024    &   0.000   \\
 817.25    &   0.000    &   0.000    &   0.000    &   0.000    &  98.978    &   0.000   \\
 822.25    &   0.002    &   0.000    &   0.000    &   0.000    &  98.937    &   0.000   \\
 827.25    &   0.002    &   0.000    &   0.000    &   0.000    &  98.821    &   0.001   \\
 832.25    &   0.002    &   0.000    &   0.000    &   0.000    &  98.248    &   0.020   \\
 837.25    &   0.002    &   0.000    &   0.000    &   0.000    &  92.816    &   4.992   \\
 842.25    &   0.002    &   0.000    &   0.000    &   0.000    &  10.020    &  82.218   \\
 847.25    &   0.001    &   0.000    &   0.000    &   0.000    &   0.031    &  97.964   \\
 852.25    &   0.001    &   0.000    &   0.000    &   0.000    &   0.002    &  99.124   \\
 857.25    &   0.001    &   0.000    &   0.000    &   0.000    &   0.001    &  98.764   \\
 862.25    &   0.000    &   0.000    &   0.000    &   0.000    &   0.001    &  99.076   \\
 867.25    &   0.000    &   0.000    &   0.000    &   0.000    &   0.001    &  98.628   \\
 872.25    &   0.000    &   0.000    &   0.000    &   0.001    &   0.001    &  98.586   \\
 877.25    &   0.000    &   0.000    &   0.000    &   0.001    &   0.001    &  98.683   \\
 882.25    &   0.000    &   0.000    &   0.000    &   0.002    &   0.000    &  99.010   \\
 887.25    &   0.000    &   0.000    &   0.000    &   0.001    &   0.000    &  98.841   \\
 892.25    &   0.000    &   0.000    &   0.000    &   0.000    &   0.000    &  98.842   \\
 897.25    &   0.000    &   0.000    &   0.000    &   0.000    &   0.001    &  98.987   \\
 902.25    &   0.000    &   0.000    &   0.000    &   0.001    &   0.000    &  98.857   \\
 907.25    &   0.000    &   0.000    &   0.000    &   0.000    &   0.000    &  99.079   \\
 912.25    &   0.000    &   0.000    &   0.000    &   0.000    &   0.000    &  99.054   \\
 917.25    &   0.000    &   0.000    &   0.000    &   0.001    &   0.000    &  99.204   \\
 922.25    &   0.000    &   0.000    &   0.000    &   0.001    &   0.000    &  99.068   \\
 927.25    &   0.000    &   0.000    &   0.000    &   0.000    &   0.000    &  99.056   \\
 932.25    &   0.000    &   0.000    &   0.000    &   0.000    &   0.000    &  99.095   \\
 937.25    &   0.000    &   0.000    &   0.000    &   0.000    &   0.000    &  99.234   \\
 942.25    &   0.000    &   0.000    &   0.000    &   0.000    &   0.000    &  99.058   \\
 947.25    &   0.000    &   0.000    &   0.000    &   0.000    &   0.000    &  99.110   \\
 952.25    &   0.001    &   0.000    &   0.000    &   0.000    &   0.000    &  99.154   \\
 957.25    &   0.002    &   0.000    &   0.000    &   0.000    &   0.000    &  99.155   \\
 962.25    &   0.005    &   0.000    &   0.000    &   0.000    &   0.000    &  99.077   \\
 967.25    &   0.004    &   0.000    &   0.000    &   0.000    &   0.000    &  99.118   \\
 972.25    &   0.002    &   0.000    &   0.000    &   0.000    &   0.000    &  98.900   \\
 977.25    &   0.002    &   0.000    &   0.000    &   0.000    &   0.000    &  98.465   \\
 982.25    &   0.001    &   0.000    &   0.000    &   0.000    &   0.000    &  97.255   \\
 987.25    &   0.002    &   0.000    &   0.000    &   0.000    &   0.000    &  88.720   \\
 992.25    &   0.004    &   0.000    &   0.000    &   0.000    &   0.000    &  14.360   \\
 997.25    &   0.007    &   0.000    &   0.000    &   0.000    &   0.000    &   0.158   \\
1002.25    &   0.005    &   0.000    &   0.000    &   0.001    &   0.000    &   0.014   \\
1007.25    &   0.002    &   0.000    &   0.000    &   0.001    &   0.000    &   0.005   \\
1012.25    &   0.001    &   0.000    &   0.000    &   0.003    &   0.000    &   0.002   \\
1017.25    &   0.001    &   0.000    &   0.000    &   0.002    &   0.000    &   0.001   \\
1022.25    &   0.001    &   0.000    &   0.000    &   0.000    &   0.000    &   0.001   \\

\end{tabular}
    }
    \end{table}

\begin{table}
  \caption[]{Relative percentile transmission of the BlackGEM system from the top
    of the atmosphere at zenith, in 5nm bins. A full resolution version is provided electronically. \label{tab:transmission}}
  \tiny{
  \begin{tabular}{lrrrrrr}
      \hline
$\lambda$~(nm)&   $u_{BG}$&$g_{BG}$&$q_{BG}$&$r_{BG}$&$i_{BG}$&$z_{BG}$\\ \hline

   292.250    &   0.000    &   0.000    &   0.000    &   0.000    &   0.000    &   0.000   \\
 297.250    &   0.000    &   0.000    &   0.000    &   0.000    &   0.000    &   0.000   \\
 302.250    &   0.000    &   0.000    &   0.000    &   0.000    &   0.000    &   0.000   \\
 307.250    &   0.000    &   0.000    &   0.000    &   0.000    &   0.000    &   0.000   \\
 312.250    &   0.000    &   0.000    &   0.000    &   0.000    &   0.000    &   0.000   \\
 317.250    &   0.000    &   0.000    &   0.000    &   0.000    &   0.000    &   0.000   \\
 322.250    &   0.000    &   0.000    &   0.000    &   0.000    &   0.000    &   0.000   \\
 327.250    &   0.000    &   0.000    &   0.000    &   0.000    &   0.000    &   0.000   \\
 332.250    &   0.000    &   0.000    &   0.000    &   0.000    &   0.000    &   0.000   \\
 337.250    &   0.000    &   0.000    &   0.000    &   0.000    &   0.000    &   0.000   \\
 342.250    &   0.010    &   0.000    &   0.000    &   0.000    &   0.000    &   0.000   \\
 347.250    &   3.750    &   0.000    &   0.000    &   0.000    &   0.000    &   0.000   \\
 352.250    &  17.390    &   0.000    &   0.000    &   0.000    &   0.000    &   0.000   \\
 357.250    &  20.450    &   0.000    &   0.000    &   0.000    &   0.000    &   0.000   \\
 362.250    &  22.770    &   0.000    &   0.000    &   0.000    &   0.000    &   0.000   \\
 367.250    &  25.070    &   0.000    &   0.000    &   0.000    &   0.000    &   0.000   \\
 372.250    &  27.010    &   0.000    &   0.000    &   0.000    &   0.000    &   0.000   \\
 377.250    &  28.920    &   0.000    &   0.000    &   0.000    &   0.000    &   0.000   \\
 382.250    &  30.770    &   0.000    &   0.000    &   0.000    &   0.000    &   0.000   \\
 387.250    &  32.450    &   0.000    &   0.000    &   0.000    &   0.000    &   0.000   \\
 392.250    &  34.360    &   0.000    &   0.000    &   0.000    &   0.000    &   0.000   \\
 397.250    &  36.370    &   0.000    &   0.000    &   0.000    &   0.000    &   0.000   \\
 402.250    &  38.170    &   0.000    &   0.000    &   0.000    &   0.000    &   0.000   \\
 407.250    &  23.790    &   0.190    &   0.000    &   0.000    &   0.000    &   0.000   \\
 412.250    &   0.010    &  31.040    &   0.000    &   0.000    &   0.000    &   0.000   \\
 417.250    &   0.000    &  43.920    &   0.000    &   0.000    &   0.000    &   0.000   \\
 422.250    &   0.000    &  45.850    &   0.000    &   0.000    &   0.000    &   0.000   \\
 427.250    &   0.000    &  47.870    &   0.000    &   0.000    &   0.000    &   0.000   \\
 432.250    &   0.000    &  49.770    &   0.000    &   0.000    &   0.000    &   0.000   \\
 437.250    &   0.000    &  51.500    &   2.920    &   0.000    &   0.000    &   0.000   \\
 442.250    &   0.000    &  53.060    &  44.320    &   0.000    &   0.000    &   0.000   \\
 447.250    &   0.000    &  54.460    &  53.740    &   0.000    &   0.000    &   0.000   \\
 452.250    &   0.000    &  55.620    &  55.210    &   0.000    &   0.000    &   0.000   \\
 457.250    &   0.000    &  56.810    &  56.330    &   0.000    &   0.000    &   0.000   \\
 462.250    &   0.000    &  57.790    &  57.520    &   0.000    &   0.000    &   0.000   \\
 467.250    &   0.000    &  58.670    &  58.530    &   0.000    &   0.000    &   0.000   \\
 472.250    &   0.000    &  59.550    &  59.360    &   0.000    &   0.000    &   0.000   \\
 477.250    &   0.000    &  60.260    &  59.960    &   0.000    &   0.000    &   0.000   \\
 482.250    &   0.000    &  61.150    &  61.000    &   0.000    &   0.000    &   0.000   \\
 487.250    &   0.000    &  62.230    &  61.950    &   0.000    &   0.000    &   0.000   \\
 492.250    &   0.000    &  62.920    &  62.600    &   0.000    &   0.000    &   0.000   \\
 497.250    &   0.000    &  63.560    &  63.300    &   0.000    &   0.000    &   0.000   \\
 502.250    &   0.000    &  64.170    &  63.790    &   0.000    &   0.000    &   0.000   \\
 507.250    &   0.000    &  64.510    &  63.950    &   0.000    &   0.000    &   0.000   \\
 512.250    &   0.000    &  65.160    &  64.840    &   0.000    &   0.000    &   0.000   \\
 517.250    &   0.000    &  65.790    &  64.920    &   0.000    &   0.000    &   0.000   \\
 522.250    &   0.000    &  66.010    &  65.590    &   0.000    &   0.000    &   0.000   \\
 527.250    &   0.000    &  66.030    &  65.910    &   0.000    &   0.000    &   0.000   \\
 532.250    &   0.000    &  66.540    &  66.360    &   0.000    &   0.000    &   0.000   \\
 537.250    &   0.000    &  66.910    &  66.760    &   0.000    &   0.000    &   0.000   \\
 542.250    &   0.000    &  66.710    &  66.900    &   0.000    &   0.000    &   0.000   \\
 547.250    &   0.000    &  54.490    &  67.330    &   0.000    &   0.000    &   0.000   \\
 552.250    &   0.000    &   0.830    &  67.640    &   0.000    &   0.000    &   0.000   \\
 557.250    &   0.000    &   0.000    &  67.490    &   0.160    &   0.000    &   0.000   \\
 562.250    &   0.000    &   0.000    &  67.740    &  44.310    &   0.000    &   0.000   \\
 567.250    &   0.000    &   0.000    &  67.890    &  68.270    &   0.000    &   0.000   \\
 572.250    &   0.000    &   0.000    &  67.550    &  68.150    &   0.000    &   0.000   \\
 577.250    &   0.000    &   0.000    &  67.470    &  68.650    &   0.000    &   0.000   \\
 582.250    &   0.000    &   0.000    &  68.420    &  69.420    &   0.000    &   0.000   \\
 587.250    &   0.000    &   0.000    &  68.710    &  69.620    &   0.000    &   0.000   \\
 592.250    &   0.000    &   0.000    &  68.620    &  69.440    &   0.000    &   0.000   \\
 597.250    &   0.000    &   0.000    &  68.890    &  69.800    &   0.000    &   0.000   \\
 602.250    &   0.000    &   0.000    &  69.370    &  70.230    &   0.000    &   0.000   \\
 607.250    &   0.000    &   0.000    &  69.720    &  70.540    &   0.000    &   0.000   \\
 612.250    &   0.000    &   0.000    &  70.110    &  70.910    &   0.000    &   0.000   \\
 617.250    &   0.000    &   0.000    &  70.600    &  71.220    &   0.000    &   0.000   \\
 622.250    &   0.000    &   0.000    &  70.830    &  71.350    &   0.000    &   0.000   \\
 627.250    &   0.000    &   0.000    &  70.130    &  70.690    &   0.000    &   0.000   \\
 632.250    &   0.000    &   0.000    &  70.610    &  71.290    &   0.000    &   0.000   \\
 637.250    &   0.000    &   0.000    &  71.370    &  72.010    &   0.000    &   0.000   \\
 642.250    &   0.000    &   0.000    &  71.710    &  72.360    &   0.000    &   0.000   \\
 647.250    &   0.000    &   0.000    &  71.530    &  72.040    &   0.000    &   0.000   \\
 652.250    &   0.000    &   0.000    &  71.990    &  72.280    &   0.000    &   0.000   \\   
  \end{tabular}
 }
 \end{table}    
\addtocounter{table}{-1}
\begin{table}
  \caption[]{, continued}
    \tiny{
    \begin{tabular}{lrrrrrr}
      \hline
      $\lambda$ (nm)&   $u_{BG}$&$g_{BG}$&$q_{BG}$&$r_{BG}$&$i_{BG}$&$z_{BG}$\\ \hline
657.250    &   0.000    &   0.000    &  72.240    &  72.500    &   0.000    &   0.000   \\
 662.250    &   0.000    &   0.000    &  72.430    &  72.890    &   0.000    &   0.000   \\
 667.250    &   0.000    &   0.000    &  72.350    &  72.970    &   0.000    &   0.000   \\
 672.250    &   0.000    &   0.000    &  72.430    &  72.540    &   0.000    &   0.000   \\
 677.250    &   0.000    &   0.000    &  72.390    &  72.880    &   0.000    &   0.000   \\
 682.250    &   0.000    &   0.000    &  72.300    &  71.710    &   0.000    &   0.000   \\
 687.250    &   0.000    &   0.000    &  64.590    &  38.990    &   4.600    &   0.000   \\
 692.250    &   0.000    &   0.000    &  68.980    &   0.120    &  64.220    &   0.000   \\
 697.250    &   0.000    &   0.000    &  71.020    &   0.000    &  71.460    &   0.000   \\
 702.250    &   0.000    &   0.000    &  70.740    &   0.000    &  71.500    &   0.000   \\
 707.250    &   0.000    &   0.000    &  70.930    &   0.000    &  71.780    &   0.000   \\
 712.250    &   0.000    &   0.000    &  69.460    &   0.000    &  71.670    &   0.000   \\
 717.250    &   0.000    &   0.000    &  58.090    &   0.000    &  67.180    &   0.000   \\
 722.250    &   0.000    &   0.000    &   5.350    &   0.000    &  66.960    &   0.000   \\
 727.250    &   0.000    &   0.000    &   0.000    &   0.000    &  66.800    &   0.000   \\
 732.250    &   0.000    &   0.000    &   0.000    &   0.000    &  68.520    &   0.000   \\
 737.250    &   0.000    &   0.000    &   0.000    &   0.000    &  69.360    &   0.000   \\
 742.250    &   0.000    &   0.000    &   0.000    &   0.000    &  69.580    &   0.000   \\
 747.250    &   0.000    &   0.000    &   0.000    &   0.000    &  69.200    &   0.000   \\
 752.250    &   0.000    &   0.000    &   0.000    &   0.000    &  68.790    &   0.000   \\
 757.250    &   0.000    &   0.000    &   0.000    &   0.000    &  65.090    &   0.000   \\
 762.250    &   0.000    &   0.000    &   0.000    &   0.000    &  34.620    &   0.000   \\
 767.250    &   0.000    &   0.000    &   0.000    &   0.000    &  60.020    &   0.000   \\
 772.250    &   0.000    &   0.000    &   0.000    &   0.000    &  66.630    &   0.000   \\
 777.250    &   0.000    &   0.000    &   0.000    &   0.000    &  66.140    &   0.000   \\
 782.250    &   0.000    &   0.000    &   0.000    &   0.000    &  65.550    &   0.000   \\
 787.250    &   0.000    &   0.000    &   0.000    &   0.000    &  64.630    &   0.000   \\
 792.250    &   0.000    &   0.000    &   0.000    &   0.000    &  63.780    &   0.000   \\
 797.250    &   0.000    &   0.000    &   0.000    &   0.000    &  62.980    &   0.000   \\
 802.250    &   0.000    &   0.000    &   0.000    &   0.000    &  61.670    &   0.000   \\
 807.250    &   0.000    &   0.000    &   0.000    &   0.000    &  61.140    &   0.000   \\
 812.250    &   0.000    &   0.000    &   0.000    &   0.000    &  58.710    &   0.000   \\
 817.250    &   0.000    &   0.000    &   0.000    &   0.000    &  54.350    &   0.000   \\
 822.250    &   0.000    &   0.000    &   0.000    &   0.000    &  54.600    &   0.000   \\
 827.250    &   0.000    &   0.000    &   0.000    &   0.000    &  54.910    &   0.000   \\
 832.250    &   0.000    &   0.000    &   0.000    &   0.000    &  54.510    &   0.000   \\
 837.250    &   0.000    &   0.000    &   0.000    &   0.000    &  51.930    &   2.790   \\
 842.250    &   0.000    &   0.000    &   0.000    &   0.000    &   5.590    &  45.470   \\
 847.250    &   0.000    &   0.000    &   0.000    &   0.000    &   0.010    &  53.390   \\
 852.250    &   0.000    &   0.000    &   0.000    &   0.000    &   0.000    &  52.920   \\
 857.250    &   0.000    &   0.000    &   0.000    &   0.000    &   0.000    &  51.690   \\
 862.250    &   0.000    &   0.000    &   0.000    &   0.000    &   0.000    &  50.730   \\
 867.250    &   0.000    &   0.000    &   0.000    &   0.000    &   0.000    &  49.330   \\
 872.250    &   0.000    &   0.000    &   0.000    &   0.000    &   0.000    &  48.080   \\
 877.250    &   0.000    &   0.000    &   0.000    &   0.000    &   0.000    &  46.840   \\
 882.250    &   0.000    &   0.000    &   0.000    &   0.000    &   0.000    &  45.630   \\
 887.250    &   0.000    &   0.000    &   0.000    &   0.000    &   0.000    &  44.140   \\
 892.250    &   0.000    &   0.000    &   0.000    &   0.000    &   0.000    &  42.400   \\
 897.250    &   0.000    &   0.000    &   0.000    &   0.000    &   0.000    &  37.390   \\
 902.250    &   0.000    &   0.000    &   0.000    &   0.000    &   0.000    &  36.540   \\
 907.250    &   0.000    &   0.000    &   0.000    &   0.000    &   0.000    &  35.430   \\
 912.250    &   0.000    &   0.000    &   0.000    &   0.000    &   0.000    &  33.590   \\
 917.250    &   0.000    &   0.000    &   0.000    &   0.000    &   0.000    &  32.400   \\
 922.250    &   0.000    &   0.000    &   0.000    &   0.000    &   0.000    &  32.990   \\
 927.250    &   0.000    &   0.000    &   0.000    &   0.000    &   0.000    &  30.220   \\
 932.250    &   0.000    &   0.000    &   0.000    &   0.000    &   0.000    &  18.960   \\
 937.250    &   0.000    &   0.000    &   0.000    &   0.000    &   0.000    &  19.320   \\
 942.250    &   0.000    &   0.000    &   0.000    &   0.000    &   0.000    &  19.800   \\
 947.250    &   0.000    &   0.000    &   0.000    &   0.000    &   0.000    &  18.770   \\
 952.250    &   0.000    &   0.000    &   0.000    &   0.000    &   0.000    &  18.480   \\
 957.250    &   0.000    &   0.000    &   0.000    &   0.000    &   0.000    &  18.320   \\
 962.250    &   0.000    &   0.000    &   0.000    &   0.000    &   0.000    &  18.530   \\
 967.250    &   0.000    &   0.000    &   0.000    &   0.000    &   0.000    &  19.520   \\
 972.250    &   0.000    &   0.000    &   0.000    &   0.000    &   0.000    &  18.640   \\
 977.250    &   0.000    &   0.000    &   0.000    &   0.000    &   0.000    &  17.110   \\
 982.250    &   0.000    &   0.000    &   0.000    &   0.000    &   0.000    &  16.420   \\
 987.250    &   0.000    &   0.000    &   0.000    &   0.000    &   0.000    &  14.190   \\
 992.250    &   0.000    &   0.000    &   0.000    &   0.000    &   0.000    &   2.160   \\
 997.250    &   0.000    &   0.000    &   0.000    &   0.000    &   0.000    &   0.010   \\
1002.250    &   0.000    &   0.000    &   0.000    &   0.000    &   0.000    &   0.000   \\
1007.250    &   0.000    &   0.000    &   0.000    &   0.000    &   0.000    &   0.000   \\
1012.250    &   0.000    &   0.000    &   0.000    &   0.000    &   0.000    &   0.000   \\
1017.250    &   0.000    &   0.000    &   0.000    &   0.000    &   0.000    &   0.000   \\
1022.250    &   0.000    &   0.000    &   0.000    &   0.000    &   0.000    &   0.000   \\
\end{tabular}
    }
    \end{table}

\end{document}